\documentclass[12pt]{iopart}
\usepackage{url}
\usepackage{dsfont}
\usepackage{amsmath,amssymb}
\usepackage{color}
\usepackage[utf8]{inputenc}
\usepackage{graphicx}
\usepackage{bm}
\usepackage{blkarray}
\usepackage{hyperref}
\begin{document}

\title[Grassmann TRG]{Tensor Renormalization Group for fermions}

\author{
    Shinichiro Akiyama
    \footnote{
        \href{mailto:akiyama@ccs.tsukuba.ac.jp}{akiyama@ccs.tsukuba.ac.jp}\\
        UTCCS-P-152
    }
    ,
}
\address{Center for Computational Sciences, University of Tsukuba, Tsukuba, Ibaraki 305-8577, Japan}
\address{Graduate School of Science, The University of Tokyo, Bunkyo-ku, Tokyo 113-0033, Japan}

\author{Yannick Meurice,}
\address{Department of Physics and Astronomy, The University of Iowa, Iowa City, Iowa 52242, USA}

\author{Ryo Sakai}
\address{\textit{Jij Inc.}, Bunkyo-ku, Tokyo 113-0031, Japan}

\def\rev{Topical Review}

\begin{abstract}
We review the basic ideas of the Tensor Renormalization Group method and show how they can be applied for lattice field theory models involving relativistic fermions and Grassmann variables in arbitrary dimensions. 
We discuss recent progress for entanglement filtering, 
loop optimization, bond-weighting techniques and matrix product decompositions for Grassmann tensor networks. The new methods are tested with two-dimensional Wilson--Majorana fermions and multi-flavor 
Gross--Neveu models. We show that the methods can also be applied to the fermionic Hubbard model in 1+1 and 2+1 dimensions. 
\end{abstract}

\maketitle
\newpage
\contentsline {section}{\numberline {1}Introduction}{3}{section.1}%
\contentsline {section}{\numberline {2}Formalism for the Grassmann TRG}{5}{section.2}%
\contentsline {subsection}{\numberline {2.1}Grassmann tensor network representation}{5}{subsection.2.1}%
\contentsline {subsection}{\numberline {2.2}Exact contraction}{10}{subsection.2.2}%
\contentsline {subsection}{\numberline {2.3}Model-independent notation}{14}{subsection.2.3}%
\contentsline {subsection}{\numberline {2.4}Extension to lattice gauge theories}{16}{subsection.2.4}%
\contentsline {subsection}{\numberline {2.5}Approximate contraction by the Levin-Nave TRG}{20}{subsection.2.5}%
\contentsline {subsection}{\numberline {2.6}Approximate contraction by the Higher-Order TRG}{24}{subsection.2.6}%
\contentsline {subsection}{\numberline {2.7}Practical remarks}{27}{subsection.2.7}%
\contentsline {section}{\numberline {3}Examples of numerical calculations}{30}{section.3}%
\contentsline {subsection}{\numberline {3.1}Wilson--Majorana fermions}{30}{subsection.3.1}%
\contentsline {subsection}{\numberline {3.2}The Schwinger model}{31}{subsection.3.2}%
\contentsline {section}{\numberline {4}Improved TRG methods for fermions}{33}{section.4}%
\contentsline {subsection}{\numberline {4.1}Corner double line structure on tensor network}{33}{subsection.4.1}%
\contentsline {subsection}{\numberline {4.2}Removal of CDL from network}{33}{subsection.4.2}%
\contentsline {subsection}{\numberline {4.3}Bond-weighting technique}{35}{subsection.4.3}%
\contentsline {subsection}{\numberline {4.4}Multilayered tensor network formulations for $N_{f}$-flavor fermions}{38}{subsection.4.4}%
\contentsline {section}{\numberline {5}Relativistic models with fermion interactions}{40}{section.5}%
\contentsline {subsection}{\numberline {5.1}Gross--Neveu model}{40}{subsection.5.1}%
\contentsline {subsection}{\numberline {5.2}QCD in the infinite-coupling limit}{41}{subsection.5.2}%
\contentsline {subsection}{\numberline {5.3}Nambu--Jona-Lasinio model}{43}{subsection.5.3}%
\contentsline {subsection}{\numberline {5.4}$\mathcal {N}=1$ Wess--Zumino model}{44}{subsection.5.4}%
\contentsline {subsection}{\numberline {5.5}Non-abelian lattice gauge theories coupled to fermions}{45}{subsection.5.5}%
\contentsline {section}{\numberline {6}The Hubbard model}{46}{section.6}%
\contentsline {subsection}{\numberline {6.1}(1+1)-dimensional model}{48}{subsection.6.1}%
\contentsline {subsection}{\numberline {6.2}(2+1)-dimensional model}{51}{subsection.6.2}%
\contentsline {section}{\numberline {7}Conclusions}{51}{section.7}%
\contentsline {section}{\numberline {8}Acknowledgments}{52}{section.8}%

\newpage
\section{Introduction}

The renormalization group (RG) approach of lattice models has been crucial to identify their universal critical behavior, construct their phase diagram and understanding their continuum limits. The general idea \cite{leo66,wilson73pr,wilson74rmp} consists in integrating over some of the microscopic degrees of freedom in order to obtain an effective theory with a larger lattice spacing and repeat the procedure until one reaches a macroscopic size. This leads to an RG map connecting effective theories at increasing lattice spacing. The fixed points and relevant directions of the RG maps have universal properties such as similar critical exponents observed in very different microscopic setups (e.g., magnets and solids). 

The generic properties of RG maps are very well-understood  \cite{wilson73pr,wilson74rmp,cardy1996scaling}. However, the numerical implementation of the partial integration over some microscopic degrees of freedom (coarse-graining) can be challenging. This requires to parameterize some ``space of theories" in terms of effective couplings and find numerical methods to calculate the ``new" couplings in terms of the ``old" couplings. This provides the RG map. Simplified RG maps such as various majority rules, bond moving~\cite{Migdal:1975zf,kadanoff76} approximate recursions~\cite{wilson73pr}, hierarchical approximations~\cite{dyson68,baker72,hmreview}, local potential approximations~\cite{berges2000,bervillier2007,bervillier2013}, can be easily implemented numerically and demonstrate the generic validity of the RG approach. However, these approximations provide critical exponents which are different from the exponents of the approximated models and it is difficult to systematically improve such approximations. 

The basic RG ideas were incorporated in variational algorithms designed to construct the ground state wavefunction of Hamiltonians for one-dimensional spatial lattices 
often called the density matrix renormalization group (DMRG) method~\cite{white92,schollwock2005,uli2011}.
This led to the development of tensor network ansatzes
(e.g. matrix product states (MPS)) 
for these wavefunctions. It became clear that the success of the method was linked to its ability to handle efficiently the entanglement entropy in one spatial dimension~\cite{schollwock2005,vidal2007,cirac2009,schollwock2011,orus2013,2015PhRvL.115r0405E,silvi2017,hv2017,montangero2018,ran2020,RevModPhys.93.045003}. 
The tensor technology was also used to reformulate and coarse-grain classical lattice models~\cite{nishinoctm,Levin:2006jai,Gu:2009dr,PhysRevLett.103.160601,Gu:2010yh,Gu:2013gba,2012PhRvB..86d5139X,efratirmp,prb87}. 
This new approach is called the tensor renormalization group (TRG). 
The TRG approach allows us to interpret and quantitatively realize the traditional RG ideas in the language of the tensor network~\cite{2015PhRvL.115r0405E,Lyu:2021qlw,PhysRevB.104.165132,Ueda:2022nur,PhysRevB.108.024413,Huang:2023vwp,guo2023tensor,Ueda:2023ihr}.

Stimulated by a certain number of interdisciplinary workshops (see for instance, Ref.~\cite{Meurice:2011wy}), tensorial methods became of interest to the lattice gauge theory community. MPS were used for the Schwinger model~\cite{Byrnes:2002nv,Banuls:2013jaa,buyens2014,buyens2016,Funcke:2019zna,Dempsey:2022nys,Okuda:2022hsq,Honda:2022edn,Itou:2023img}, non-Abelian gauge theories~\cite{kuhn2015,banuls2017,Hayata:2023pkw,Liu:2023lsr}, and the
$O(3)$ nonlinear sigma model~\cite{bruckmann2018}. Tensor network techniques for lattice gauge theories are also discussed in  Refs.~\cite{celi2014,silvi2016,silvi2019,rico2014,pichler2016,zohar2015b} and reviewed in~\cite{banuls2018,cinchy2019}.  
At the same time, it was shown that character expansions used in the context of the strong coupling expansion~\cite{balian75,drouffe92} could be used for TRG approaches of most models studied in lattice gauge theory~\cite{Liu:2013nsa,pre89,prd89,Zou:2014rha,Akiyama:2023fyk}. 
This led to a new unified way to understand global, local, continuous, and discrete symmetries~\cite{meurice2019,meurice2020} more generally to the development of Tensor Lattice Field Theory (TLFT) recently reviewed in Ref.~\cite{Meurice:2020pxc}.

Fermionic tensor network studies with applications for the Fermi-Hubbard model were discussed in Refs.~\cite{PhysRevA.80.042333,corboz2010}.  
The application of the TRG method to fermionic systems relying on Grassmann numbers was proposed by Gu, Verstraete, and Wen in Refs.~\cite{Gu:2010yh,Gu:2013gba} in the context of the tensor product states (TPS)~\cite{2001PThPh.105..409N,2003PThPh.110..691G} and projected entangled pair states (PEPS)~\cite{verstraete2004}.
TPS and PEPS are known as variational wavefunctions that can efficiently represent the ground state of a gapped local Hamiltonian in higher dimensions.
They extended TPS to fermionic systems and proposed to construct the TPS using the Grassmann variables based on the idea of the path-integral formalism of fermionic systems.
This generic variational wavefunction is referred to as the Grassmann TPS (GTPS).
The TRG, which explicitly includes the Grassmann variables, was devised as a method for approximate contraction of the Grassmann tensor network derived from the GTPS.
Note that Ref.~\cite{Gu:2010yh} showed that the fermionic PEPS (fPEPS) proposed in Ref.~\cite{Kraus:2010zak} can be classified as a special subclass of GTPS. 
fPEPS and Grassmann TPS can be considered as an origin of the current Grassmann TRG approach~\cite{Kraus:2010zak,Gu:2010yh,Gu:2013gba}.
Note that several kinds of higher-dimensional tensor network ansatzes with variational methods, some of which are combined with the Monte Carlo method, have recently been applied to various lattice gauge theories~\cite{Zohar:2015jnb,Zohar:2017yxl,Zapp:2017fcr,Felser:2019xyv,Magnifico:2020bqt,Felser:2020ntd,Emonts:2020drm,Robaina:2020aqh,Emonts:2022yom,Bender:2023gwr,Cataldi:2023xki,Emonts:2023ttz}. For a set of lectures notes on the applications of PEPS in the context of lattice gauge theory see Ref. \cite{emontspepslectures}.

In this \rev, we discuss recent TRG applications for fermionic models. 
The starting point is a path integral involving Grassmann variables. 
We will consider models in Euclidean spacetime and the statistical weights in the path integral will have the generic form ${\rm e}^{-S}$ for some classical action $S$. The first step of the tensorial approach consists in expanding the exponential of the non-local parts of $S$ (attached to links or plaquettes of the lattice) 
in terms of discrete tensor indices and then perform the integration over the original variables. With this procedure, the path integral becomes a sum of product of local tensors with their indices contracted. For bosonic variables, the tensors are just ordinary functions of the coupling that can be collected in a straightforward manner. However, for Grassmann variables, one needs to keep track of the signs resulting from various orderings. As we will show in Sec. \ref{sec:formalism}, this can be handled by introducing auxiliary Grassmann variables which will be incorporated in the tensors. We then proceed to coarse-grain the reformulation of the path integral by performing approximately partial contractions following the early TRG method of Levin-Nave 
(illustrated in Fig. \ref{fig:schematic_levin_nave_trg}) and their HOTRG version (illustrated in Fig. \ref{fig:schematic_hotrg}). In Sec. \ref{sec:examples}, we review simple examples of numerical calculations for free Wilson-Majorana fermions (equivalent to the Ising model) and 1+1 QED (the Schwinger model).

An important aspect of the TRG coarse-graining is that it can in principle be performed exactly~\cite{prb87,Liu:2013nsa}. However, in practice, the computational cost still scales exponentially with the size of the system because despite the partial integration over some of the microscopic degrees of freedom, effective tensors with more indices are generated. The scaling is less severe, but nevertheless exponential. For this reason and also in order to get RG maps relating same size tensors, truncations are necessary. It has been argued that the inverse of the truncation size can be considered as a relevant direction which can interfere with the study of fixed points~\cite{Vanhecke:2019pez}.
It has also been known~\cite{Levin:2006jai,Gu:2009dr} that short-range entanglement can remain present during TRG iterations and generate unphysical fixed points. 
Improvements of this situation for fermionic theories are discussed in Sec. \ref{sec:improvements}. 
We introduce the corner double line (CDL) fixed point and then discuss methods to remove them. 
This includes the tensor network renormalization (TNR)~\cite{2015PhRvL.115r0405E}, the loop-TNR~\cite{yang2017loop}, and the gilt-TNR~\cite{Hauru:2017tne} algorithm. 

Applications for relativistic models are presented in Sec.~\ref{sec:relativistic} which covers the Gross--Neveu model~\cite{Takeda:2014vwa,Akiyama:2023lvr}, QCD at infinite coupling~\cite{Bloch:2022vqz}, the Nambu--Jona-Lasinio model~\cite{Akiyama:2020soe}, the $\mathcal{N}=1$ Wess--Zumino model~\cite{Kadoh:2018hqq} and non-abelian gauge theories with fermion~\cite{Asaduzzaman:2023pyz}. Finally, the Fermi-Hubbard model which has been extensively studied in condensed matter community~\cite{1994RvMP...66..763D,RevModPhys.68.13,Hal_Tasaki_1998,PhysRevX.5.041041,doi:10.1146/annurev-conmatphys-031620-102024,doi:10.1146/annurev-conmatphys-090921-033948,Ostmeyer:2022gwi} is considered in Sec.~\ref{sec:hubbard} following the approach of  Ref.~\cite{Akiyama:2021xxr} in 1+1 dimensions and Ref.~\cite{Akiyama:2021glo} in 2+1 dimensions.

\section{Formalism for the Grassmann TRG}
\label{sec:formalism}

We begin by explaining how to express fermionic path integrals in the language of tensor networks.
Because of the nilpotency of the Grassmann variables, they can be straightforwardly rewritten with finite-dimensional tensors.
We show how to construct the exact contraction between these fundamental tensors, which is necessary to reproduce original path integrals.
However, the exact contraction is usually prohibited in practice because it requires exponentially large computational resources.
One of the promising approaches is the TRG which carries out these contractions approximately.
Although the TRG algorithms are often used to compute the partition functions without the Grassmann variables, we will see that any TRG algorithm can be applied to fermionic path integrals.

\subsection{Grassmann tensor network representation}
\label{subsec:gtn_rep}


For the time being in this section, we use the following simple quadratic model to explain how to derive the Grassmann tensor network representation.
\begin{align}
\label{eq:action_ex}
    S
    =
    -t
    \sum_{n\in\Lambda}
    \sum_{\nu=1}^{2}
    \left[
        \bar{\psi}(n+\hat{\nu})\psi(n)
        +
        \bar{\psi}(n)\psi(n+\hat{\nu})
    \right]
    +m
    \sum_{n}
    \bar{\psi}(n)\psi(n)
    ,
\end{align}
where we assume that $\psi(n)$ and $\bar{\psi}(n)$ are single-component Grassmann fields for simplicity.
We consider the model on a two-dimensional square lattice $\Lambda$ with periodic boundary conditions.
The path integral on the lattice is defined via
\begin{align}
\label{eq:z_ex}
    Z=\int
    \prod_{n}
    {\rm d}\psi(n)
    {\rm d}\bar{\psi}(n)
    ~
    {\rm e}^{-S}
    .
\end{align}
Our goal is to rewrite Eq.~\eqref{eq:z_ex} introducing local tensors.
There are multiple ways to define these local tensors so that the tensor network representation is in general not unique.
This situation is the same as in constructing tensor network representations for theories that do not include fermions.

\begin{figure}[htbp]
    \centering
    \begin{minipage}{0.90\hsize}
        \includegraphics[width=\hsize]{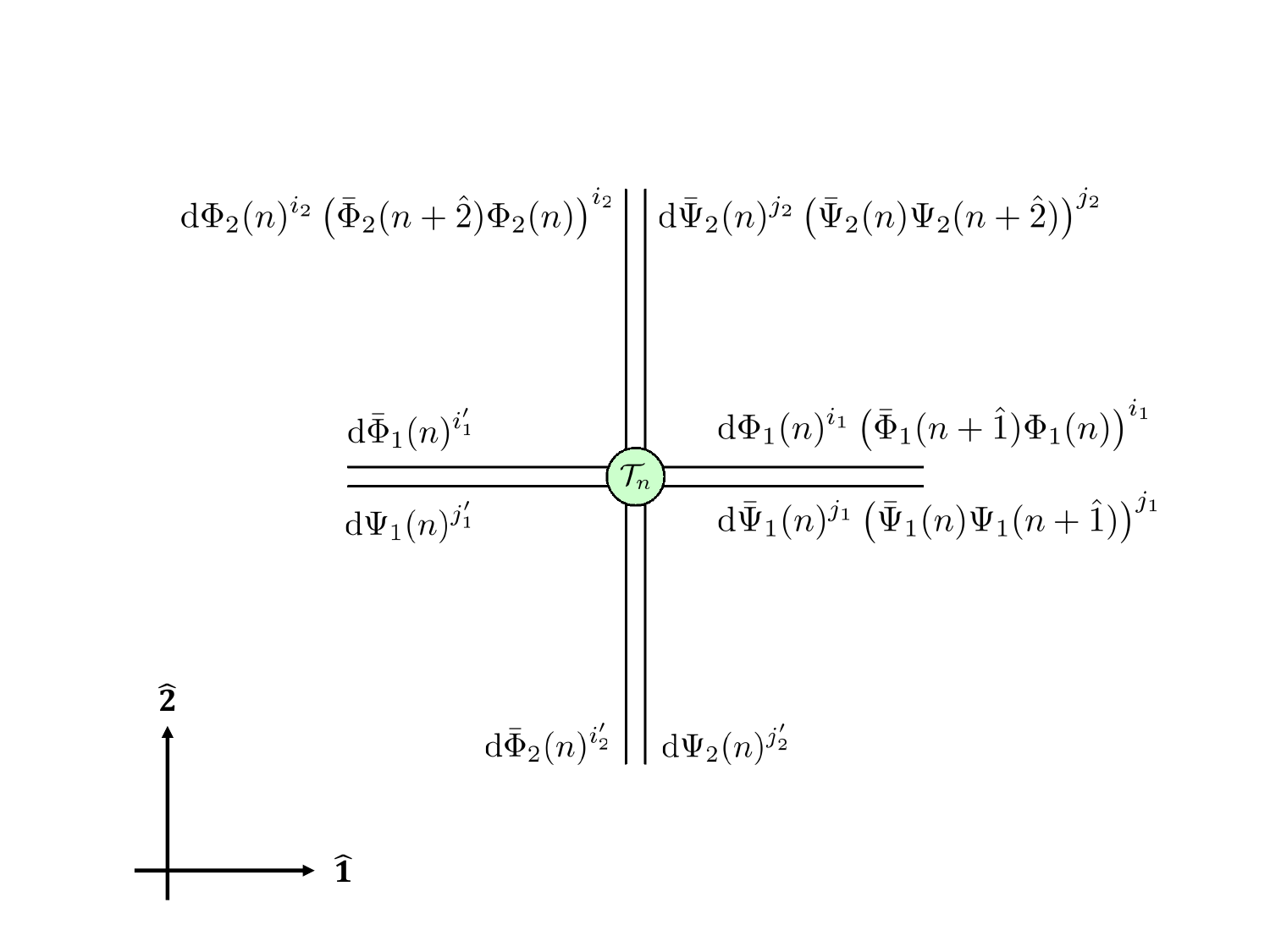}
    \end{minipage}
    \begin{minipage}{0.90\hsize}
        \includegraphics[width=\hsize]{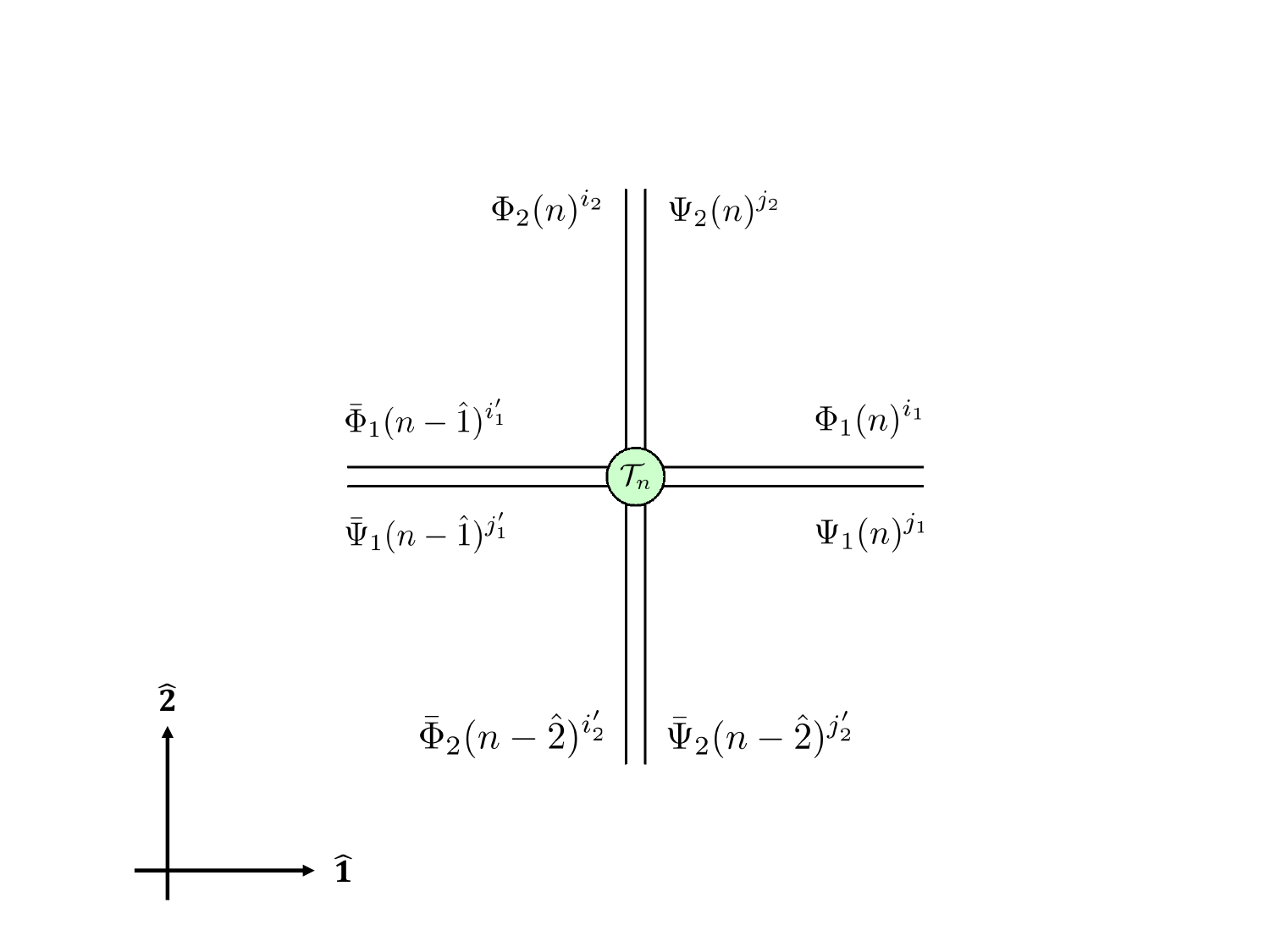}
    \end{minipage}
    \caption{
    Graphical representation of the fundamental tensor $\mathcal{T}_{n}$.
    (Top) The fundamental tensor in Eq.~\eqref{eq:ty_fund}.
    Each line denotes the single-component auxiliary Grassmann measure and the occupation number.
    Grassmann-even factors are associated with forward lines.
    (Bottom) The fundamental tensor in Eq.~\eqref{eq:ak_fund}.
    Each line denotes the single-component auxiliary Grassmann variable.
    }
  \label{fig:fund_tensor}
\end{figure}

One methodology was given by Shimizu and Kuramashi~\cite{Shimizu:2014uva} and subsequently refined by Takeda and Yoshimura by explicitly introducing auxiliary Grassmann variables~\cite{Takeda:2014vwa}.
Here, we review the formalism in Ref.~\cite{Takeda:2014vwa}.
Firstly, the hopping terms are decomposed introducing auxiliary Grassmann variables as
\begin{align}
\label{eq:ty_id_1}
    &{\rm e}^{t\bar{\psi}(n+\hat{\nu})\psi(n)}=
    \nonumber\\
    &\sum_{i_{\nu}(n)=0}^{1}
    \left(\int
        \sqrt{t}\bar{\psi}(n+\hat{\nu}){\rm d}\bar{\Phi}_{\nu}(n+\hat{\nu})
    \right)^{i_{\nu}(n)}
    \left(\int
        \sqrt{t}\psi(n){\rm d}\Phi_{\nu}(n)
    \right)^{i_{\nu}(n)}
    \left(
        \bar{\Phi}_{\nu}(n+\hat{\nu})\Phi_{\nu}(n)
    \right)^{i_{\nu}(n)}
    ,
\end{align}
\begin{align}
\label{eq:ty_id_2}
    &{\rm e}^{t\bar{\psi}(n)\psi(n+\hat{\nu})}=
    \nonumber\\
    &\sum_{j_{\nu}(n)=0}^{1}
    \left(\int 
        \sqrt{t}\bar{\psi}(n){\rm d}\bar{\Psi}_{\nu}(n)
    \right)^{j_{\nu}(n)}
    \left(\int
        \sqrt{t}\psi(n+\hat{\nu}){\rm d}\Psi_{\nu}(n+\hat{\nu})
    \right)^{j_{\nu}(n)}
    \left(
        \bar{\Psi}_{\nu}(n)\Psi_{\nu}(n+\hat{\nu})
    \right)^{j_{\nu}(n)}
    ,
\end{align}
where $\Psi_{\nu}$, $\bar{\Psi}_{\nu}$, $\Phi_{\nu}$, and $\bar{\Phi}_{\nu}$ are single-component Grassmann variables along the $\nu$-directional link.
The bit $i_{\nu}(n)$ ($j_{\nu}(n)$) labels the Taylor expansion of the forward (backward) hopping term on the link $(n,n+\hat{\nu})$.
These bits are nothing but the occupation numbers.
We use capital Greek letters for the auxiliary Grassmann variables 
($\Psi_{\nu}$, $\bar{\Psi}_{\nu}$, etc.) as opposed to the original microscopic Grassman variables which are lowercase ($\psi$, $\bar{\psi}$).
In the rest of this section, in order to have more compact notations, we will drop the $\int$'s in Eqs.~\eqref{eq:ty_id_1} and ~\eqref{eq:ty_id_2}.
In other words, when $i_{\nu}(n)=1$ or $j_{\nu}(n)=1$, the differentials of the auxiliary Grassmann variables such as ${\rm d}\Psi_{\nu}$, should be understood as 
$\int {\rm d}\Psi_{\nu}$.
In the right-hand sides of Eqs.~\eqref{eq:ty_id_1} and \eqref{eq:ty_id_2}, the original fields living on different sites are decomposed into different Grassmann-even pairs.
Therefore, $\psi(n)$ and $\bar{\psi}(n)$ in Eq.~\eqref{eq:z_ex} can be easily integrated at each lattice site $n$ independently.
At each lattice site $n$, we consider the following integral
\begin{align}
\label{eq:ty_pre_tensor}
    &\mathcal{T}_{n;(i_{1}j_{1})(i_{2}j_{2})(i'_{1}j'_{1})(i'_{2}j'_{2})}
    \nonumber\\
    &=
    \int
    {\rm d}\psi
    {\rm d}\bar{\psi}
    ~
    {\rm e}^{-m\bar{\psi}\psi}
    \prod_{\nu}
    \left(
        \sqrt{t}\psi{\rm d}\Phi_{\nu}(n)
    \right)^{i_{\nu}}
    \left(
        \sqrt{t}\bar{\psi}{\rm d}\bar{\Psi}_{\nu}(n)
    \right)^{j_{\nu}}
    \left(
        \sqrt{t}\bar{\psi}{\rm d}\bar{\Phi}_{\nu}(n)
    \right)^{i'_{\nu}}
    \left(
        \sqrt{t}\psi{\rm d}\Psi_{\nu}(n)
    \right)^{j'_{\nu}}
    \nonumber\\
    &\times
    \left(
        \bar{\Phi}_{\nu}(n+\hat{\nu})\Phi_{\nu}(n)
    \right)^{i_{\nu}}
    \left(
        \bar{\Psi}_{\nu}(n)\Psi_{\nu}(n+\hat{\nu})
    \right)^{j_{\nu}}
    ,
\end{align}
where we have introduced several shorthand notations such as $i_{\nu}:=i_{\nu}(n)$, $j_{\nu}:=j_{\nu}(n)$, $i'_{\nu}:=i_{\nu}(n-\hat{\nu})$, and $j'_{\nu}:=j_{\nu}(n-\hat{\nu})$.
We regard Eq.~\eqref{eq:ty_pre_tensor} as a fundamental tensor describing the path integral in Eq.~\eqref{eq:z_ex}.
Eq.~\eqref{eq:ty_pre_tensor} can be written as
\begin{align}
\label{eq:ty_fund}
    \mathcal{T}_{n;(i_{1}j_{1})(i_{2}j_{2})(i'_{1}j'_{1})(i'_{2}j'_{2})}
    =
    T_{n;(i_{1}j_{1})(i_{2}j_{2})(i'_{1}j'_{1})(i'_{2}j'_{2})}
    \mathcal{G}_{n;(i_{1}j_{1})(i_{2}j_{2})(i'_{1}j'_{1})(i'_{2}j'_{2})}
    ,
\end{align}
with
\begin{align}
\label{eq:ty_coeff}
    T_{n;(i_{1}j_{1})(i_{2}j_{2})(i'_{1}j'_{1})(i'_{2}j'_{2})}
    &=
    \sqrt{t}^{\sum_{\nu}(i_{\nu}+j_{\nu}+i'_{\nu}+j'_{\nu})}
    (-1)^{j_{1}(i_{2}+j'_{1}+j'_{2})+j_{2}(j'_{1}+j'_{2})+i'_{1}j'_{2}}
    \nonumber\\
    &\times
    \left[
        -m
        \delta_{i_{1}+i_{2}+j'_{1}+j'_{2},0}
        \delta_{i'_{1}+i'_{2}+j_{1}+j_{2},0}
        +
        \delta_{i_{1}+i_{2}+j'_{1}+j'_{2},1}
        \delta_{i'_{1}+i'_{2}+j_{1}+j_{2},1}
    \right]
    ,
\end{align}
\begin{align}
\label{eq:ty_gpart}
    &\mathcal{G}_{n;(i_{1}j_{1})(i_{2}j_{2})(i'_{1}j'_{1})(i'_{2}j'_{2})}
    \nonumber\\
    &=
    {\rm d}\Phi_{1}(n)^{i_{1}}
    {\rm d}\bar{\Psi}_{1}(n)^{j_{1}}
    {\rm d}\Phi_{2}(n)^{i_{2}}
    {\rm d}\bar{\Psi}_{2}(n)^{j_{2}}
    {\rm d}\Psi_{1}(n)^{j'_{1}}
    {\rm d}\bar{\Phi}_{1}(n)^{i'_{1}}
    {\rm d}\Psi_{2}(n)^{j'_{2}}
    {\rm d}\bar{\Phi}_{2}(n)^{i'_{2}}
    \nonumber\\
    &\times
    \prod_{\nu}
    \left(
        \bar{\Phi}_{\nu}(n+\hat{\nu})\Phi_{\nu}(n)
    \right)^{i_{\nu}}
    \left(
        \bar{\Psi}_{\nu}(n)\Psi_{\nu}(n+\hat{\nu})
    \right)^{j_{\nu}}
    .
\end{align}
The Kronecker deltas in Eq.~\eqref{eq:ty_coeff} imply that 
\begin{align}
    \label{eq:g_even}    
    i_{1}+j_{1}+i_{2}+j_{2}+i'_{1}+j'_{1}+i'_{2}+j'_{2} \mod 2
    =
    0.  
\end{align}
In other words, the fundamental tensor $\mathcal{T}_{n}$ in Eq.~\eqref{eq:ty_fund} is always Grassmann-even.
Following Ref.~\cite{Takeda:2014vwa}, we call $T_{n}$ in Eq.~\eqref{eq:ty_coeff} as the bosonic part of $\mathcal{T}_{n}$ and $\mathcal{G}_{n}$ as the Grassmann part.
Note the order of the Grassmann measures in Eq.~\eqref{eq:ty_gpart}.
Although $\Phi_{\nu}(n)^{i_{\nu}}$ and $\bar{\Psi}_{\nu}(n)^{j_{\nu}}$~($\nu=1,2$) can be integrated within Eq.~\eqref{eq:ty_gpart}, the integration shall not be performed at this stage.
The graphical representation of Eq.~\eqref{eq:ty_fund} is provided in Figure~\ref{fig:fund_tensor} (notice the left-right asymmetry). 
$Z$ in Eq.~\eqref{eq:z_ex} is reproduced by summing over all bits and integrating over all auxiliary Grassmann variables.
This situation is symbolically expressed as
\begin{align}
\label{eq:ty_z}
    Z=
    \sum_{\{i_{1},j_{1},i_{2},j_{2}\}}
    \int\prod_{n}\mathcal{T}_{n}
    ,
\end{align}
where
\begin{align}
    \sum_{\{i_{1},j_{1},i_{2},j_{2}\}}
    =
    \prod_{n,\nu}
    \sum_{i_{\nu}(n)=0}^{1}
    \sum_{j_{\nu}(n)=0}^{1}
    ,
\end{align}
and $\int$'s are over the auxiliary Grassmann variables, which is implicit in Eq.~\eqref{eq:ty_gpart}.

This kind of Grassmann tensor network formulation has been widely applied: the Schwinger model~\cite{Shimizu:2014uva,Shimizu:2014fsa,Shimizu:2017onf}
\footnote{
Although the previous study in Ref.~\cite{Shimizu:2014uva} does not appear to introduce auxiliary Grassmann variables explicitly, it gives a formulation that is equivalent to introducing auxiliary variables.
Consequently, their fundamental tensor in Eq.~(31) in Ref.~\cite{Shimizu:2014uva} has the same structure with Eq.~\eqref{eq:ty_fund}.
}, the Gross--Neveu model~\cite{Takeda:2014vwa,Asaduzzaman:2022pnw}, free Wilson fermions~\cite{Sakai:2017jwp,Yoshimura:2017jpk}, the $\mathcal{N}=1$ Wess--Zumino model~\cite{Kadoh:2018hqq}, the Nambu--Jona-Lasinio model~\cite{Akiyama:2020soe}, Wilson--Majorana fermions~\cite{Asaduzzaman:2022pnw}, infinite-coupling QCD~\cite{Bloch:2022vqz}, and $SU(2)$ lattice gauge theory with reduced staggered fermions~\cite{Asaduzzaman:2023pyz}.

There are alternative ways to formulate the tensor network with Grassmann variables.
Meurice introduced a multilinear combination of Grassman variables to define a fundamental tensor~\cite{Meurice:2018fky} and Bao summarized the contraction, decomposition, and conjugation for the tensors associated with the Grassmann variables~\cite{Bao:2019hfc}.
Akiyama and Kadoh then gave a general methodology to derive the tensor network representation for fermionic path integrals based on the concrete definition of the Grassmann tensor~\cite{Akiyama:2020sfo}.
Here, we follow the formalism provided in Ref.~\cite{Akiyama:2020sfo}.

They also decompose the hopping terms but use
\begin{align}
\label{eq:ak_id_1} 
    {\rm e}^{t\bar{\psi}(n+\hat{\nu})\psi(n)}=
    \int
    {\rm d}\bar{\Phi}_{\nu}(n){\rm d}\Phi_{\nu}(n)
    ~{\rm e}^{-\bar{\Phi}_{\nu}(n)\Phi_{\nu}(n)}
    ~{\rm e}^{\sqrt{t}\bar{\psi}(n+\hat{\nu})\bar{\Phi}_{\nu}(n)}
    ~{\rm e}^{\sqrt{t}\psi(n)\Phi_{\nu}(n)}
    ,
\end{align}
\begin{align}
\label{eq:ak_id_2} 
    {\rm e}^{t\bar{\psi}(n)\psi(n+\hat{\nu})}=
    \int
    {\rm d}\bar{\Psi}_{\nu}(n){\rm d}\Psi_{\nu}(n)
    ~{\rm e}^{-\bar{\Psi}_{\nu}(n)\Psi_{\nu}(n)}
    ~{\rm e}^{\sqrt{t}\bar{\psi}(n)\Psi_{\nu}(n)}
    ~{\rm e}^{-\sqrt{t}\psi(n+\hat{\nu})\bar{\Psi}_{\nu}(n)}
    ,
\end{align}
instead of Eqs.~\eqref{eq:ty_id_1} and \eqref{eq:ty_id_2}.
Again, $\Psi_{\nu}$, $\bar{\Psi}_{\nu}$, $\Phi_{\nu}$, and $\bar{\Phi}_{\nu}$ are the single-component Grassmann variables.
We are now ready to carry out the integral over $\psi(n)$ and $\bar{\psi}(n)$ in Eq.~\eqref{eq:z_ex} at each site $n$ independently as before.
The fundamental tensor is defined via
\begin{align}
\label{eq:ak_pre_tensor}
    \mathcal{T}_{n}
    =
    \int
    {\rm d}\psi
    {\rm d}\bar{\psi}
    ~
    {\rm e}^{-m\bar{\psi}\psi}
    \prod_{\nu}
    {\rm e}^{\sqrt{t}\bar{\psi}\bar{\Phi}_{\nu}(n-\hat{\nu})}
    {\rm e}^{\sqrt{t}\bar{\psi}\Psi_{\nu}(n)}
    {\rm e}^{\sqrt{t}\psi\Phi_{\nu}(n)}
    {\rm e}^{-\sqrt{t}\psi\bar{\Psi}_{\nu}(n-\hat{\nu})}
    .
\end{align}
Expanding the integrand, one obtains
\begin{align}
\label{eq:ak_fund}    
    \mathcal{T}_{n;
        \Phi_{1}\Psi_{1}
        \Phi_{2}\Psi_{2}
        \bar{\Psi}_{1}\bar{\Phi}_{1}
        \bar{\Psi}_{2}\bar{\Phi}_{2}
    }
    =
    \left(
        \prod_{\nu}
        \sum_{i_{\nu},j_{\nu},i'_{\nu},j'_{\nu}}
    \right)
    T_{n;(i_{1}j_{1})(i_{2}j_{2})(i'_{1}j'_{1})(i'_{2}j'_{2})}
    \Phi_{1}^{i_{1}}\Psi_{1}^{j_{1}}
    \Phi_{2}^{i_{2}}\Psi_{2}^{j_{2}}
    \bar{\Psi}_{1}^{j'_{1}}\bar{\Phi}_{1}^{i'_{1}}
    \bar{\Psi}_{2}^{j'_{2}}\bar{\Phi}_{2}^{i'_{2}}
    ,
\end{align}
with
\begin{align}
\label{eq:ak_coeff}
    T_{n;(i_{1}j_{1})(i_{2}j_{2})(i'_{1}j'_{1})(i'_{2}j'_{2})}
    &=
    (-1)^{\sum_{\nu}j'_{\nu}}
    \sqrt{t}^{\sum_{\nu}(i_{\nu}+j_{\nu}+i'_{\nu}+j'_{\nu})}
    (-1)^{
        j_{1}(i_{2}+j'_{1}+j'_{2})
        +j_{2}(j'_{1}+j'_{2})
        +i'_{1}j'_{2}
    }
    \nonumber\\
    &\times
    \left[
        -m
        \delta_{i_{1}+i_{2}+j'_{1}+j'_{2},0}
        \delta_{i'_{1}+i'_{2}+j_{1}+j_{2},0}
        +
        \delta_{i_{1}+i_{2}+j'_{1}+j'_{2},1}
        \delta_{i'_{1}+i'_{2}+j_{1}+j_{2},1}
    \right]
    .
\end{align}
As for Eq.~\eqref{eq:ty_coeff}, Eq.~\eqref{eq:ak_coeff} shows explicitly that the fundamental tensor in Eq.~\eqref{eq:ak_fund} is Grassmann-even.
Compared with the previous formulation based on Ref.~\cite{Takeda:2014vwa}, 
all the bits introduced by the Taylor expansion are summed within the fundamental tensor $\mathcal{T}_{n}$.
Instead, we can identify the auxiliary Grassmann variables as the indices of $\mathcal{T}_{n}$.
This is why we have introduced the notation as in the left-hand side of Eq.~\eqref{eq:ak_fund}.
Following Ref.~\cite{Akiyama:2020sfo}, we refer $\mathcal{T}_{n}$ in Eq.~\eqref{eq:ak_fund} as the Grassmann tensor and $T_{n}$ in Eq.~\eqref{eq:ak_coeff} as the coefficient tensor of $\mathcal{T}_{n}$.
The difference between the bosonic part in Eq.~\eqref{eq:ty_coeff} and the coefficient tensor in Eq.~\eqref{eq:ak_coeff} is just the sign factor $(-1)^{\sum_{\nu}j'_{\nu}}$ in Eq.~\eqref{eq:ak_coeff}, which originates from the negative sign in the last exponential factor in Eq.~\eqref{eq:ak_id_2}.
The graphical expression of Eq.~\eqref{eq:ak_fund} is shown in Figure~\ref{fig:fund_tensor}.
The path integral is reproduced from the fundamental tensor $\mathcal{T}_{n}$ in Eq.~\eqref{eq:ak_fund}.
Introducing the following abbreviation,
\begin{align}
\label{eq:def_ak_int}
    \int_{\bar{\Phi},\Phi}
    =
    \int
    {\rm d}\bar{\Phi}
    {\rm d}\Phi
    ~{\rm e}^{-\bar{\Phi}\Phi}
    ,
\end{align}
and
\begin{align}
\label{eq:ak_gtr}
    {\rm gTr}[~\cdot~]
    =
    \left(
        \prod_{n,\nu}
        \int_{\bar{\Phi}_{\nu}(n),\Phi_{\nu}(n)}
        \int_{\bar{\Psi}_{\nu}(n),\Psi_{\nu}(n)}
    \right)
    [~\cdot~]
    ,
\end{align}
$Z$ in Eq.~\eqref{eq:z_ex} is given by
\begin{align}
\label{eq:ak_z}
    Z
    =
    {\rm gTr}
    \left[
        \prod_{n}
        \mathcal{T}_{n}
    \right]
    .
\end{align}
The symbol ``gTr" in Eq.~\eqref{eq:ak_gtr} means the Grassmann tensor trace that is analogous to the tensor trace symbol ``tTr" common in the tensor network formulation in the spin systems.

This formulation has been applied recently for the free Wilson and staggered fermions~\cite{Akiyama:2020sfo}, Hubbard models~\cite{Akiyama:2021xxr,Akiyama:2021glo}, $N_{f}=1,2,3$ Gross--Neveu model~\cite{Akiyama:2022pse,Akiyama:2023lvr}, $\mathbb{Z}_{n}$ and $U(1)$ gauge theories with $N_{f}=1,2,4$ Wilson fermions~\cite{Yosprakob:2023tyr}, and several public codes for the Grassmann TRG methods~\cite{Akiyama:2023rih,Yosprakob:2023flr}.

\subsection{Exact contraction}
\label{subsec:exact_contraction}

\begin{figure}[htbp]
    \centering
    \begin{minipage}{0.90\hsize}
        \includegraphics[width=\hsize]{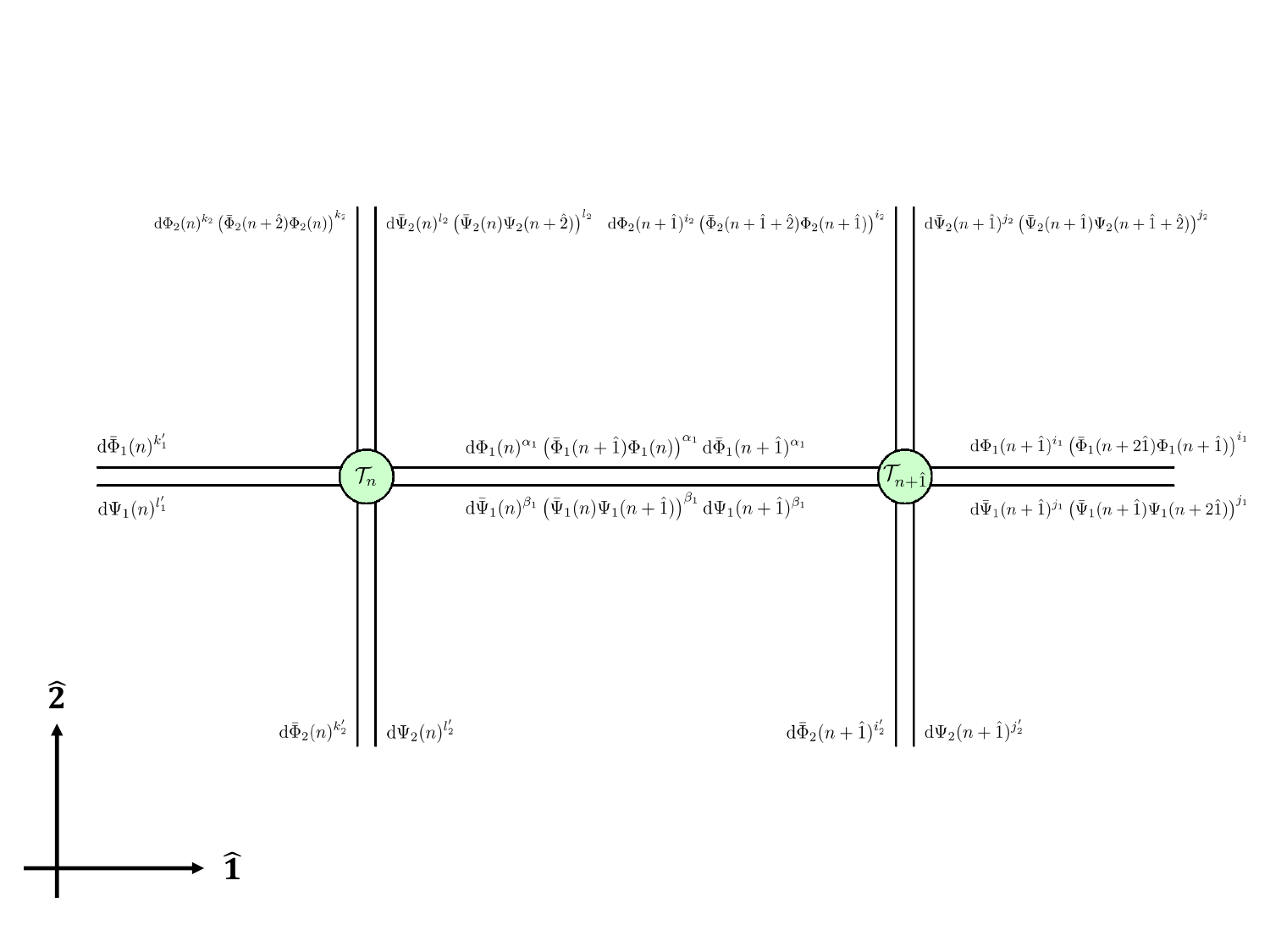}
    \end{minipage}
    \begin{minipage}{0.90\hsize}
        \includegraphics[width=\hsize]{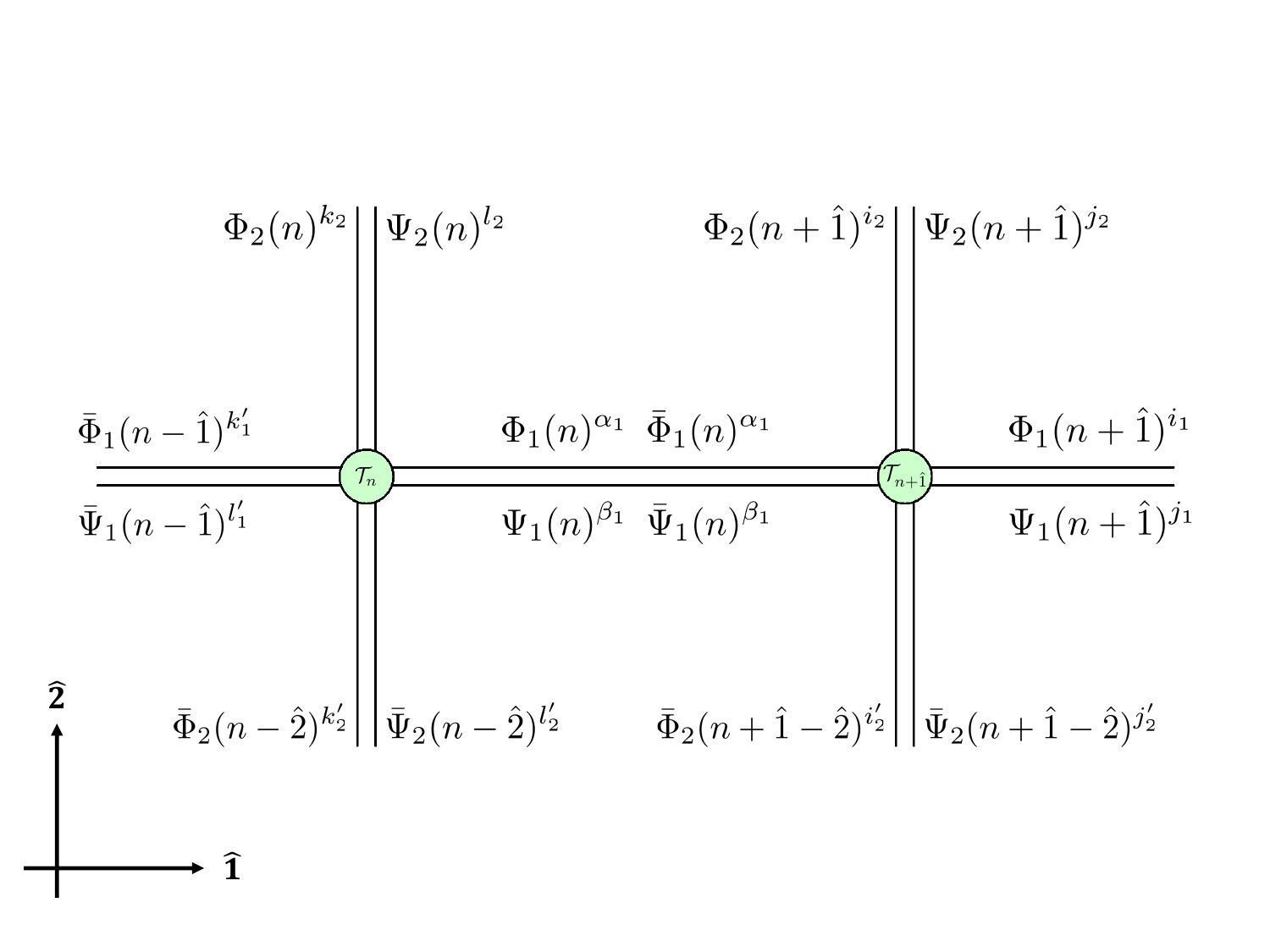}
    \end{minipage}
    \caption{
    Graphical representation of the exact contraction between $\mathcal{T}_{n+\hat{1}}$ and $\mathcal{T}_{n}$.
    (Top) Illustration of Eq.~\eqref{eq:ty_exact_cont}. Internal lines denote the integration over the auxiliary Grassmann variables and summation over $\alpha_{1}$ and $\beta_{1}$.
    (Bottom) Illustration of Eq.~\eqref{eq:ak_exact_cont}. Internal lines denote the weighted integrals $\int_{\bar{\Phi}_{1}(n),\Phi_{1}(n)}$ and $\int_{\bar{\Psi}_{1}(n),\Psi_{1}(n)}$ defined by Eq.~\eqref{eq:def_ak_int}.
    We have used the common Greek indices for the contracted auxiliary Grassmann variables because of Eq.~\eqref{eq:id_delta}.
    }
  \label{fig:contraciton}
\end{figure}

Let us figure out how to carry out the contraction between the fundamental tensors derived above.
As an example, we consider the contraction between $\mathcal{T}_{n}$ and $\mathcal{T}_{n+\hat{1}}$, which reproduces the hopping terms on the link $(n,n+\hat{1})$.
When we use the fundamental tensor given in Eq.~\eqref{eq:ty_fund}, the contraction is defined by
\begin{align}
\label{eq:ty_exact_cont}
    \sum\int
    \mathcal{T}_{n+\hat{1}}
    \mathcal{T}_{n}
    &=
    \sum_{\alpha_{1},\beta_{1}}
    T_{n+\hat{1};(i_{1}j_{1})(i_{2}j_{2})(\alpha_{1}\beta_{1})(i'_{2}j'_{2})}
    T_{n;(\alpha_{1}\beta_{1})(k_{2}l_{2})(k'_{1}l'_{1})(k'_{2}l'_{2})}
    \nonumber\\
    &\times
    \int
    \mathcal{G}_{n+\hat{1};(i_{1}j_{1})(i_{2}j_{2})(\alpha_{1}\beta_{1})(i'_{2}j'_{2})}
    \mathcal{G}_{n;(\alpha_{1}\beta_{1})(k_{2}l_{2})(k'_{1}l'_{1})(k'_{2}l'_{2})}
    .
\end{align}
Note that $\int$ on the right-hand side means the integration over the auxiliary Grassmann variables labeled by repeated Greek indices.
Firstly, we integrate out $(\bar{\Phi}_{1}(n+\hat{1})\Phi_{1}(n))^{\alpha_{1}}$ and $(\bar{\Psi}_{1}(n)\Psi_{1}(n+\hat{1}))^{\beta_{1}}$ originating from $\mathcal{G}_{n}$, which results in
\begin{align}
\label{eq:ty_gcont}
    &\int
    \mathcal{G}_{n+\hat{1};(i_{1}j_{1})(i_{2}j_{2})(\alpha_{1}\beta_{1})(i'_{2}j'_{2})}
    \mathcal{G}_{n;(\alpha_{1}\beta_{1})(k_{2}l_{2})(k'_{1}l'_{1})(k'_{2}l'_{2})}
    \nonumber\\
    &=
    (-1)^{(\alpha_{1}+\beta_{1})(i'_{2}+j'_{2})+\alpha_{1}}
    \mathcal{Q}_{
        (i_{1}j_{1})
        (i_{2}j_{2})(i'_{2}j'_{2})
        (k_{2}l_{2})
        (k'_{1}l'_{1})(k'_{2}l'_{2})
    },
\end{align}
with
\begin{align}
\label{eq:ty_new_g}
    \mathcal{Q}_{
        (i_{1}j_{1})
        (i_{2}j_{2})(i'_{2}j'_{2})
        (k_{2}l_{2})
        (k'_{1}l'_{1})(k'_{2}l'_{2})
    }
    &=
    {\rm d}\Phi_{1}^{i_{1}}{\rm d}\bar{\Phi}_{1}^{j_{1}}
    {\rm d}\Psi_{2}^{i_{2}}{\rm d}\bar{\Psi}_{2}^{j_{2}}
    {\rm d}\Psi_{2}^{j'_{2}}{\rm d}\bar{\Phi}_{2}^{i'_{2}}
    {\rm d}\Psi_{2}^{k_{2}}{\rm d}\bar{\Psi}_{2}^{l_{2}}
    {\rm d}\Psi_{1}^{l'_{1}}{\rm d}\bar{\Phi}_{1}^{k'_{1}}
    {\rm d}\Psi_{2}^{l'_{2}}{\rm d}\bar{\Phi}_{2}^{k'_{2}}
    \nonumber\\
    &\times
    \left(
        \bar{\Phi}_{1}\Phi_{1}
    \right)^{i_{1}}
    \left(
        \bar{\Psi}_{1}\Psi_{1}
    \right)^{j_{1}}
    \left(
        \bar{\Phi}_{2}\Phi_{2}
    \right)^{i_{2}}
    \left(
        \bar{\Psi}_{2}\Psi_{2}
    \right)^{j_{2}}
    \left(
        \bar{\Phi}_{2}\Phi_{2}
    \right)^{k_{2}}
    \left(
        \bar{\Psi}_{2}\Psi_{2}
    \right)^{l_{2}}
    .
\end{align}
We have omitted the site dependence in the auxiliary Grassmann variables in Eq.~\eqref{eq:ty_new_g} because it can be read from the bit indices immediately.
The sign in Eq.~\eqref{eq:ty_gcont} is then taken into account by contracting the bosonic parts.
As a result of Eq.~\eqref{eq:ty_exact_cont}, we obtain
\begin{align}
\label{eq:ty_result_1}
    \sum\int
    \mathcal{T}_{n+\hat{1}}
    \mathcal{T}_{n}
    &=
    M_{
        (i_{1}j_{1})(i_{2}j_{2})(i'_{2}j'_{2})
        (k_{2}l_{2})(k'_{1}l'_{1})(k'_{2}l'_{2})
    }
    \mathcal{Q}_{
        (i_{1}j_{1})
        (i_{2}j_{2})(i'_{2}j'_{2})
        (k_{2}l_{2})
        (k'_{1}l'_{1})(k'_{2}l'_{2})
    }
    ,
\end{align}
where
\begin{align}
\label{eq:ty_t_p}
    &M_{
        (i_{1}j_{1})(i_{2}j_{2})(i'_{2}j'_{2})
        (k_{2}l_{2})(k'_{1}l'_{1})(k'_{2}l'_{2})
    }
    \nonumber\\
    &=
    \sum_{\alpha_{1},\beta_{1}}
    (-1)^{(\alpha_{1}+\beta_{1})(i'_{2}+j'_{2})+\alpha_{1}}
    T_{n+\hat{1};(i_{1}j_{1})(i_{2}j_{2})(\alpha_{1}\beta_{1})(i'_{2}j'_{2})}
    T_{n;(\alpha_{1}\beta_{1})(k_{2}l_{2})(k'_{1}l'_{1})(k'_{2}l'_{2})}
    .
\end{align}
It is important to realize that the auxiliary Grassmann variables should be integrated before the bosonic parts are contracted when we use the fundamental tensor given in Eq.~\eqref{eq:ty_fund}.
Figure~\ref{fig:contraciton} illustrates the exact contraction in Eq.~\eqref{eq:ty_exact_cont},  explicitly showing the site dependence of auxiliary Grassmann variables.

Note that the sign factor in Eq.~\eqref{eq:ty_gcont} can be further simplified by rearranging the Grassmann measures in Eq.~\eqref{eq:ty_new_g}.
Instead of Eq.~\eqref{eq:ty_result_1}, one can find
\begin{align}
    \sum\int
    \mathcal{T}_{n+\hat{1}}
    \mathcal{T}_{n}
    &=
    M'_{
        (i_{1}j_{1})
        (i_{2}j_{2})(k_{2}l_{2})
        (k'_{1}l'_{1})
        (i'_{2}j'_{2})(k'_{2}l'_{2})
    }
    \mathcal{Q}'_{
        (i_{1}j_{1})
        (i_{2}j_{2})(k_{2}l_{2})
        (k'_{1}l'_{1})
        (k'_{2}l'_{2})(i'_{2}j'_{2})
    }
    ,
\end{align}
where
\begin{align}
\label{eq:ty_t_d}
    M'_{
        (i_{1}j_{1})
        (i_{2}j_{2})(k_{2}l_{2})
        (k'_{1}l'_{1})
        (i'_{2}j'_{2})(k'_{2}l'_{2})
    }
    =
    \sum_{\alpha_{1},\beta_{1}}
    (-1)^{\alpha_{1}}
    T_{n+\hat{1};(i_{1}j_{1})(i_{2}j_{2})(\alpha_{1}\beta_{1})(i'_{2}j'_{2})}
    T_{n;(\alpha_{1}\beta_{1})(k_{2}l_{2})(k'_{1}l'_{1})(k'_{2}l'_{2})}
    ,
\end{align}
\begin{align}
\label{eq:g_part_d}
    \mathcal{Q}'_{
        (i_{1}j_{1})
        (i_{2}j_{2})(k_{2}l_{2})
        (k'_{1}l'_{1})
        (k'_{2}l'_{2})(i'_{2}j'_{2})
    }
    &=
    {\rm d}\Phi_{1}^{i_{1}}{\rm d}\bar{\Phi}_{1}^{j_{1}}
    {\rm d}\Psi_{2}^{i_{2}}{\rm d}\bar{\Psi}_{2}^{j_{2}}
    {\rm d}\Psi_{2}^{k_{2}}{\rm d}\bar{\Psi}_{2}^{l_{2}}
    {\rm d}\Psi_{1}^{l'_{1}}{\rm d}\bar{\Phi}_{1}^{k'_{1}}
    {\rm d}\Psi_{2}^{l'_{2}}{\rm d}\bar{\Phi}_{2}^{k'_{2}}
    {\rm d}\Psi_{2}^{j'_{2}}{\rm d}\bar{\Phi}_{2}^{i'_{2}}
    \nonumber\\
    &\times
    \left(
        \bar{\Phi}_{1}\Phi_{1}
    \right)^{i_{1}}
    \left(
        \bar{\Psi}_{1}\Psi_{1}
    \right)^{j_{1}}
    \left(
        \bar{\Phi}_{2}\Phi_{2}
    \right)^{i_{2}}
    \left(
        \bar{\Psi}_{2}\Psi_{2}
    \right)^{j_{2}}
    \left(
        \bar{\Phi}_{2}\Phi_{2}
    \right)^{k_{2}}
    \left(
        \bar{\Psi}_{2}\Psi_{2}
    \right)^{l_{2}}
    .
\end{align}
The identity used here is 
\begin{align}
\label{eq:trick_1}
    (-1)^{(i'_{2}+j'_{2})(k_{2}+l_{2}+k'_{1}+l'_{1}+k'_{2}+l'_{2})}
    =
    (-1)^{(i'_{2}+j'_{2})(\alpha_{1}+\beta_{1})}
    ,
\end{align}
because the fundamental tensor $\mathcal{T}_{n}$ is Grassmann-even; $T_{n;(\alpha_{1}\beta_{1})(k_{2}l_{2})(k'_{1}l'_{1})(k'_{2}l'_{2})}$ takes a non-zero value only if
\begin{align}
\label{eq:trick_2}
    \alpha_{1}+\beta_{1}+k_{2}+l_{2}+k'_{1}+l'_{1}+k'_{2}+l'_{2} \mod 2
    =
    0
\end{align}
is satisfied.
This kind of identity is useful to simplify the sign factor arising in the exact contraction between the fundamental tensors.

Let us consider the same contraction but with Eq.~\eqref{eq:ak_fund}.
Since $\mathcal{T}_{n+\hat{1}}$ and $\mathcal{T}_{n}$ are connected via $\Phi_{1}(n)$, $\bar{\Phi}_{1}(n)$, $\Psi_{1}(n)$, and $\bar{\Psi}_{1}(n)$, they should be integrated. The exact contraction gives
\begin{align}
\label{eq:ak_exact_cont}
    &\int_{\bar{\Phi}_{1}(n),\Phi_{1}(n)}
    \int_{\bar{\Psi}_{1}(n),\Psi_{1}(n)}
    \mathcal{T}_{n+\hat{1};
        \Phi_{1}\Psi_{1}
        \Phi_{2}\Psi_{2}
        \bar{\Psi}_{1}(n)\bar{\Phi}_{1}(n)
        \bar{\Psi}_{2}\bar{\Phi}_{2}
    }
    \mathcal{T}_{n;
        \Phi_{1}(n)\Psi_{1}(n)
        \Phi_{2}\Psi_{2}
        \bar{\Psi}_{1}\bar{\Phi}_{1}
        \bar{\Psi}_{2}\bar{\Phi}_{2}
    }
    \nonumber\\
    &=
    \sum
    M_{
        (i_{1}j_{1})(i_{2}j_{2})(i'_{2}j'_{2})
        (k_{2}l_{2})(k'_{1}l'_{1})(k'_{2}l'_{2})
    }
    \Phi_{1}^{i_{1}}\Psi_{1}^{j_{1}}
    \Phi_{2}^{i_{2}}\Psi_{2}^{j_{2}}
    \bar{\Psi}_{2}^{j'_{2}}\bar{\Phi}_{2}^{i'_{2}}
    \Phi_{2}^{k_{2}}\Psi_{2}^{l_{2}}
    \bar{\Psi}_{1}^{l'_{1}}\bar{\Phi}_{1}^{k'_{1}}
    \bar{\Psi}_{2}^{l'_{2}}\bar{\Phi}_{2}^{k'_{2}}
    ,
\end{align}
with
\begin{align}
\label{eq:ak_t_p}
    &M_{
        (i_{1}j_{1})(i_{2}j_{2})(i'_{2}j'_{2})
        (k_{2}l_{2})(k'_{1}l'_{1})(k'_{2}l'_{2})
    }
    \nonumber\\
    &=
    \sum_{\alpha_{1},\beta_{1}}
    (-1)^{(\alpha_{1}+\beta_{1})(i'_{2}+j'_{2})+\alpha_{1}+\beta_{1}}
    T_{n+\hat{1};(i_{1}j_{1})(i_{2}j_{2})(\alpha_{1}\beta_{1})(i'_{2}j'_{2})}
    T_{n;(\alpha_{1}\beta_{1})(k_{2}l_{2})(k'_{1}l'_{1})(k'_{2}l'_{2})}
    .
\end{align}
The right-hand side of Eq.~\eqref{eq:ak_exact_cont} defines a new Grassmann tensor.
Here, the contraction between the coefficient tensors should be understood as a result of the integrals on the auxiliary Grassmann variables. From Eq.~\eqref{eq:def_ak_int}, we find that 
\begin{align}
\label{eq:id_delta}
    \int_{\bar{\Theta},\Theta}
    \Theta^{i}\bar{\Theta}^{j}
    =
    \delta_{ij}
    ,
\end{align}
where $\Theta$ and $\bar{\Theta}$ are the Grassmann variables.
The integration of the auxiliary variables naturally introduces contractions between the corresponding coefficient tensors in this formalism.
Figure~\ref{fig:contraciton} illustrates the exact contraction in the left-hand side of Eq.~\eqref{eq:ak_exact_cont},  without omitting the site dependence of auxiliary Grassmann variables.

Rearranging the auxiliary Grassmann variables, the right-hand side of Eq.~\eqref{eq:ak_exact_cont} can be
\begin{align}
    \sum
    M'_{
        (i_{1}j_{1})
        (i_{2}j_{2})(k_{2}l_{2})
        (k'_{1}l'_{1})
        (k'_{2}l'_{2})(i'_{2}j'_{2})
    }
    \Phi_{1}^{i_{1}}\Psi_{1}^{j_{1}}
    \Phi_{2}^{i_{2}}\Psi_{2}^{j_{2}}
    \Phi_{2}^{k_{2}}\Psi_{2}^{l_{2}}
    \bar{\Psi}_{1}^{l'_{1}}\bar{\Phi}_{1}^{k'_{1}}
    \bar{\Psi}_{2}^{l'_{2}}\bar{\Phi}_{2}^{k'_{2}}
    \bar{\Psi}_{2}^{j'_{2}}\bar{\Phi}_{2}^{i'_{2}}
    ,
\end{align}
where 
\begin{align}
\label{eq:ak_t_d}
    &M'_{
        (i_{1}j_{1})
        (i_{2}j_{2})(k_{2}l_{2})
        (k'_{1}l'_{1})
        (k'_{2}l'_{2})(i'_{2}j'_{2})
    }
    =
    \sum_{\alpha_{1},\beta_{1}}
    (-1)^{\alpha_{1}+\beta_{1}}
    T_{n+\hat{1};(i_{1}j_{1})(i_{2}j_{2})(\alpha_{1}\beta_{1})(i'_{2}j'_{2})}
    T_{n;(\alpha_{1}\beta_{1})(k_{2}l_{2})(k'_{1}l'_{1})(k'_{2}l'_{2})}
    .
\end{align}
The same identity as in Eqs.~\eqref{eq:trick_1} and \eqref{eq:trick_2} has been utilized.

Comparing Eq.~\eqref{eq:ty_t_p} and Eq.~\eqref{eq:ak_t_p}, or Eq.~\eqref{eq:ty_t_d} and Eq.~\eqref{eq:ak_t_d}, we see that the difference between the signs obtained in the two formulations is $(-1)^{\beta_{1}}$.
This can be explained by the different ways to decompose the hopping terms employed in the two formulations.
In Eqs.~\eqref{eq:ty_id_1} and \eqref{eq:ty_id_2}, the last factors built just by auxiliary Grassmann variables inherit the structure of original hopping terms.
On the other hand, there is no such difference between Eqs.~\eqref{eq:ak_id_1} and \eqref{eq:ak_id_2}. 
The corresponding sign factor $(-1)^{j'_{1}}$ has already been included in the coefficient tensor as in Eq.~\eqref{eq:ak_coeff}.

If we performed the exact contractions using the methods described so far, the path integral shown in Eqs.~\eqref{eq:ty_z} and \eqref{eq:ak_z} could be obtained exactly.
However, in practice, it is impossible to keep carrying out the exact contractions when we consider the model in an arbitrarily large volume.
This is because the number of fundamental tensors in Eqs.~\eqref{eq:ty_z} and \eqref{eq:ak_z} is the same as the number of lattice sites in $\Lambda$.
Therefore, we need to consider performing the contraction not exactly but approximately.
Several TRG schemes are reviewed in Sec.~\ref{subsec:gtrg} and \ref{subsec:ghotrg}.

\subsection{Model-independent notation}
\label{subsec:multi_notation}

In this subsection, we introduce a new notation for the fundamental tensor
\begin{align}
\label{eq:ty_multi_fund}
    \mathcal{T}_{n}
    =
    T_{n;xtx't'}
    \mathcal{G}_{n;xtx't'}
    ,
\end{align}
instead of Eq.~\eqref{eq:ty_fund}. The variable 
$x$ is defined by two bits via $x=(i_{1}j_{1})$ and $t$, $x'$, and $t'$ are defined similarly.
$\mathcal{G}_{n;xtx't'}$ is defined by the right-hand side of Eq.~\eqref{eq:ty_gpart}.
The expression in Eq.~\eqref{eq:ty_multi_fund} may seem fine because it does not depend on the details of our model, the number of components in the original fermionic field or the structure of hopping terms, and it only depends on the lattice geometry. 
However, this notation can be problematic when considering their exact contractions.
As we have observed in Eq.~\eqref{eq:ty_t_p}, or Eq.~\eqref{eq:ty_t_d}, the sign factor from the auxiliary Grassmann integrals does distinguish the forward and backward hopping terms in the formulation in Ref.~\cite{Takeda:2014vwa}.
One of the ways to resolve this issue is to modify Eq.~\eqref{eq:ty_id_2} as
\begin{align}
    &{\rm e}^{t\bar{\psi}(n)\psi(n+\hat{\nu})}=
    \nonumber\\
    &\sum_{j_{\nu}(n)=0}^{1}
    \left(
        \sqrt{t}\bar{\psi}(n){\rm d}\bar{\Psi}_{\nu}(n)
    \right)^{j_{\nu}(n)}
    \left(
        \sqrt{t}\psi(n+\hat{\nu}){\rm d}\Psi_{\nu}(n+\hat{\nu})
    \right)^{j_{\nu}(n)}
    \left(
        -\Psi_{\nu}(n+\hat{\nu})\bar{\Psi}_{\nu}(n)
    \right)^{j_{\nu}(n)}
    ,
\end{align}
and absorbing the extra factor $(-1)^{j_{\nu}(n)}$ into the bosonic tensor in Eq.~\eqref{eq:ty_coeff}.
Note that this modification makes the bosonic tensor in Eq.~\eqref{eq:ty_coeff} exactly the same as the coefficient tensor in Eq.~\eqref{eq:ak_coeff}.
From now on, we identify the expression in Eq.~\eqref{eq:ty_multi_fund} with this modification.
The exact contraction demonstrated in Sec.~\ref{subsec:exact_contraction} is then denoted by
\begin{align}
\label{eq:ty_exact_mod}
    \sum\int
    \mathcal{T}_{n+\hat{1}}\mathcal{T}_{n}
    =
    \sum_{\alpha}
    T_{n+\hat{1};xt_{1}\alpha t'_{1}}
    T_{n;\alpha t_{2}x't'_{2}}
    \int
    \mathcal{G}_{n+\hat{1};xt_{1}\alpha t'_{1}}
    \mathcal{G}_{n;\alpha t_{2}x't'_{2}}
    .
\end{align}
Now, we introduce the Grassmann parity function $f_{x}$ for $x=(i_{1}j_{1})$ such that 
\begin{align}
\label{eq:sign_func}
    f_{x}
    =
    i_{1}+j_{1}
    ~
    \mod 2
    ,
\end{align}
and $f_{t}$, $f_{x'}$, and $f_{t'}$ are done in the same way.
Using these parity functions, Eq.~\eqref{eq:ty_exact_mod} is evaluated as
\begin{align}
    \sum\int
    \mathcal{T}_{n+\hat{1}}\mathcal{T}_{n}
    =
    M_{xt_{1}t_{2}x't'_{1}t'_{2}}
    \mathcal{Q}_{xt_{1}t_{2}x't'_{2}t'_{1}}
    ,
\end{align}
where 
\begin{align}
\label{eq:common_coeff}
    M_{xt_{1}t_{2}x't'_{1}t'_{2}}
    =
    \sum_{\alpha}
    (-1)^{f_{\alpha}}
    T_{n+\hat{1};xt_{1}\alpha t'_{1}}
    T_{n;\alpha t_{2}x't'_{2}},
\end{align}
and 
\begin{align}
    \mathcal{Q}_{
        xt_{1}t_{2}x't'_{2}t'_{1}
    }
    &=
    {\rm d}\Phi_{1}^{i_{1}}{\rm d}\bar{\Phi}_{1}^{j_{1}}
    {\rm d}\Psi_{2}^{i_{2}}{\rm d}\bar{\Psi}_{2}^{j_{2}}
    {\rm d}\Psi_{2}^{k_{2}}{\rm d}\bar{\Psi}_{2}^{l_{2}}
    {\rm d}\Psi_{1}^{l'_{1}}{\rm d}\bar{\Phi}_{1}^{k'_{1}}
    {\rm d}\Psi_{2}^{l'_{2}}{\rm d}\bar{\Phi}_{2}^{k'_{2}}
    {\rm d}\Psi_{2}^{j'_{2}}{\rm d}\bar{\Phi}_{2}^{i'_{2}}
    \nonumber\\
    &\times
    \left(
        \bar{\Phi}_{1}\Phi_{1}
    \right)^{i_{1}}
    \left(
        \Psi_{1}\bar{\Psi}_{1}
    \right)^{j_{1}}
    \left(
        \bar{\Phi}_{2}\Phi_{2}
    \right)^{i_{2}}
    \left(
        \Psi_{2}\bar{\Psi}_{2}
    \right)^{j_{2}}
    \left(
        \bar{\Phi}_{2}\Phi_{2}
    \right)^{k_{2}}
    \left(
        \Psi_{2}\bar{\Psi}_{2}
    \right)^{l_{2}}
    .
\end{align}
It should be emphasized that the exact contraction between the fundamental tensors expressed as in Eq.~\eqref{eq:ty_multi_fund} is straightforwardly generalized to other models just modifying the definition of the Grassmann parity function in Eq.~\eqref{eq:sign_func}.

Similarly, we introduce the following expression for Eq.~\eqref{eq:ak_fund},
\begin{align}
\label{eq:ak_multi_fund}
    \mathcal{T}_{n;XT\bar{X}\bar{T}}
    =
    \sum_{x,t,x',t'}
    T_{n;xtx't'}
    X^{x}T^{t}\bar{X}^{x'}\bar{T}^{t'}
    .
\end{align}
As in Eq.~\eqref{eq:ty_multi_fund}, $x$ is defined by two bits via $x=(i_{1}j_{1})$ and $t$, $x'$, and $t'$ are defined similarly.
$X$ can be regarded as a two-component auxiliary Grassmann variables via $X=(\Phi_{1},\Psi_{1})$ and $X^{x}$ as $X^{x}=\Phi_{1}^{i_{1}}\Psi_{1}^{j_{1}}$.
$\bar{X}$ can also be regarded like $\bar{X}=(\bar{\Phi}_{1},\bar{\Psi}_{1})$, but $\bar{X}^{x}$ should be understood as $\bar{X}^{x}=\bar{\Psi}_{1}^{j'_{1}}\bar{\Phi}_{1}^{i'_{1}}$.
$T~(\bar{T})$ and $T^{t}~(\bar{T}^{t'})$ are defined in the same way.
Eq.~\eqref{eq:ak_exact_cont} can be equivalently expressed as
\begin{align}
\label{eq:ak_tt}
    &\int_{\bar{\Theta},\Theta}
    \mathcal{T}_{n+\hat{1};XT_{1}\bar{\Theta}\bar{T}_{1}}
    \mathcal{T}_{n;\Theta T_{2}\bar{X}\bar{T}_{2}}
    =
    \sum_{x,t_{1},t_{2},x',t'_{1},t'_{2}}
    M_{xt_{1}t_{2}x't'_{1}t'_{2}}
    X^{x}
    T_{1}^{t'_{1}}T_{2}^{t'_{2}}
    \bar{X}^{x'}
    \bar{T}_{2}^{t'_{2}}\bar{T}_{1}^{t'_{1}}
    ,
\end{align}
where the coefficient tensor $M_{xt_{1}t_{2}x't'_{1}t'_{2}}$ is defined in the exactly same way with Eq.~\eqref{eq:common_coeff} using the Grassmann parity function defined in Eq.~\eqref{eq:sign_func}.
Note that $\Theta$ and $\bar{\Theta}$ in Eq.~\eqref{eq:ak_tt} are two-component Grassmann variables and $\int_{\bar{\Theta},\Theta}$ in the left-hand side is defined by
\begin{align}
    \int_{\bar{\Theta},\Theta}
    =
    \prod_{i}
    \int{\rm d}\bar{\Theta}_{i}{\rm d}\Theta_{i}~{\rm e}^{-\bar{\Theta}_{i}\Theta_{i}}
    ,
\end{align}
which is a natural extension of Eq.~\eqref{eq:def_ak_int}.
We can also see that $f_{x}$ in Eq.~\eqref{eq:sign_func} counts the Grassmann parity of $X^{x}$.
The right-hand side of Eq.~\eqref{eq:ak_tt} is the Grassmann tensor that can be written as $\mathcal{M}_{XT_{1}T_{2}\bar{X}\bar{T}_{2}\bar{T}_{1}}$ in our notation.
Extend the definition of the Grassmann parity function in Eq.~\eqref{eq:sign_func}, and the notation in Eq.~\eqref{eq:ak_multi_fund} immediately allows us to deal with the case where the original fermionic model is described by the multi-component Grassmann variables.

\begin{figure}[htbp]
  	\centering
	\includegraphics[width=1\hsize]{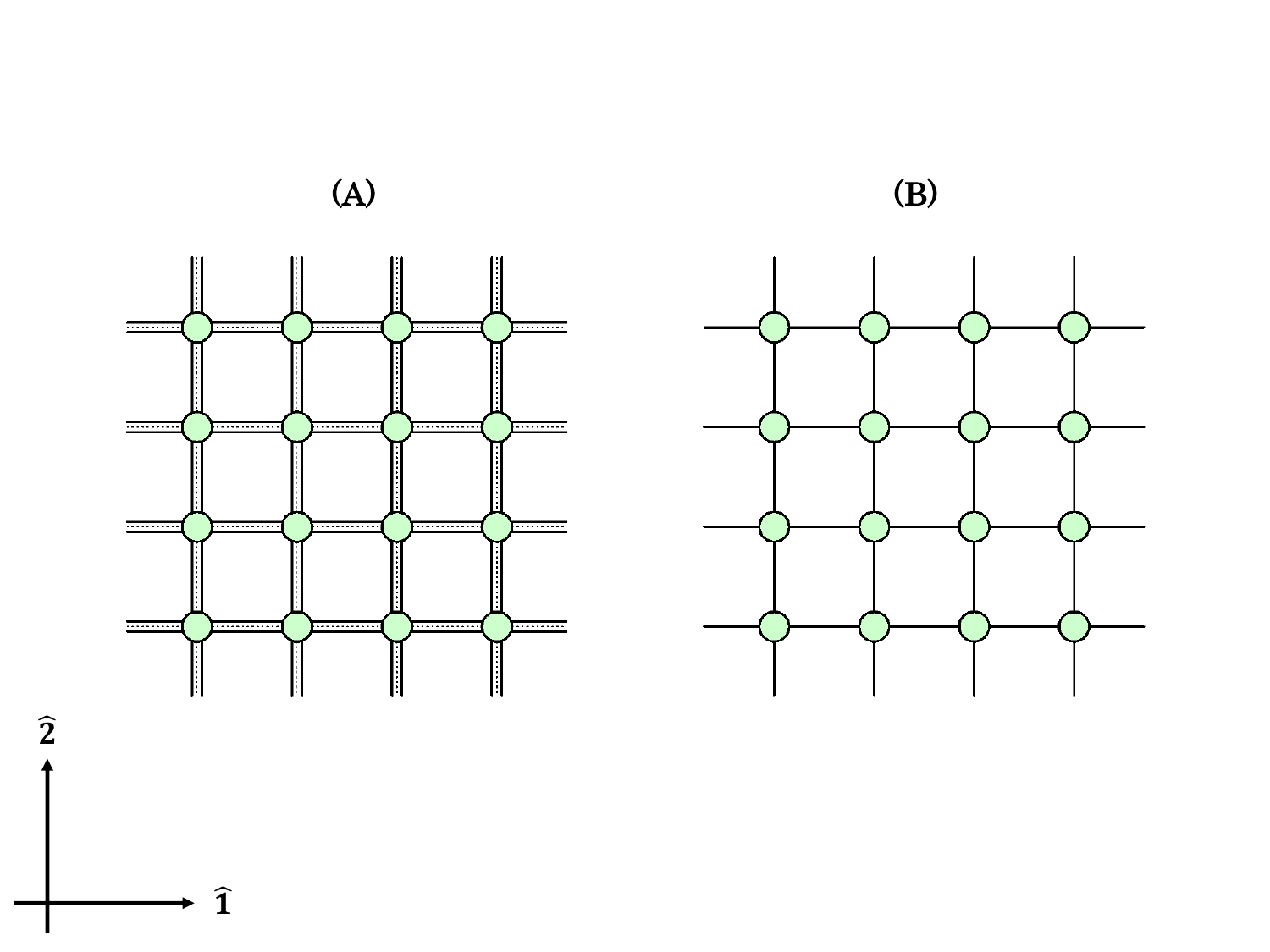}
	\caption{
	Diagrammatic representation of Grassmann tensor network on a two-dimensional square lattice.
        (A) Grassmann tensor network representation of Eq.~\eqref{eq:z_ex}.
        Interpretation of each line is given in Figure~\ref{fig:fund_tensor}.
        Background dotted lines denote the real-space lattice.
        (B) Model-independent description of two-dimensional Grassmann tensor network.
        The shape of each fundamental tensor is determined only by the lattice geometry.
	}
  	\label{fig:gtn}
\end{figure}

So far, we have confirmed that the bosonic tensor derived by Ref.~\cite{Takeda:2014vwa} and the coefficient tensor by Ref.~\cite{Akiyama:2020sfo} can be equivalent if we slightly modify the formulation of Ref.~\cite{Takeda:2014vwa}.
It has also been confirmed that the exact contraction can be described similarly using the same Grassmann parity function with either formulation.
Figure~\ref{fig:gtn} graphically shows the Grassmann tensor network representation of Eq.~\eqref{eq:z_ex}.
Figure~\ref{fig:gtn}~(A) can be interpreted in two ways following the notation rule in Figures~\ref{fig:fund_tensor} and \ref{fig:contraciton}.
Figure~\ref{fig:gtn}~(B) assumes the model-independent notation.
The following demonstration always assumes the formalism in Ref.~\cite{Akiyama:2020sfo} with the notation introduced in this Sec.~\ref{subsec:multi_notation}.

\subsection{Extension to lattice gauge theories}

\begin{figure}[htbp]
  	\centering
	\includegraphics[width=1\hsize]{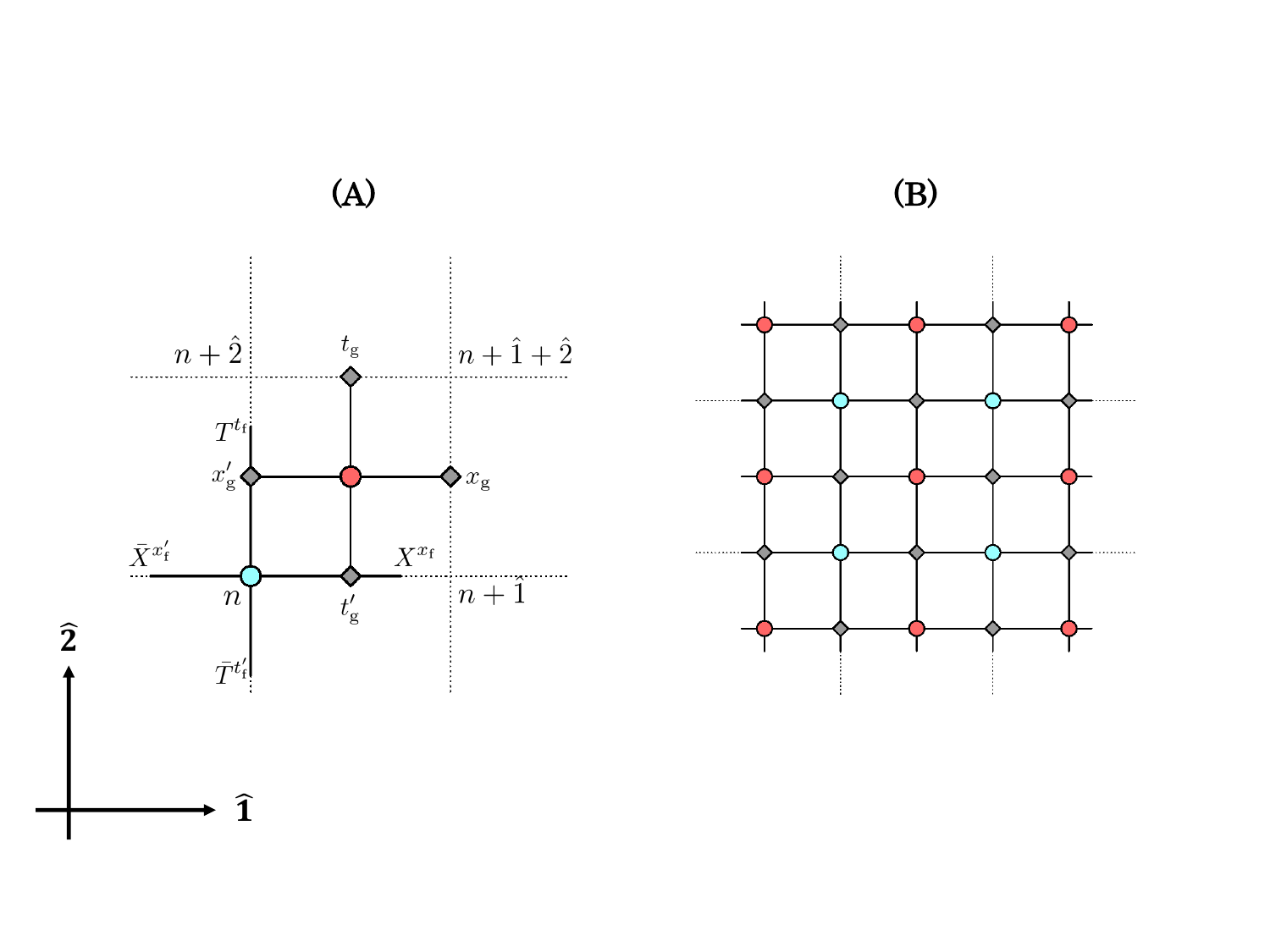}
	\caption{
	Diagrammatic representation of Grassmann tensor network formulation for two-dimensional lattice gauge theories.
        Background dotted lines show a real-space square lattice.
        (A) Structure of the fundamental tensor defined in Eq.~\eqref{eq:ak_fund_mix}.
        The blue symbol is located on the site and denotes the right-hand side of Eq.~\eqref{eq:ak_gtensor_lgt}.
        Note that the coefficient tensor depends on $n_{1}$ as shown in Eq.~\eqref{eq:tf_staggered}.
        The red symbol is located on the plaquette and denotes the four-leg tensor $T^{({\rm g})}_{x_{\rm g}t_{\rm g}x'_{\rm g}t'_{\rm g}}$ in Eq.~\eqref{eq:gauge_pl_tensor}.
        Each diamond shows the link variable.
        (B) Grassmann tensor network in Eq.~\eqref{eq:gtn_zn}.
        Each internal line represents the contraction between auxiliary Grassmann variables and integration over the shared link variable.
	}
  	\label{fig:gauge_theory}
\end{figure}

Current Grassmann tensor network formulations are easily combined with the usual tensor network formulations for pure gauge or bosonic theories.
See Ref.~\cite{Liu:2013nsa} and several reviews~\cite{Meurice:2020pxc,Okunishi:2021but,Akiyama:2021nhe,Kadoh:2022loj} for more details on how to construct fundamental tensors in these theories.
Here, let us consider the two-dimensional $\mathbb{Z}_{N}$ gauge theory on a periodic square lattice defined by
\begin{align}
    S=S_{\rm g}+S_{\rm f}
    ,
\end{align}
as an example.
We assume the standard Wilson action,
\begin{align}
\label{eq:s_g_zn}
    S_{\rm g}
    =
    -\beta\sum_{n\in\Lambda}\Re\left[
        U_{1}(n)U_{2}(n+\hat{1})U^{*}_{1}(n+\hat{2})U^{*}_{2}(n)
    \right]
    ,
\end{align}
and the staggered fermion action,
\begin{align}
\label{eq:s_f_zn}
    S_{\rm f}
    =
    \sum_{n,\nu}
    \frac{\eta_{\nu}(n)}{2}\left[
        \bar{\chi}(n)U_{\nu}(n)\chi(n+\hat{\nu})
        -
        \bar{\chi}(n+\hat{\nu})U^{*}_{\nu}(n)\chi(n)
    \right]
    +
    m
    \sum_{n}\bar{\chi}(n)\chi(n)
    .
\end{align}
$\beta$ and $m$ denote the inverse gauge coupling and mass, respectively.
The staggered sign function $\eta_{\nu}(n)$ is defined by $\eta_{1}(n)=1$ and $\eta_{2}(n)=(-1)^{n_{1}}$ at each site $n=(n_{1},n_{2})$.
Our goal is to derive the Grassmann tensor network representation for the following path integral,
\begin{align}
\label{eq:z_zn}
    Z=
    \int
    \prod_{n,\nu}
    {\rm d}U_{\nu}(n)
    \prod_{n}
    {\rm d}\chi(n)
    {\rm d}\bar{\chi}(n)
    ~
    {\rm e}^{-S}
    .
\end{align}
We parameterize $U_{\nu}(n)$ by an integer $q_{\nu}(n)$ mod $N$ via $U_{\nu}(n)=\exp\left[2\pi{\rm i}q_{\nu}(n)/N\right]$.
Following Sec.~\ref{subsec:gtn_rep}, one can immediately obtain a Grassmann tensor with some integer indices.
We begin with decomposing the hopping terms via
\begin{align}
\label{eq:ak_lgt_1} 
    {\rm e}^{\eta_{\nu}(n)\bar{\chi}(n+\hat{\nu})U^{*}_{\nu}(n)\chi(n)/2}=
    \int
    {\rm d}\bar{\Phi}_{\nu}(n){\rm d}\Phi_{\nu}(n)
    ~{\rm e}^{-\bar{\Phi}_{\nu}(n)\Phi_{\nu}(n)}
    ~{\rm e}^{\bar{\chi}(n+\hat{\nu})\bar{\Phi}_{\nu}(n)/\sqrt{2}}
    ~{\rm e}^{\eta_{\nu}(n)U^{*}_{\nu}(n)\chi(n)\Phi_{\nu}(n)/\sqrt{2}}
    ,
\end{align}
\begin{align}
\label{eq:ak_lgt_2} 
    {\rm e}^{-\eta_{\nu}(n)\bar{\chi}(n)U_{\nu}(n)\chi(n+\hat{\nu})/2}=
    \int
    {\rm d}\bar{\Psi}_{\nu}(n){\rm d}\Psi_{\nu}(n)
    ~{\rm e}^{-\bar{\Psi}_{\nu}(n)\Psi_{\nu}(n)}
    ~{\rm e}^{\eta_{\nu}(n)U_{\nu}(n)\bar{\chi}(n)\Psi_{\nu}(n)/\sqrt{2}}
    ~{\rm e}^{\chi(n+\hat{\nu})\bar{\Psi}_{\nu}(n)/\sqrt{2}}
    .
\end{align}
Integrating out the original staggered fields at each site, one obtains
\begin{align}
\label{eq:ak_gtensor_lgt}
    &\int{\rm d}\chi{\rm d}\bar{\chi}
    ~{\rm e}^{-m\bar{\chi}\chi}
    \prod_{\nu}
    {\rm e}^{\bar{\chi}\bar{\Phi}_{\nu}(n-\hat{\nu})/\sqrt{2}}
    {\rm e}^{\eta_{\nu}(n)U_{\nu}(n)\bar{\chi}\Psi_{\nu}(n)/\sqrt{2}}
    {\rm e}^{\eta_{\nu}(n)U^{*}_{\nu}(n)\chi\Phi_{\nu}(n)/\sqrt{2}}
    {\rm e}^{\chi\bar{\Psi}_{\nu}(n-\hat{\nu})/\sqrt{2}}
    \nonumber\\
    &=
    \sum_{x_{\rm f},t_{\rm f},x'_{\rm f},t'_{\rm f}}
    T^{({\rm f})}_{
        x_{\rm f}t_{\rm f}x'_{\rm f}t'_{\rm f},
        q_{2}(n)q_{1}(n)
    }
    X^{x_{\rm f}}
    T^{t_{\rm f}}
    \bar{X}^{x'_{\rm f}}
    \bar{T}^{t'_{\rm f}}
    ,
\end{align}
where we have employed the model-independent notation in Sec.~\ref{subsec:multi_notation} and the coefficient tensor is given by
\begin{align}
\label{eq:tf_staggered}
    T^{({\rm f})}_{
        x_{\rm f}t_{\rm f}x'_{\rm f}t'_{\rm f},
        q_{2}(n)q_{1}(n)
    }
    &=
    T^{({\rm f})}_{
        (i_{1}j_{1})(i_{2}j_{2})(i'_{1}j'_{1})(i'_{2}j'_{2}),
        q_{2}(n)q_{1}(n)
    }
    \nonumber\\
    &=
    \sqrt{2}^{-\sum_{\nu}(i_{\nu}+j_{\nu}+i'_{\nu}+j'_{\nu})}
    (-1)^{
        j_{1}(i_{2}+j'_{1}+j'_{2})
        +j_{2}(j'_{1}+j'_{2})
        +i'_{1}j'_{2}
        +n_{1}(i_{2}+j_{2})
    }
    {\rm e}^{
        \frac{2\pi{\rm i}}{N}
        \sum_{\nu}q_{\nu}(n)\left(j_{\nu}-i_{\nu}\right)
    }
    \nonumber\\
    &\times
    \left[
        -m
        \delta_{i_{1}+i_{2}+j'_{1}+j'_{2},0}
        \delta_{i'_{1}+i'_{2}+j_{1}+j_{2},0}
        +
        \delta_{i_{1}+i_{2}+j'_{1}+j'_{2},1}
        \delta_{i'_{1}+i'_{2}+j_{1}+j_{2},1}
    \right]
    .
\end{align}
Note that $X^{x_{\rm f}}$, $T^{t_{\rm f}}$, $\bar{X}^{x'_{\rm f}}$, and $\bar{T}^{t'_{\rm f}}$ in the right-hand side of Eq.~\eqref{eq:ak_gtensor_lgt} have been defined in the same way with Eq.~\eqref{eq:ak_multi_fund}. 
We also introduce a four-leg tensor to describe the plaquette interaction term in ${\rm e}^{-S_{\rm g}}$ via
\begin{align}
\label{eq:gauge_pl_tensor}
    T^{({\rm g})}_{
        q_{2}(n+\hat{1})
        q_{1}(n+\hat{2})
        q_{2}(n)
        q_{1}(n)
    }
    =
    \exp\left[
        \beta\cos\left\{
            \frac{2\pi}{N}
            \left(
                q_{1}(n)+q_{2}(n+\hat{1})-q_{1}(n+\hat{2})-q_{2}(n)
            \right)
        \right\}
    \right]
    ,
\end{align}
following Ref.~\cite{Kuramashi:2019cgs}.
Regarding $q_{\nu}$'s as tensor subscripts, we now define the fundamental tensor associated at the lattice site $n$ as
\begin{align}
\label{eq:ak_fund_mix}
    \mathcal{T}_{n;XT\bar{X}\bar{T},x_{\rm g}t_{\rm g}x'_{\rm g}t'_{\rm g}}
    =
    \sum_{x_{\rm f},t_{\rm f},x'_{\rm f},t'_{\rm f}}
    T_{n;x_{\rm f}t_{\rm f}x'_{\rm f}t'_{\rm f},x_{\rm g}t_{\rm g}x'_{\rm g}t'_{\rm g}}
    X^{x_{\rm f}}
    T^{t_{\rm f}}
    \bar{X}^{x'_{\rm f}}
    \bar{T}^{t'_{\rm f}}
    ,
\end{align}
where
\begin{align}
    T_{n;x_{\rm f}t_{\rm f}x'_{\rm f}t'_{\rm f},x_{\rm g}t_{\rm g}x'_{\rm g}t'_{\rm g}}
    =
    T^{({\rm f})}_{x_{\rm f}t_{\rm f}x'_{\rm f}t'_{\rm f},x_{\rm g}t_{\rm g}}
    \cdot
    T^{({\rm g})}_{x_{\rm g}t_{\rm g}x'_{\rm g}t'_{\rm g}}
    .
\end{align}
The path integral in Eq.~\eqref{eq:z_zn} is now represented by
\begin{align}
\label{eq:gtn_zn}
    Z=
    {\rm gTr}
    \left[
        \prod_{n}
        \mathcal{T}_{n}
    \right]
    .
\end{align}
Here, ${\rm ``gTr"}$ stands not only for the integrations over all auxiliary Grassmann variables but also for the summations over all integers corresponding to the link variables.
Analogously, we refer $\mathcal{T}_{n}$ and $T_{n}$ in Eq.~\eqref{eq:ak_fund_mix} to the Grassmann tensor and its coefficient tensor, respectively.
See Figure~\ref{fig:gauge_theory} for the diagrammatic explanation.

Let us briefly see how the exact contraction is carried out between these fundamental tensors.
As in Sec.~\ref{subsec:exact_contraction}, we consider the contraction between the fundamental tensors at $n+\hat{1}$ and $n$ for demonstration.
Since a link variable is shared between these two fundamental tensors, the exact contraction should be
\begin{align}
    \sum_{\alpha_{\rm g}}
    \int_{\bar{\Theta},\Theta}
    \mathcal{T}_{n+\hat{1};
        XT_{1}\bar{\Theta}\bar{T}_{1},
        x_{\rm g}{t_{\rm g}}_{1}\alpha_{\rm g}{t'_{\rm g}}_{1}
    }
    \mathcal{T}_{n;
        \Theta T_{2}\bar{X}\bar{T}_{2},
        \alpha_{\rm g}{t_{\rm g}}_{2}x'_{\rm g}{t'_{\rm g}}_{2}
    }
    .
\end{align}
One will immediately find that the above contraction results in a new Grassmann tensor such that,
\begin{align}
    &\mathcal{M}_{
        XT_{1}T_{2}\bar{X}\bar{T}_{2}\bar{T}_{1},
        x_{\rm g}
        {t_{\rm g}}_{1}{t_{\rm g}}_{2}
        x'_{\rm g}
        {t'_{\rm g}}_{1}{t'_{\rm g}}_{2}
    }
    \nonumber\\
    &=
    \sum_{x_{\rm f},{t_{\rm f}}_{1},{t_{\rm f}}_{2},x'_{\rm f},{t'_{\rm f}}_{1},{t'_{\rm f}}_{2}}
    M_{
        x_{\rm f}
        {t_{\rm f}}_{1}{t_{\rm f}}_{2}
        x'_{\rm f}
        {t'_{\rm f}}_{1},{t'_{\rm f}}_{2},
        x_{\rm g}
        {t_{\rm g}}_{1}{t_{\rm g}}_{2}
        x'_{\rm g}
        {t'_{\rm g}}_{1}{t'_{\rm g}}_{2}
    }
    X^{x_{\rm f}}
    T_{1}^{{t_{\rm f}}_{1}}
    T_{2}^{{t_{\rm f}}_{2}}
    \bar{X}^{x'_{\rm f}}
    \bar{T}_{2}^{{t'_{\rm f}}_{2}}
    \bar{T}_{1}^{{t'_{\rm f}}_{1}}
    ,
\end{align}
whose coefficient tensor is defined by
\begin{align}
\label{eq:ak_coeff_new}
    M_{
        x_{\rm f}
        {t_{\rm f}}_{1}{t_{\rm f}}_{2}
        x'_{\rm f}
        {t'_{\rm f}}_{1},{t'_{\rm f}}_{2},
        x_{\rm g}
        {t_{\rm g}}_{1}{t_{\rm g}}_{2}
        x'_{\rm g}
        {t'_{\rm g}}_{1}{t'_{\rm g}}_{2}
    }
    =
    \sum_{\alpha_{\rm f},\alpha_{\rm g}}
    (-1)^{f_{\alpha_{\rm f}}}
    T_{n+\hat{1};
        x_{\rm f}{t_{\rm f}}_{1}\alpha_{\rm f}{t'_{\rm f}}_{1},
        x_{\rm g}{t_{\rm g}}_{1}\alpha_{\rm g}{t'_{\rm g}}_{1}
    }
    T_{n;
        \alpha_{\rm f}{t_{\rm f}}_{2}x'_{\rm f}{t'_{\rm f}}_{2},
        \alpha_{\rm g}{t_{\rm g}}_{2}x'_{\rm g}{t'_{\rm g}}_{2}
    }
    .
\end{align}
The Grassmann parity function $f_{\alpha_{\rm f}}$ is given in the same way with Eq.~\eqref{eq:sign_func}.
Introducing a super index $p=(p_{\rm f},p_{\rm g})$ for $p=x,t,x',t'$, Eq.~\eqref{eq:ak_coeff_new} reads 
\begin{align}
\label{eq:new_coeff_super}
    M_{
        x
        t_{1}t_{2}
        x'
        t'_{1}t'_{2}
    }
    =
    \sum_{\alpha}
    (-1)^{F_{\alpha}}
    T_{n+\hat{1};
        xt_{1}\alpha t'_{1}
    }
    T_{n;
        \alpha t_{2}x't'_{2}
    }
    ,
\end{align}
where we have defined a new parity function for the super index $\alpha=(\alpha_{\rm f},\alpha_{\rm g})$ by
\begin{align}
    F_{\alpha}=f_{\alpha_{\rm f}}.
\end{align}
The function $F_{\alpha}$ tells us the Grassmann parity of the super index $\alpha$: when $F_{\alpha}=0~(1)$, the super index $\alpha$ is describing the Grassmann-even~(odd) contribution.

Therefore, we reach the important conclusion that the structure of the Grassmann tensor network and the contraction rule are not affected by the gauge fields.
The resulting coefficient tensor in Eq.~\eqref{eq:new_coeff_super} has the same expression as Eq.~\eqref{eq:common_coeff}, which was the coefficient tensor for the pure fermionic model in Eq.~\eqref{eq:action_ex}.
The exact contractions among the Grassmann tensors result in those among the coefficient tensors, with or without the lattice gauge fields.
One can use the tensor network diagram in Figure~\ref{fig:gtn}~(B) instead of Figure~\ref{fig:gauge_theory}~(B) to represent Eq.~\eqref{eq:gtn_zn}.
Although we have assumed $\mathbb{Z}_{N}$ as a gauge group for simplicity, the above construction can be combined with other tensor network formulations for various lattice gauge theories including the non-Abelian fields~\cite{Liu:2013nsa,Fukuma:2021cni,Hirasawa:2021qvh,Dittrich:2014mxa,Kuwahara:2022ubg,Luo:2022eje,Akiyama:2022eip}.

\subsection{Approximate contraction by the Levin-Nave TRG}
\label{subsec:gtrg}

The usual TRG algorithms are intended to be applied to the classical systems without any Grassmann variables. 
Still, any TRG algorithm can be utilized to evaluate the Grassmann path integrals.
As we have seen, the contractions between the Grassmann tensors always result in those between the corresponding coefficient tensors with some sign factors arising from the integrals over the auxiliary Grassmann variables.
Therefore, we can use the TRG algorithms to carry out the contractions among the coefficient tensors approximately.
In this subsection, we demonstrate how to extend the Levin-Nave TRG~\cite{Levin:2006jai} for the Grassmann tensor networks. The case of the HOTRG~\cite{2012PhRvB..86d5139X} is discussed in the next subsection. 

\begin{figure}[htbp]
  	\centering
	\includegraphics[width=1\hsize]{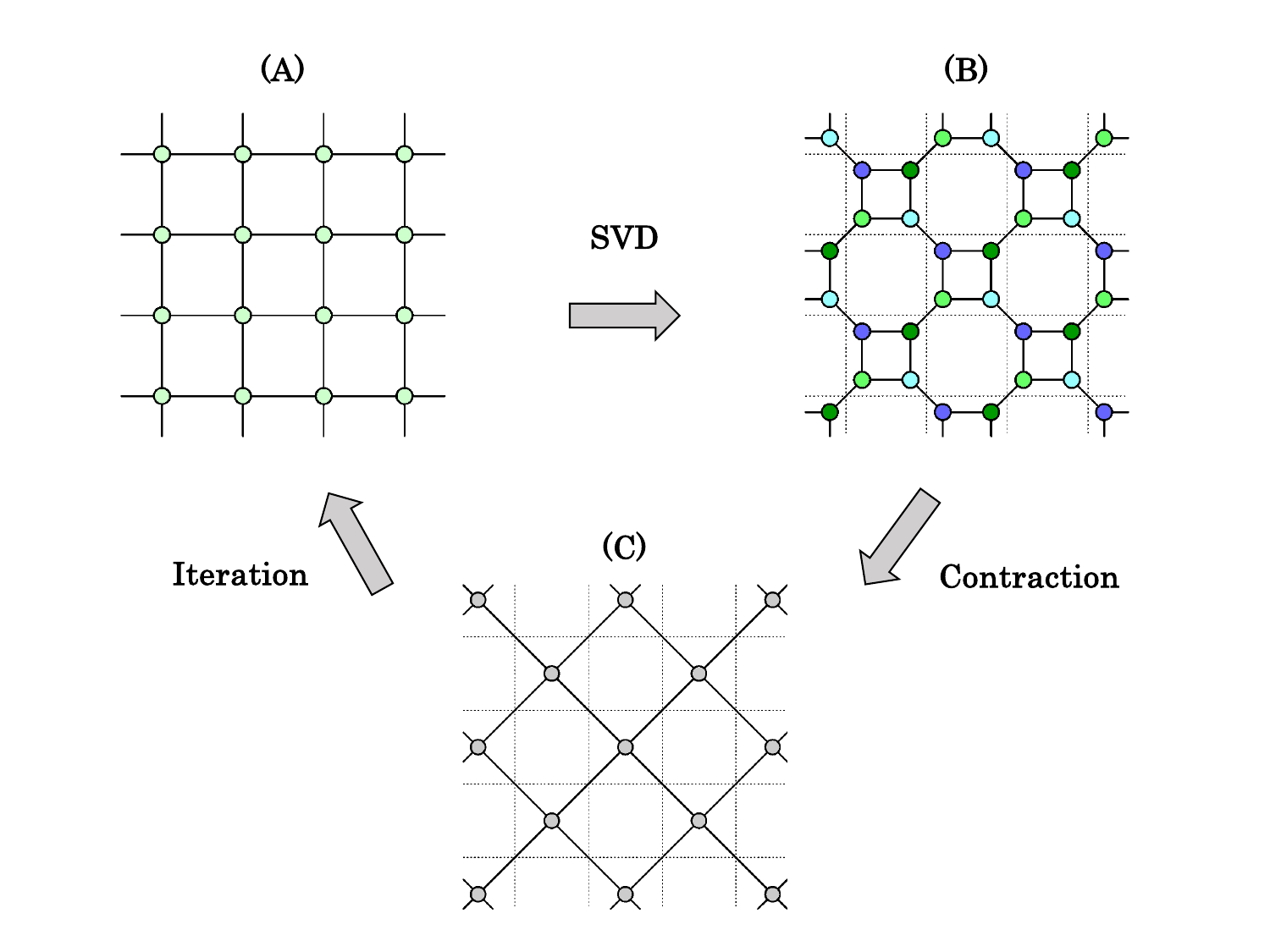}
	\caption{
	Schematic illustration of the Levin-Nave TRG algorithm.
        Background dotted lines show a real-space square lattice.
        (A) Initial tensor network on the lattice.
        (B) Two kinds of SVD as shown in Eqs.~\eqref{eq:lntrg_svd_e} and \eqref{eq:lntrg_svd_o}.
        (C) New tensor network by contracting four three-leg tensors.
	}
  	\label{fig:schematic_levin_nave_trg}
\end{figure}

Let us begin with reviewing the original Levin-Nave TRG.
The algorithm aims to evaluate the partition function or path integral represented by the tensor network
\begin{align}
\label{eq:ttr_z}
    Z=
    {\rm tTr}
    \left[
        \prod_{n\in\Lambda}T_{n}
    \right]
    ,
\end{align}
where we assume the two-dimensional periodic square lattice $\Lambda$ and $T_{n}$ is a four-leg tensor defined on the site $n$.
The algorithm employs the singular value decomposition (SVD) to decimate the four-leg tensor $T_{n}$ into three-leg tensors:
\begin{align}
\label{eq:lntrg_svd_e}
    T_{n;xtx't'}
    \simeq
    \sum_{a=1}^{D_{\rm LNTRG}}
    A_{n;xta}B_{n;ax't'}
    ,
\end{align}
\begin{align}
\label{eq:lntrg_svd_o}
    T_{n;xtx't'}
    \simeq
    \sum_{a=1}^{D_{\rm LNTRG}}
    C_{n;xt'a}D_{n;ax't}
    .
\end{align}
Each three-leg tensor is defined as a unitary matrix multiplied by the square root of its singular value.
In the above expressions, we have assumed that the singular values are in descending order.
The truncation parameter $D_{\rm LNTRG}$ is called the bond dimension.
Since we have used the SVD, Eqs.~\eqref{eq:lntrg_svd_e} and \eqref{eq:lntrg_svd_o} give the best approximation in terms of the Frobenious norm of $T_{n}$ under the fixed bond dimension.
Then, we define a new four-leg tensor via
\begin{align}
\label{eq:lntrg_update}
    T_{n';xtx't'}
    =
    \sum_{x_{1},x_{2},t_{1},t_{2}}
    C_{n+\hat{2};x_{2}t_{1}x}
    A_{n;x_{1}t_{1}t}
    D_{n+\hat{1};x'x_{1}t_{2}}
    B_{n+\hat{1}+\hat{2};t'x_{2}t_{2}}
    ,
\end{align}
which approximates the original $Z$ as
\begin{align}
\label{eq:ttr_z_new}
    Z\simeq
    {\rm tTr}
    \left[
        \prod_{n'\in\Lambda'}T_{n'}
    \right]
    .
\end{align}
Therefore, $T_{n'}$ is a new fundamental tensor defined on a site $n'$ in the coarse-grained lattice $\Lambda'$.
Since Eq.~\eqref{eq:ttr_z_new} describes the tensor network whose geometry is the same as that in Eq.~\eqref{eq:ttr_z}, we can easily repeat the above decimation procedure.
Repeating this procedure $N$ times, $2^{N}$ original fundamental tensors are approximately contracted.
Thanks to this property, the algorithm allows us to evaluate the partition function in the thermodynamic limit.
The Levin-Nave TRG requires the $O\left(D_{\rm LNTRG}^{6}\right)$ complexity and the $O\left(D_{\rm LNTRG}^{4}\right)$ memory cost.
Note that the cost can further be reduced to the $O\left(D_{\rm LNTRG}^{5}\right)$ complexity and the $O\left(D_{\rm LNTRG}^{3}\right)$ memory using the randomized SVD~\cite{2018PhRvE..97c3310M}.
The algorithm is graphically summarized in Figure~\ref{fig:schematic_levin_nave_trg}.

We now need to define the SVD for the Grassmann tensor to extend the algorithm for the Grassmann tensor network.
Suppose $\mathcal{O}_{\Phi\Psi}$ is a Grassmann tensor defined via
\begin{align}
\label{eq:def_grassmann_m}
    \mathcal{O}_{\Phi\Psi}
    =
    \sum_{i,j}
    O_{ij}
    \Phi^{i}\Psi^{j}
    ,
\end{align}
where we have assumed the notation in Sec.~\ref{subsec:multi_notation}, say $i=(i_
{1}\cdots i_{m})$ and $j=(j_{1}\cdots j_{n})$ with $\Phi^{i}=\Phi_{1}^{i_{1}}\cdots\Phi_{m}^{i_{m}}$ and $\Psi^{j}=\Psi_{1}^{j_{1}}\cdots\Psi_{n}^{j_{n}}$. We also assume that $\mathcal{O}$ is Grassmann-even, in other words, the matrix elements are zero unless the sum of all the indices ($i$ and $j$) is even.
Since the coefficient tensor is normal, we can apply the SVD for $O_{ij}$,
\begin{align}
\label{eq:svd_m}
    O_{ij}
    =
    \sum_{a=1}^{\min(2^{m},2^{n})}
    U_{ia}\sigma_{a}V^{\dag}_{aj}
    =
    \sum_{a,b}
    A_{ia}\delta_{ab}B_{bj}
    .
\end{align}
In the second equality, we set $A=U\sqrt{\sigma}$ and $B=\sqrt{\sigma}V^{\dag}$.
Recalling the identity in Eq.~\eqref{eq:id_delta}, we can express $\delta_{ab}$ by introducing new auxiliary Grassmann variables.
Plugging Eq.~\eqref{eq:svd_m} into Eq.~\eqref{eq:def_grassmann_m}, we now have the SVD for the Grassmann tensor $\mathcal{O}$ as
\begin{align}
\label{eq:def_gsvd}
    \mathcal{O}_{\Phi\Psi}
    =
    \int_{\bar{\Xi},\Xi}
    \mathcal{A}_{\Phi\Xi}
    \mathcal{B}_{\bar{\Xi}\Psi}
    ,
\end{align}
where $\Xi$ and $\bar{\Xi}$ are the $\min(m,n)$-component auxiliary Grassmann variables introduced via Eq.~\eqref{eq:id_delta}.
The Grassmann tensors $\mathcal{A}$ and $\mathcal{B}$ in the right-hand side have $A$ and $B$ in Eq.~\eqref{eq:svd_m} as their coefficient tensors.
Since we have assumed that $\mathcal{O}$ is Grassmann-even, the resulting $\mathcal{A}$ and $\mathcal{B}$ are also Grassmann-even, as we will see in Sec.~\ref{subsec:remark}.
It is no exaggeration to say that the SVD for a Grassmann tensor is the SVD for its coefficient tensor.
There is no difficulty in introducing the low-rank approximation based on the SVD.
In the following, we use the approximation
\begin{align}
\label{eq:def_gsvd_trunc}
    \mathcal{O}_{\Phi\Psi}
    \simeq
    \int_{\bar{\Xi},\Xi}^{D_{\rm LNTRG}}
    \mathcal{A}_{\Phi\Xi}
    \mathcal{B}_{\bar{\Xi}\Psi}
    ,
\end{align}
which means that the SVD of the coefficient tensor is truncated up to the bond dimension $D_{\rm LNTRG}$ as
\begin{align}
\label{eq:def_gsvd_coeff_trunc}
    O_{ij}
    \simeq
    \sum_{a,b=1}^{D_{\rm LNTRG}}
    A_{ia}\delta_{ab}B_{bj}
    .
\end{align}
We will make some practical remarks on the SVD of the Grassmann tensor in Sec.~\ref{subsec:remark}

\begin{figure}[htbp]
  	\centering
	\includegraphics[width=1\hsize]{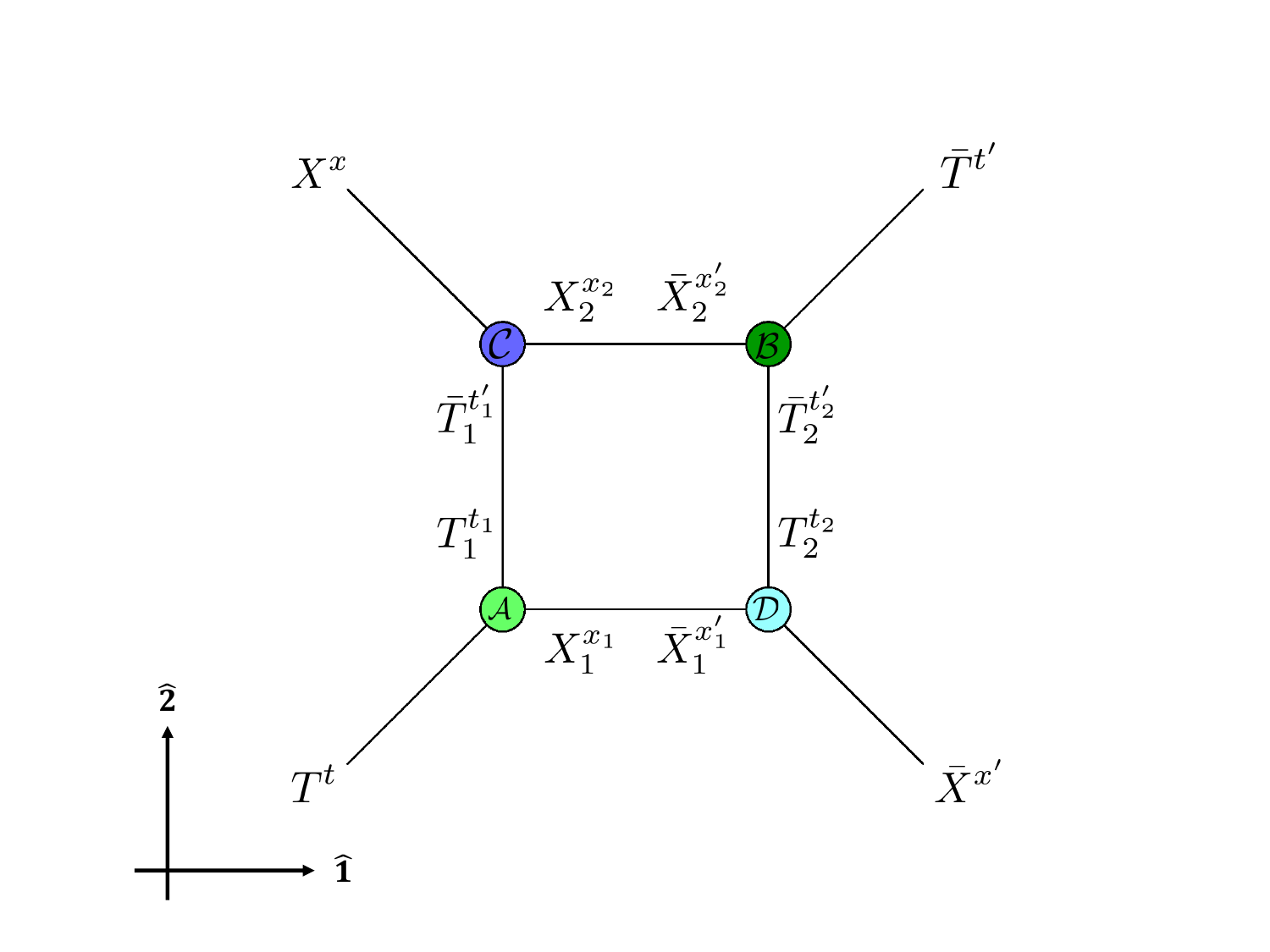}
	\caption{
	Diagrammatic representation of Eq.~\eqref{eq:g_tensor_new}.
	}
  	\label{fig:ln_contract}
\end{figure}

We now extend the algorithm for evaluating 
\begin{align}
\label{eq:gtr_z_old}
    Z
    =
    {\rm gTr}
    \left[
        \prod_{n\in\Lambda}
        \mathcal{T}_{n}
    \right]
    ,
\end{align}
with the Grassmann tensor $\mathcal{T}_{n;XT\bar{X}\bar{T}}$.
We can formally write the SVD for the Grassmann tensor corresponding to Eqs.~\eqref{eq:lntrg_svd_e} and \eqref{eq:lntrg_svd_o} as
\begin{align}
\label{eq:gsvd_1}
    \mathcal{T}_{n;XT\bar{X}\bar{T}}
    \simeq
    \int_{\bar{\Xi},\Xi}^{D_{\rm LNTRG}}
    \mathcal{A}_{n;XT\Xi}
    \mathcal{B}_{n;\bar{\Xi}\bar{X}\bar{T}}
    ,
\end{align}
\begin{align}
\label{eq:gsvd_2}
    \mathcal{T}_{n;XT\bar{X}\bar{T}}
    \simeq
    \int_{\bar{\Xi},\Xi}^{D_{\rm LNTRG}}
    \mathcal{C}_{n;X\bar{T}\Xi}
    \mathcal{D}_{n;\bar{\Xi}\bar{X}T}
    .
\end{align}
When $\mathcal{T}_{n}$ is Grassmann-even, then $\mathcal{A}$, $\mathcal{B}$, $\mathcal{C}$, and $\mathcal{D}$ are also Grassmann-even.
Therefore, we can easily define a new Grassmann tensor from the contractions among them,
\begin{align}
\label{eq:g_tensor_new}
    \mathcal{T}_{n';XT\bar{X}\bar{T}}
    =
    \int_{\bar{X}_{1},X_{1}}
    \int_{\bar{X}_{2},X_{2}}
    \int_{\bar{T}_{1},T_{1}}
    \int_{\bar{T}_{2},T_{2}}
    \mathcal{C}_{n+\hat{2};X_{2}\bar{T}_{1}X}
    \mathcal{A}_{n;X_{1}T_{1}T}
    \mathcal{D}_{n+\hat{1};\bar{X}\bar{X}_{1}T_{2}}
    \mathcal{B}_{n+\hat{1}+\hat{2};\bar{T}\bar{X}_{2}\bar{T}_{2}}
    .
\end{align}
The above equation is analogous to Eq.~\eqref{eq:lntrg_update} in the usual Levin-Nave TRG.
Eq.~\eqref{eq:g_tensor_new} is diagrammatically shown in Figure~\ref{fig:ln_contract}.
The new Grassmann tensor approximates the path integral by
\begin{align}
\label{eq:gtr_z_new}
    Z
    \simeq
    {\rm gTr}
    \left[
        \prod_{n'\in\Lambda'}
        \mathcal{T}_{n'}
    \right]
    .
\end{align}
The resulting Grassmann tensor network in Eq.~\eqref{eq:gtr_z_new} is identical to the previous one in Eq.~\eqref{eq:gtr_z_old}, including the ordering of the Grassmann measures in ${\rm gTr}$.
Therefore, one can easily repeat the decimation procedure toward the thermodynamic limit as in the usual Levin-Nave TRG.
We can understand that the SVD of the Grassmann tensor introduces new auxiliary Grassmann variables on the coarse-grained lattice $\Lambda'$.
Therefore, the algorithm for the Grassmann tensor network perfectly corresponds to that for the normal tensor network as shown in Figure~\ref{fig:schematic_levin_nave_trg}.
In terms of coefficient tensors, several sign factors should be included in Eqs.~\eqref{eq:gsvd_2} and \eqref{eq:g_tensor_new} as we will see in Sec.~\ref{subsec:remark}.

\subsection{Approximate contraction by the Higher-Order TRG}
\label{subsec:ghotrg}

\begin{figure}[htbp]
  	\centering
	\includegraphics[width=1\hsize]{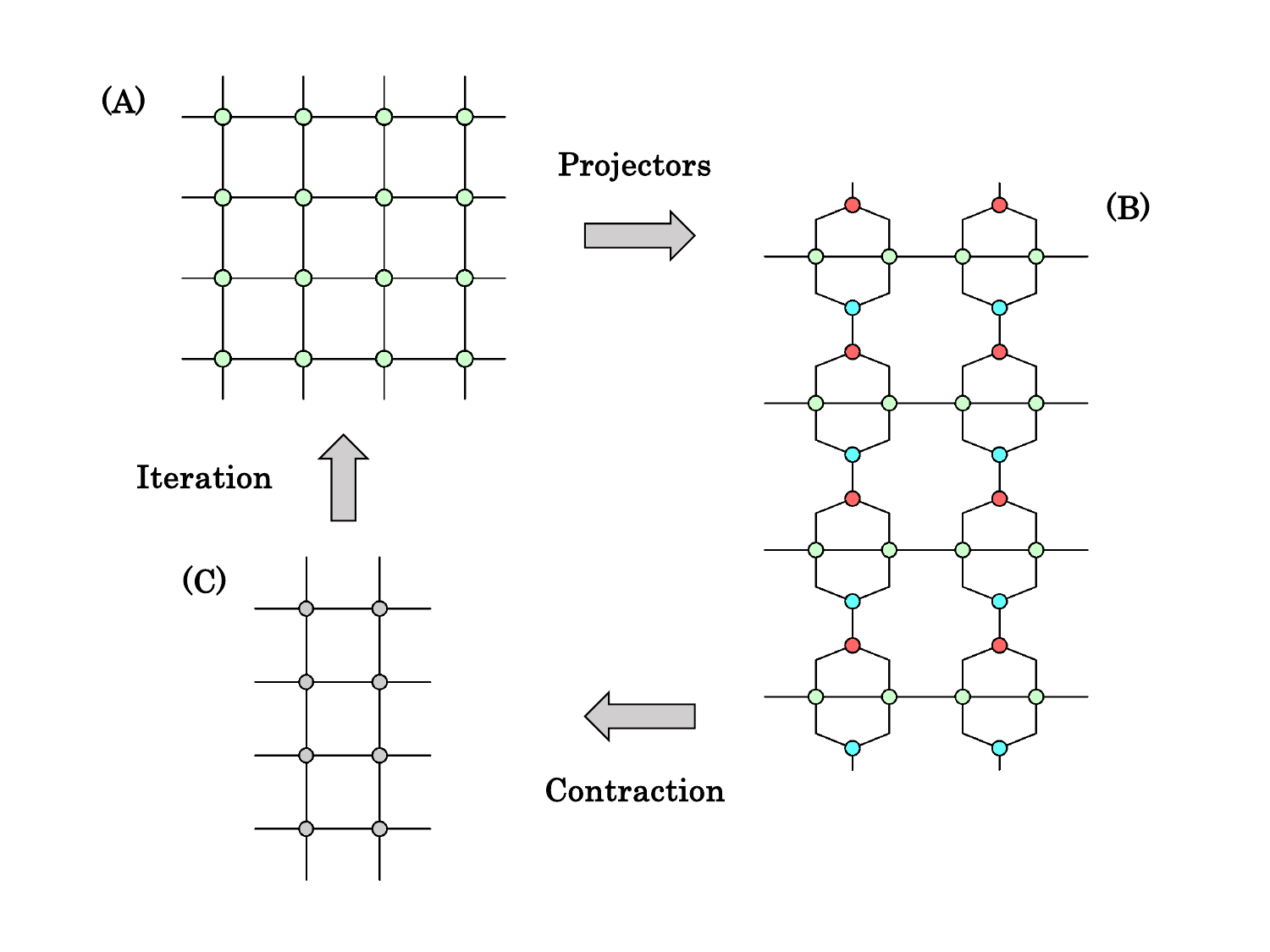}
	\caption{
	Schematic illustration of the HOTRG algorithm.
        (A) Initial tensor network on a square lattice.
        (B) Inset projectors into the network.
        Red and blue symbols show $P$ and $Q$ in Eq.~\eqref{eq:hotrg_update}, respectively.
        (C) New tensor network by contracting adjacent two fundamental tensors with two projectors.
	}
  	\label{fig:schematic_hotrg}
\end{figure}

Let us next focus on the HOTRG algorithm that applies to any $d$-dimensional system. 
The computational complexity scales with $O\left(D_{\rm HOTRG}^{4d-1}\right)$ and the memory cost does with $O\left(D_{\rm HOTRG}^{2d}\right)$, where $D_{\rm HOTRG}$ is the bond dimension in the HOTRG.
For simplicity, we consider the two-dimensional tensor network on a periodic square lattice again.
The original algorithm aims to evaluate Eq.~\eqref{eq:ttr_z} not decimating each local fundamental tensor but inserting the projectors that describe the coarse-graining transformation in the tensor-network language.
Unlike the Levin-Nave TRG, the HOTRG performs the contraction between two adjacent fundamental tensors along each direction sequentially.

We begin with reviewing the normal HOTRG.
As an example, we consider the contraction between $T_{n+\hat{1}}$ and $T_{n}$.
Introducing
\begin{align}
\label{eq:double_t}
    M_{xt_{1}t_{2}x't'_{1}t'_{2}}
    =
    \sum_{\alpha}
    T_{n+\hat{1};xt_{1}\alpha t'_{1}}
    T_{n;\alpha t_{2}x't'_{2}}
    ,
\end{align}
the HOTRG provides the coarse-graining transformation such that
\begin{align}
\label{eq:hotrg_update}
    T_{n';xtx't'}
    =
    \sum_{t_{1},t_{2},t'_{1},t'_{2}}
    P_{t_{1}t_{2}t}
    M_{xt_{1}t_{2}x't'_{1}t'_{2}}
    Q_{t't'_{1}t'_{2}}
    ,
\end{align}
where the three-leg tensors $P$ and $Q$ are projectors to decimate the degrees of freedom.
They play roles to map the original degrees of freedom $(t^{(')}_{1}t^{(')}_{2})$ to the coarse-grained one $t^{(')}$, whose size is restricted by $D_{\rm HOTRG}$, the bond dimension in the algorithm.
We can also regard that the HOTRG is inserting the following four-leg tensor,
\begin{align}
\label{eq:pseudo_id}
    W_{t_{1}t_{2}t'_{1}t'_{2}}
    =
    \sum_{t=1}^{D_{\rm HOTRG}}
    P_{t_{1}t_{2}t}Q_{tt'_{1}t'_{2}}
    ,
\end{align}
into the tensor network. 
With sufficiently large $D_{\rm HOTRG}$, $W_{t_{1}t_{2}t'_{1}t'_{2}}$ should be equivalent to $\delta_{t_{1},t'_{1}}\delta_{t_{2},t'_{2}}$ and the algorithm gives the exact contraction.
$T_{n'}$ generates the approximated tensor network representation of the partition function.
The coarse-graining transformation in the HOTRG can be easily repeated. 
As in the case of the Levin-Nave TRG, the HOTRG allows us to contract $2^{N}$ original fundamental tensors approximately, just in $N$ times of iteration.
Figure~\ref{fig:schematic_hotrg} graphically demonstrates the procedure of the algorithm.

There are several ways to determine $P$ and $Q$.
Here, we just follow the original proposal in Ref.~\cite{2012PhRvB..86d5139X} assuming the translational symmetry for the tensor network in Eq.~\eqref{eq:ttr_z}.
We consider the following two reduced density matrices,
\begin{align}  
\label{eq:rho_np}
    \rho_{(t_{1}t_{2})(\tilde{t}_{1}\tilde{t}_{2})}
    =
    \sum_{x,x',t'_{1},t'_{2}}
    M_{xt_{1}t_{2}x't'_{1}t'_{2}}
    M_{x\tilde{t}_{1}\tilde{t}_{2}x't'_{1}t'_{2}}^{*}
    ,
\end{align}
\begin{align}  
\label{eq:rho_p}
    \rho_{(t'_{1}t'_{2})(\tilde{t}'_{1}\tilde{t}'_{2})}
    =
    \sum_{x,t_{1},t_{2},x'}
    M_{xt_{1}t_{2}x't'_{1}t'_{2}}
    M_{xt_{1}t_{2}x'\tilde{t}'_{1}\tilde{t}'_{2}}^{*}
    .
\end{align}
These matrices can be decomposed as
\begin{align}
\label{eq:svd_np}
    \rho_{(t_{1}t_{2})(\tilde{t}_{1}\tilde{t}_{2})}
    =
    \sum_{i}
    U_{(t_{1}t_{2})i}\lambda_{i}U^{\dag}_{i(\tilde{t}_{1}\tilde{t}_{2})}
    ,
\end{align}
\begin{align}
\label{eq:svd_p}
    \rho'_{(t'_{1}t'_{2})(\tilde{t}'_{1}\tilde{t}'_{2})}
    =
    \sum_{i}
    V_{(t'_{1}t'_{2})i}\lambda'_{i}V^{\dag}_{i(\tilde{t}'_{1}\tilde{t}'_{2})}
    ,
\end{align}
where $U$ and $V$ are unitary matrices and $\lambda$ and $\lambda'$ denote the singular-value matrices with descending order.
Defining the following quantities,
\begin{align}
    \epsilon^{(')}
    =
    \sum_{i>D_{\rm HOTRG}}\lambda^{(')}_{i}
    ,
\end{align}
we choose $P_{t_{1}t_{2}t}=U^{*}_{(t_{1}t_{2})t}$ and $Q_{t't'_{1}t'_{2}}=U_{(t'_{1}t'_{2})t'}$ if $\epsilon<\epsilon'$ and choose $P_{t_{1}t_{2}t}=V_{(t_{1}t_{2})t}$ and $Q_{t't'_{1}t'_{2}}=V^{*}_{(t'_{1}t'_{2})t'}$ if $\epsilon>\epsilon'$.
This procedure is equivalent to the higher-order SVD (HOSVD), which is a tensorial extension of the SVD as explained in Ref.~\cite{2012PhRvB..86d5139X}.
See Refs.~\cite{wang2011cluster,PhysRevLett.113.046402,PhysRevB.100.035449} for other ways to derive the optimal $P$ and $Q$ without assuming the translational invariance on the tensor network.

The HOTRG algorithm is straightforwardly extended to evaluate the Grassmann tensor network in Eq.~\eqref{eq:gtr_z_old}.
Eq.~\eqref{eq:double_t} is correspondingly denoted by
\begin{align}
    \mathcal{M}_{XT_{1}T_{2}\bar{X}\bar{T}_{2}\bar{T}_{1}}
    =
    \int_{\bar{\Theta},\Theta}
    \mathcal{T}_{n+\hat{1};XT_{1}\bar{\Theta}\bar{T}_{1}}
    \mathcal{T}_{n;\Theta T_{2}\bar{X}\bar{T}_{2}}
    ,
\end{align}
where the coefficient tensor of $\mathcal{M}$ has been already derived in Eq.~\eqref{eq:common_coeff} or Eq.~\eqref{eq:new_coeff_super}.
The HOTRG for the Grassmann tensor network should provide us with the coarse-graining transformation such as
\begin{align}
\label{eq:ghotrg_update}
    \mathcal{T}_{n';
        XT\bar{X}\bar{T}
    }
    =
    \int_{\bar{T}_{1},T_{1}}
    \int_{\bar{T}_{2},T_{2}}
    \int_{\bar{T}'_{1},T'_{1}}
    \int_{\bar{T}'_{2},T'_{2}}
    \mathcal{P}_{\bar{T}_{2}\bar{T}_{1}T}
    \mathcal{M}_{XT_{1}T_{2}\bar{X}\bar{T}'_{2}\bar{T}'_{1}}
    \mathcal{Q}_{\bar{T}T'_{1}T'_{2}}
    ,
\end{align}
where $\mathcal{P}$ and $\mathcal{Q}$ are the Grassmann projectors defined by
\begin{align}
\label{eq:grassmann_p}
    \mathcal{P}_{\bar{T}_{2}\bar{T}_{1}T}
    =
    \sum_{t_{1},t_{2},t}
    P_{t_{1}t_{2}t}
    \bar{T}_{2}^{t_{2}}\bar{T}_{1}^{t_{1}}T^{t}
    ,
\end{align}
\begin{align}
\label{eq:grassmann_q}
    \mathcal{Q}_{\bar{T}T'_{1}T'_{2}}
    =
    \sum_{t'_{1},t'_{2},t'}
    Q_{t't'_{1}t'_{2}}
    \bar{T}^{t'}T^{'t'_{1}}_{1}T^{'t'_{2}}_{2}
    .
\end{align}
In analogy with Eq.~\eqref{eq:pseudo_id}, we can identify that the algorithm inserts
\begin{align}
\label{eq:pseudo_g_id}
    \mathcal{W}_{\bar{T}_{2}\bar{T}_{1}T'_{1}T'_{2}}
    =
    \int_{\bar{T},T}^{D_{\rm HOTRG}}
    \mathcal{P}_{\bar{T}_{2}\bar{T}_{1}T}
    \mathcal{Q}_{\bar{T}T'_{1}T'_{2}}
    ,
\end{align}
into the Grassmann tensor network. 
$\mathcal{P}$ and $\mathcal{Q}$, or $P$ and $Q$ in other words, are determined via the same procedure in the normal HOTRG.
Firstly, we define the conjugation of the Grassmann tensor by
\begin{align}
\label{eq:g_conjugate}
    (\mathcal{O}_{\Phi\Psi})^{\dag}
    =
    \mathcal{O}^{*}_{\bar{\Psi}\bar{\Phi}}
    =
    \sum_{i,j}O^{*}_{ij}\bar{\Psi}^{j}\bar{\Phi}^{i}
    ,
\end{align}
where we have assumed that $\mathcal{O}_{\Phi\Psi}$ is defined by Eq.~\eqref{eq:def_grassmann_m}.
Note that for $\Phi^{i}=\Phi_{1}^{i_{1}}\cdots\Phi_{m}^{i_{m}}$, we define $\bar{\Phi}^{i}$ in Eq.~\eqref{eq:g_conjugate} as $\bar{\Phi}^{i}=\bar{\Phi}_{m}^{i_{m}}\cdots\bar{\Phi}_{1}^{i_{1}}$.
$\bar{\Psi}^{j}$ is defined in the same way.
Then, we define the Grassmann reduced density matrices as
\begin{align}
    \varrho_{T_{1}T_{2}\bar{T}_{2}\bar{T}_{1}}
    =
    \int_{\bar{X},X}
    \int_{X',\bar{X}'}
    \int_{T'_{2},\bar{T}'_{2}}
    \int_{T'_{1},\bar{T}'_{1}}
    \mathcal{M}_{XT_{1}T_{2}\bar{X}'\bar{T}'_{2}\bar{T}'_{1}}
    \mathcal{M}^{*}_{T'_{1}T'_{2}X'\bar{T}_{2}\bar{T}_{1}\bar{X}}
    ,
\end{align}
\begin{align}
    \varrho'_{\bar{T}_{2}\bar{T}_{1}T_{1}T_{2}}
    =
    \int_{\bar{X},X}
    \int_{\bar{T}'_{1},T'_{1}}
    \int_{\bar{T}'_{2},T'_{2}}
    \int_{X',\bar{X}'}
    \mathcal{M}_{XT'_{1}T'_{2}\bar{X}'\bar{T}_{2}\bar{T}_{1}}
    \mathcal{M}^{*}_{T_{1}T_{2}X'\bar{T}'_{2}\bar{T}'_{1}\bar{X}}
    ,
\end{align}
where the ordering of the Grassmann measures are defined so that the coefficient tensors of $\varrho_{T_{1}T_{2}\bar{T}_{2}\bar{T}_{1}}$ and $\varrho'_{\bar{T}_{2}\bar{T}_{1}T_{1}T_{2}}$ are given by $\rho_{(t_{1}t_{2})(\tilde{t}_{1}\tilde{t}_{2})}$ and $\rho'_{(t'_{1}t'_{2})(\tilde{t}'_{1}\tilde{t}'_{2})}$ in Eqs.~\eqref{eq:rho_np} and \eqref{eq:rho_p}, respectively.
Note that $\varrho_{T_{1}T_{2}\bar{T}_{2}\bar{T}_{1}}$ and $\varrho'_{\bar{T}_{2}\bar{T}_{1}T_{1}T_{2}}$ are Grassmann-even since $\mathcal{M}_{XT_{1}T_{2}\bar{X}'\bar{T}'_{2}\bar{T}'_{1}}$ is Grassmann-even.
As we have demonstrated in Sec.~\ref{subsec:gtrg}, the SVD of these Grassmann reduced density matrices results in the SVD of their coefficient tensors as in Eqs.~\eqref{eq:svd_np} and \eqref{eq:svd_p}.
The coefficient tensors $P$ and $Q$ in Eqs.~\eqref{eq:grassmann_p} and \eqref{eq:grassmann_q} are then determined similarly with the normal HOTRG.
Since resulting $\mathcal{P}$ and $\mathcal{Q}$ are Grassmann-even, no extra sign factor appears in calculating the right-hand side of Eq.~\eqref{eq:ghotrg_update}.
The Grassmann tensor $\mathcal{T}_{n'}$ in Eq.~\eqref{eq:ghotrg_update} gives the coarse-grained Grassmann tensor network ${\rm gTr}
    \left[
        \prod_{n'\in\Lambda'}
        \mathcal{T}_{n'}
    \right]$ 
with the same ordering of Grassmann measures in Eq.~\eqref{eq:gtr_z_old}, as is evident from Eq.~\eqref{eq:pseudo_g_id}.
Therefore, we can iterate the coarse-graining transformation as in Fig.~\ref{fig:schematic_hotrg} even for the Grassmann tensor network, without any extra difficulty.

We have reviewed two types of TRG algorithms so far and both of them employ the local approximation based on the (HO)SVD to define the coarse-graining transformations.
By constructing a coarse-graining transformation that includes not only the local tensor(s) but also the effects of the surrounding tensors (they are usually referred to as the environment), we can construct a more accurate transformation.
This kind of improvement is called the second renormalization group (SRG)~\cite{PhysRevLett.103.160601,PhysRevB.81.174411,2012PhRvB..86d5139X}.
One of the other ways to improve the TRG algorithms is to remove the redundant loop structure in the tensor network~\cite{Gu:2009dr}.
This can be achieved by the tensor network renormalization (TNR)~\cite{2015PhRvL.115r0405E,2015arXiv150907484E}.
We will see several TNR-type algorithms in Sec.~\ref{sec:improvements}.

In addition to the TRG approach, many other algorithms perform approximate tensor contractions, and they are used according to their purpose and cost performance.
For example, corner transfer matrix renormalization group (CTMRG)~\cite{doi:10.1143/JPSJ.65.891} and time-evolving block decimation (TEBD)~\cite{PhysRevB.78.155117,PhysRevB.78.155117} are widely used in the tensor network computations based on the Hamiltonian formalism.
See Ref.~\cite{Okunishi:2021but} as a recent review.

\subsection{Practical remarks}
\label{subsec:remark}

Here, we see how to carry out the SVD of Grassmann tensors in practical computations.
We again consider the Grassmann-even tensor $\mathcal{O}_{\Phi\Psi}$ in Eq.~\eqref{eq:def_grassmann_m} as an example.
Since $\mathcal{O}_{\Phi\Psi}$ is Grassmann-even, the corresponding coefficient tensor $O_{ij}$ takes a non-zero value only when $\Phi^{i}\Psi^{j}$ is Grassmann-even.
Therefore, we can always convert $O_{ij}$ into a block-diagonalized form by the matrix elementary operations; regarding $i$ as a row and $j$ as a column, $O_{ij}$ can be 
\begin{align}
\label{eq:block_diagonalized_coeff}
    O=
    \begin{blockarray}{ccccc}
        j~{\rm:even} & j~{\rm:odd} & \\
        \begin{block}{[cc]ccc}
            O^{\rm (even)} & 0 & i~{\rm:even}  \\
            0 & O^{\rm (odd)} & i~{\rm:odd~}   \\
        \end{block}
    \end{blockarray}
    ,
\end{align}
where ``$i$~:~even~(odd)" means the row index $i$ such that $\Phi^{i}$ becomes Grassmann-even~(odd).
Similarly, ``$j$~:~even~(odd)" means the column index $j$ such that $\Psi^{j}$ becomes Grassmann-even~(odd).
Using this basis, the SVD of $O$ gives us
\begin{align}
    O_{ij}
    =
    \sum_{k,l}
    U_{ik}
    \sigma_{kl}
    V^{\dag}_{lj}
    ,
\end{align}
where $U$ and $V^{\dag}$ are unitary matrices such that
\begin{align}
\label{eq:gsvd_u}
    U=
    \begin{blockarray}{ccccc}
        k~{\rm:even} & k~{\rm:odd} & \\
        \begin{block}{[cc]ccc}
            U^{\rm (even)} & 0 & i~{\rm:even}  \\
            0 & U^{\rm (odd)} & i~{\rm:odd~}   \\
        \end{block}
    \end{blockarray}
    ,
\end{align}
\begin{align}
\label{eq:gsvd_v}
    V^{\dag}=
    \begin{blockarray}{ccccc}
        j~{\rm:even} & j~{\rm:odd} & \\
        \begin{block}{[cc]ccc}
            V^{\dag \rm (even)} & 0 & l~{\rm:even}  \\
            0 & V^{\dag \rm (odd)} & l~{\rm:odd~}   \\
        \end{block}
    \end{blockarray}
    ,
\end{align}
and $\sigma_{kl}=\sigma_{k}\delta_{kl}$ with the singular value $\sigma_{k}$ and Kronecker's delta $\delta_{kl}$ in the form of
\begin{align}
\label{eq:gsvd_sigma}
    \sigma=
    \begin{blockarray}{ccccc}
        l~{\rm:even} & l~{\rm:odd} & \\
        \begin{block}{[cc]ccc}
            \sigma^{\rm (even)} & 0 & k~{\rm:even}  \\
            0 & \sigma^{\rm (odd)} & k~{\rm:odd~}   \\
        \end{block}
    \end{blockarray}
    .
\end{align}
Recalling Eq.~\eqref{eq:id_delta} and Eq.~\eqref{eq:def_gsvd}, we can associate $k$ and $l$ with new auxiliary Grassmann variables via $\Xi^{k}$ and $\bar{\Xi}^{l}$.
In Eqs.~\eqref{eq:gsvd_u} and \eqref{eq:gsvd_sigma},  ``$k$~:~even~(odd)" should be understood as $\Xi^{k}$ is of the Grassmann-even (odd) parity.
In the same way, ``$l$~:~even~(odd)" in Eqs.~\eqref{eq:gsvd_v} and \eqref{eq:gsvd_sigma} should be understood as $\bar{\Xi}^{l}$ is Grassmann-even (odd).
Although we have two kinds of Grassmann variables $\Xi^{k}$ and $\bar{\Xi}^{l}$, Kronecker's delta $\delta_{kl}$ in $\sigma_{kl}$ enforces them to be of the same Grassmann parity.
Therefore, we can easily define the new Grassmann parity functions corresponding to the new auxiliary Grassmann variables.

In the practical TRG computations, the Grassmann parity functions play a crucial role in restoring the Grassmann calculus on your computer.
In the case of the Levin-Nave TRG explained in Sec.~\ref{subsec:gtrg}, the data that should be kept on the computer are the coefficient tensor $T_{n;xtx't'}$ and Grassmann parity functions for each subscript.
Here, we repeat the discussion from Eq.~\eqref{eq:gsvd_1} to Eq.~\eqref{eq:gtr_z_new}.
The two kinds of SVD for the Grassmann tensors in Eqs.~\eqref{eq:gsvd_1} and \eqref{eq:gsvd_2} correspond to
\begin{align}
\label{eq:lntrg_gsvd_e}
    T_{n;xtx't'}
    \simeq
    \sum_{a=1}^{D_{\rm LNTRG}}
    A_{n;xta}
    B_{n;ax't'}
    ,
\end{align}
\begin{align}
\label{eq:lntrg_gsvd_o}
    (-1)^{f_{x'}(f_{x}+f_{t})+f_{x}f_{t}}
    T_{n;xtx't'}
    \simeq
    \sum_{a=1}^{D_{\rm LNTRG}}
    C_{n;xt'a}
    D_{n;axt'}
    ,
\end{align}
where $A$, $B$, $C$ and $D$ are defined in the similar way to Eqs.~\eqref{eq:lntrg_svd_e} and \eqref{eq:lntrg_svd_o}.
Notice that the extra sign factor in Eq.~\eqref{eq:lntrg_gsvd_o} comes from the rearrangement of auxiliary Grassmann variables.
Using the block-diagonalized basis, one can immediately find the corresponding Grassmann parity for the new index $a$.
Eq.~\eqref{eq:g_tensor_new}, or the diagram in Figure~\ref{fig:ln_contract}, is now ready to be computed.
We begin with contracting $\mathcal{A}$ and $\mathcal{D}$, that gives us an intermediate Grassmann tensor,
\begin{align}
    (\mathcal{AD})_{T_{1}T\bar{X}T_{2}}
    =
    \int_{\bar{X}_{1},X_{1}}
    \mathcal{A}_{n;X_{1}T_{1}T}
    \mathcal{D}_{n+\hat{1};\bar{X}\bar{X}_{1}T_{2}}
    ,
\end{align}
which is described by the corresponding coefficient tensors as
\begin{align}
    (AD)_{t_{1}tx't_{2}}
    =
    \sum_{x_{1},x'_{1}}
    (-1)^{f_{x_{1}}(f_{t_{1}}+f_{t})+f_{x'_{1}}f_{x'}}
    A_{n;x_{1}t_{1}t}
    \delta_{x_{1}x'_{1}}
    D_{n+\hat{1};x'x'_{1}t_{2}}
    .
\end{align}
The contraction between $\mathcal{C}$ and $\mathcal{B}$ is
\begin{align}
    (\mathcal{CB})_{\bar{T}_{1}X\bar{T}\bar{T}_{2}}
    =
    \int_{\bar{X}_{2},X_{2}}
    \mathcal{C}_{n+\hat{2};X_{2}\bar{T}_{1}X}
    \mathcal{B}_{n+\hat{1}+\hat{2};\bar{T}\bar{X}_{2}\bar{T}_{2}}
    ,
\end{align}
which corresponds with
\begin{align}
    (CB)_{t'_{1}xt't'_{2}}
    =
    \sum_{x_{2},x'_{2}}
    (-1)^{f_{x_{2}}(f_{t'_{1}}+f_{x})+f_{x'_{2}}f_{t'}}
    C_{n+\hat{2};x_{2}t'_{1}x}
    \delta_{x_{2}x'_{2}}
    B_{n+\hat{1}+\hat{2};t'x'_{2}t'_{2}}
    .
\end{align}
Finally, the contraction between $(\mathcal{AD})$ and $(\mathcal{CB})$ gives a new Grassmann tensor in Eq.~\eqref{eq:g_tensor_new}; 
\begin{align}
    \mathcal{T}_{n';XT\bar{X}\bar{T}}
    =
    \int_{\bar{T}_{1},T_{1}}
    \int_{\bar{T}_{2},T_{2}}
    (\mathcal{AD})_{T_{1}T\bar{X}T_{2}}
    (\mathcal{CB})_{\bar{T}_{1}X\bar{T}\bar{T}_{2}}
    ,
\end{align}
which is restored by
\begin{align}
    T_{n';xtx't'}
    =
    (-1)^{f_{x}(f_{t}+f_{x'})}
    \sum_{t_{1},t'_{1}}
    \sum_{t_{2},t'_{2}}
    (-1)^{
        f_{t_{1}}(f_{t}+f_{x'})
        +
        f_{t'_{2}}(f_{t'_{1}}+f_{x}+f_{t'})
    }
    (AD)_{t_{1}tx't_{2}}
    \delta_{t_{1}t'_{1}}
    \delta_{t_{2}t'_{2}}
    (CB)_{t'_{1}xt't'_{2}}
    ,
\end{align}
where the sign factor inside the summations originates from reordering auxiliary Grassmann variables in $(\mathcal{AD})$ and $(\mathcal{CB})$.
The sign factor outside the summations comes from reordering auxiliary Grassmann variables after the summations.
\footnote{
One can see that the algorithm should result in the $\mathbb{Z}_{2}$-symmetric Levin-Nave TRG when we set all the Grassmann parity functions to zero.
}

The Grassmann tensor trace is also described by the coefficient tensor and parity functions. 
Suppose $\mathcal{T}^{(N)}_{n;XT\bar{X}\bar{T}}$ be a Grassmann tensor obtained by $N$ times of Levin-Nave TRG or HOTRG transformation.
The path integral defined on a lattice with $2^{N}$ sites is approximately given by ${\rm gTr}[\mathcal{T}^{(N)}_{n}]$ and one finds
\begin{align}
    {\rm gTr}\left[\mathcal{T}^{(N)}_{n}\right]
    =
    \sum_{x,t}
    (-1)^{f_{x}f_{t}}
    T^{(N)}_{n;xtxt}
    .
\end{align}
In the case of the finite-temperature system, we need to impose the anti-periodic boundary condition for the temporal direction, which can be done by using the Grassmann parity function $f_{t}$ via
\begin{align}
    {\rm gTr}\left[\mathcal{T}^{(N)}_{n}\right]
    =
    \sum_{x,t}
    (-1)^{f_{x}f_{t}+f_{t}}
    T^{(N)}_{n;xtxt}
    .
\end{align}
Note that the extra sign factor $(-1)^{f_{t}}$ is a result of replacing $T^{t}$ with $(-T)^{t}$ in $\mathcal{T}^{(N)}_{n;XT\bar{X}\bar{T}}$.

\section{Examples of numerical calculations}
\label{sec:examples}
\subsection{Wilson--Majorana fermions}
\label{sec:wilsonmajorana}
A simple example of classical action quadratic in Grassmann variables is given by the action of Wilson--Majorana fermions: 
\begin{align}
   S =
  & \frac{1}{2} \sum_{n} \bar{\eta}(n) \left( m_{\eta} + \sum_{\mu=1}^{2} \gamma_{\mu} \partial^{\mathrm{S}}_{\mu} - \frac{1}{2} \sum_{\mu=1}^{2} \partial_{\mu}\partial^{\ast}_{\mu} \right) \eta(n) \nonumber \\
   & + \frac{1}{2} \sum_{n} \bar{\chi}(n) \left( m_{\chi} + \sum_{\mu=1}^{2} \gamma_{\mu} \partial^{\mathrm{S}}_{\mu} - \frac{1}{2} \sum_{\mu=1}^{2} \partial_{\mu}\partial^{\ast}_{\mu} \right) \chi(n) \nonumber \\
  & + \sum_{n} \bar{\eta}(n) \left( \gamma_{1} \partial^{\mathrm{S}}_{1}
    - \gamma_{2} \partial^{\mathrm{S}}_{2}
    - \frac{1}{2} \partial_{1}\partial^{\ast}_{1}
    + \frac{1}{2} \partial_{2}\partial^{\ast}_{2} \right) \chi(n)
\end{align}
with two-component Majorana spinors $\eta \equiv (\eta_1,\eta_2)^{\rm{T}}$ and $\chi\equiv(\chi_1,\chi_2)^{\rm{T}}$.
The forward, the backward, and the symmetric difference operators are defined by $\partial$, $\partial^{\ast}$, and $\partial^{\mathrm{S}} = (\partial + \partial^{\ast})/2$, respectively.

An equivalence between the lattice action for Wilson--Majorana fermions and the Ising model has been shown for two-dimensional honeycomb and square lattices for any choice of the periodic/anti-periodic boundary conditions~\cite{Wolff:2020oky}.
The masses of the fermions are functions of the ``reverse temperature'' $\kappa$:
\begin{align}
  m_{\eta} = \frac{2}{\kappa}\left( \sqrt{2}-1-\kappa \right), && m_{\chi} = - \frac{2}{\kappa}\left( \sqrt{2}+1+\kappa \right).
\end{align}
The critical point of the system $\kappa_{\mathrm{c}}=\sqrt{2}-1$ is associated with vanishing $m_{\eta}$, and this value is related to the critical point of the
two-dimensional Ising model on a square lattice $\beta_{\mathrm{c}} = \tanh^{-1} \kappa_{\mathrm{c}}$.
Even in such a model where there are several species of fermions, one can follow the prescription given in the previous section to derive a tensor network representation.

\begin{figure}[htbp]
    \centering
    \includegraphics[width=0.8\hsize]{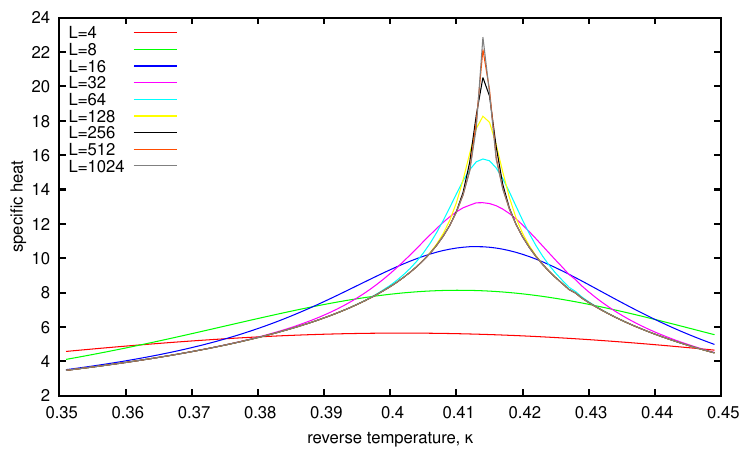}
    \caption{Specific heats of the Majorana--Wilson fermion system on several sizes of lattice. The specific heat is defined as the second derivative of free energy that is calculated by the Grassmann TRG with $D_{\mathrm{LNTRG}}=64$.}
    \label{fig:majoranawilson_specheat}
\end{figure}

One can see the equivalence to the Ising model from the specific heat of this model in Fig.~\ref{fig:majoranawilson_specheat}.
Although we omit to show a detailed finite-size scaling analysis, one can see the logarithmic divergence of the specific heat at the critical point $\kappa_{\mathrm{c}}=\sqrt{2}-1$.

In the latter section, we will show the ``renormalization flow'' of this model given by the plain and the improved coarse-graining methods.

\subsection{The Schwinger model}

In the HEP community, Shimizu and Kuramashi did a series of pioneering works for the two-dimensional Schwinger model with Wilson fermions~\cite{Shimizu:2014uva,Shimizu:2014fsa,Shimizu:2017onf}.

In Ref.~\cite{Shimizu:2014uva}, where the Grassmann TRG was firstly applied to a lattice gauge theory, the Schwinger model
\begin{align}
    S = &  -\frac{1}{2} \sum_{n} \sum_{\mu=1}^{2} \bar{\psi}(n) \left\{ \left( 1-\gamma_{\mu} \right) U_{\mu}(n) \psi(n+\hat{\mu}) + \left( 1+\gamma_{\mu} \right) U^{\dagger}_{\mu}(n-\hat{\mu}) \psi(n-\hat{\mu}) \right\} \nonumber \\
   &+ \frac{1}{2\kappa} \sum_{n} \bar{\psi}(n) \psi(n)
   - \beta \sum_{n} \cos \left( A_{1}(n) + A_{2}(n+\hat{1}) - A_{1}(n+\hat{2}) - A_{2}(n) \right),
\end{align}
where $A$ is the phase of $U(1)$ link variable,
was analyzed for some choices of the reverse coupling constant $\beta$: the strong coupling limit $\beta=0$ and finite couplings $\beta=5,10$.

\begin{figure}[htbp]
    \centering
    \includegraphics[width=0.7\hsize]{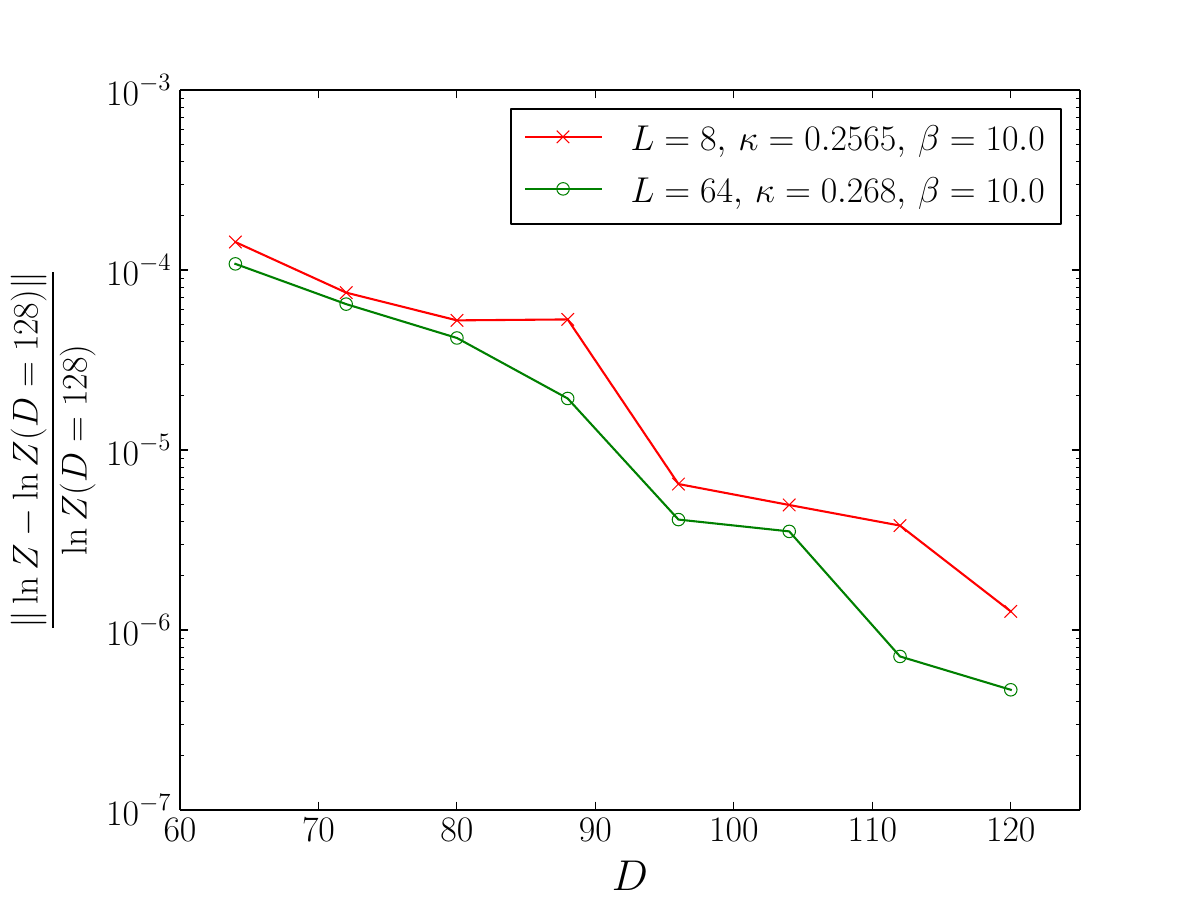}
    \caption{Adapted from Ref.~\cite{Shimizu:2014uva}. Relative errors of the free energy.}
    \label{fig:err_schwinger}
\end{figure}

They confirmed their formulation and code by checking convergences of relative errors of free energy along with increasing bond dimension (see Fig.~\ref{fig:err_schwinger}).
They defined the relative error by using the result obtained by a large bond dimension $D_{\rm LNTRG}=128$, while they fixed their bond dimension for the analyses to $D_{\rm LNTRG}=96$.

\begin{figure}[htbp]
    \centering
    \includegraphics[width=0.7\hsize]{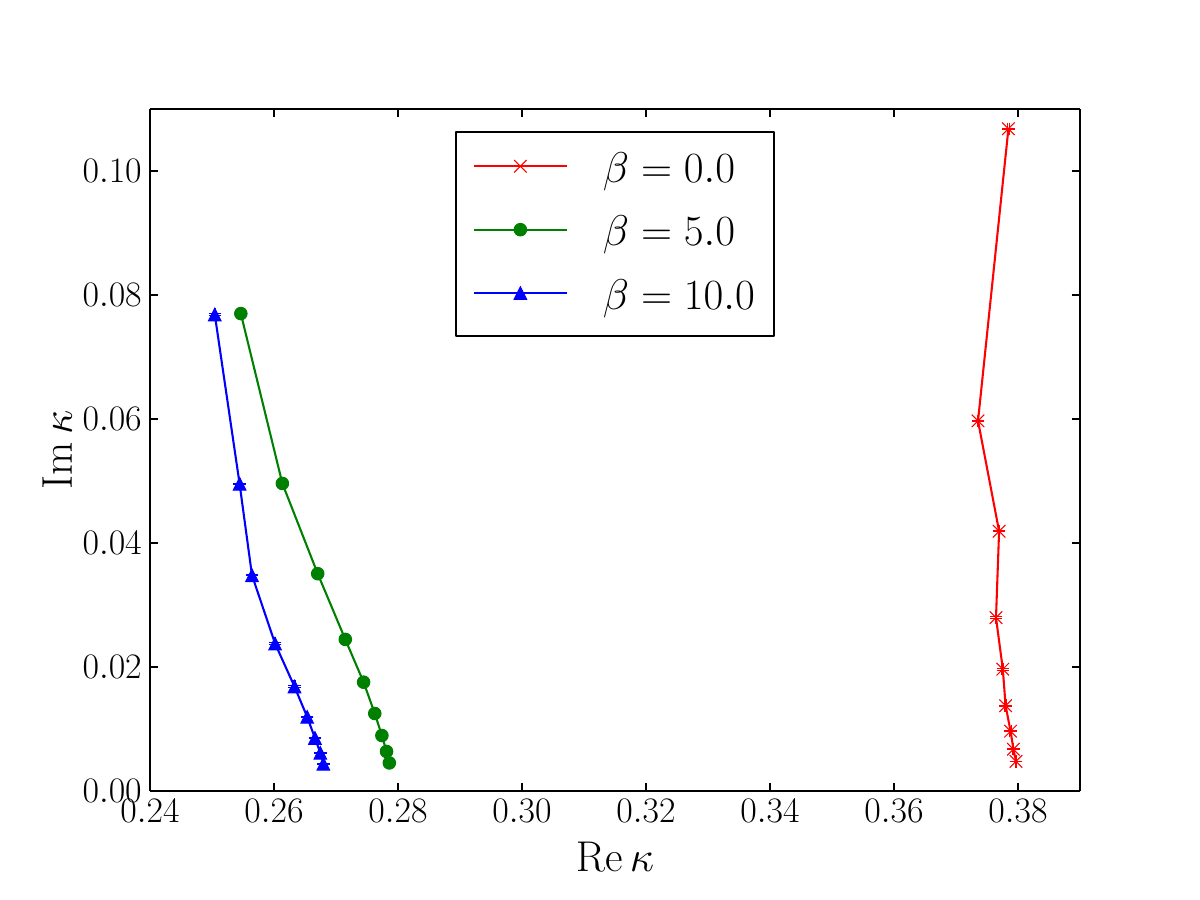}
    \caption{Adapted from Ref.~\cite{Shimizu:2014uva}. Lee--Yang zeros in the complex $\kappa$-plane.}
    \label{fig:ly_schwinger}
\end{figure}

In the paper, finite-size scaling analyses is taken place for peak heights of the chiral susceptibility and the partition function zeros in the complex $\kappa$-plane.
Using the Lee--Yang zeros, they did some reliable fittings to find out the critical exponents (see Fig.~\ref{fig:ly_schwinger}).
Note that such an investigation in the complex parameter plane fully utilizes the absence of the sign problem for the TRG approach.
Both in the strong coupling limit and for finite couplings, the critical exponents are shown to be the same as those of the two-dimensional Ising model.
Moreover, the obtained critical mass (hopping parameter) is consistent with a previous work with the eight vertex model~\cite{Gausterer:1995np}.

After this work, an analysis of the same model with the presence of $\theta$ term 
 that impressed the efficiency of the TRG approach on the community follows~\cite{Shimizu:2014fsa}.
In that work, they studied the critical behavior on $\theta=\pi$ line,
and in conclusion, on the line $\theta = \pi$, no phase transitions occur for $\kappa > \kappa_{\mathrm{c}}$, a second order phase transition that belongs to the two-dimensional Ising universality class occurs at $\kappa = \kappa_{\mathrm{c}} \simeq 0.2415$,
and first-order phase transitions occur for $\kappa < \kappa_{\mathrm{c}}$.
This is exactly the expected result.

The Schwinger model with the topological term, where the staggered discretization for the fermions is applied, was studied by a different group~\cite{Butt:2019uul}.
In their paper, a special property of the model was mentioned: a tensor network representation can be constructed without having Grassmann variables on the network.
Indeed this property was found in the context of Monte Carlo simulation in Refs.~\cite{Gattringer:2015nea,Goschl:2017kml}.

\section{Improved TRG methods for fermions}
\label{sec:improvements}

The Grassmann TRG methods introduced in Secs.~\ref{subsec:gtrg} and \ref{subsec:ghotrg} show remarkable performances as seen for some specific models in the latter sections.
However, the key component of the RG methods was the (HO)SVD that gives the best approximation for local tensors~\cite{eckart1936approximation} rather than the whole network. 
As we mentioned in the introduction, there are possible issues to describe the RG flows near critical points~\cite{Levin:2006jai,Gu:2009dr,Vanhecke:2019pez}.
This is an important motivation to seek refined RG techniques.
For tensor networks, the quality of the RG methods is often related to the ``entanglement'' of the system.
In Ref.~\cite{2015PhRvL.115r0405E}, an improved renormalization technique where (dis)entanglers are introduced to the network and are variationally tuned so that the short-range correlation is removed at each RG step was introduced.
Also, some new and less computationally demanding approaches like the loop-TNR~\cite{yang2017loop} and the gilt-TNR~\cite{Hauru:2017tne} were developed.
Such approaches are called \textit{tensor network renormalization (TNR)}.
Basically, the TNR approaches require some optimization steps on the network,
but one can easily adapt them to the Grassmann tensor networks since the treatment of the Grassmann variables is exact and factored out from the bosonic part of the network.
Note that, later in this section, we also show the Grassmann version of the bond-weighted TRG,
which represents a different notion than the TNRs.

\subsection{Corner double line structure on tensor network}

\begin{figure}[htbp]
  \centering
  \includegraphics[width=\hsize]{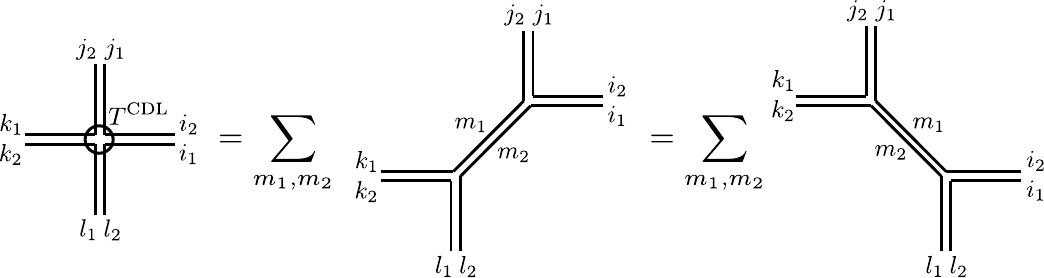}
  \caption{CDL tensor and decompositions.}
  \label{fig:cdldecomposition}
\end{figure}

\begin{figure}[htbp]
  \centering
  \includegraphics[width=0.4\hsize]{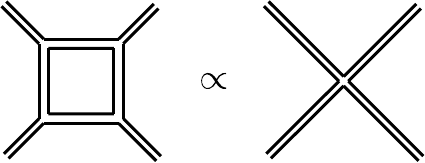}
  \caption{Coarse-graining of CDL tensor network leads to a CDL tensor network.}
  \label{fig:coarsegrainedcdl}
\end{figure}

The difficulty in approaching critical points with the TRG  has been illustrated by the corner double line (CDL) picture~\cite{Gu:2009dr}.
As a typical example, one can easily show that a toy tensor network that consists of the CDL tensor
\begin{align}
  T^{\mathrm{CDL}}_{ijkl} = \Lambda_{i_{1}l_{2}} \Lambda_{j_{1}i_{2}} \Lambda_{k_{1}j_{2}} \Lambda_{l_{1}k_{2}},
\end{align}
where $\Lambda$ can be assumed to be a diagonal matrix for simplicity,
is a fixed point of the TRG~\cite{Gu:2009dr} (see Figs.~\ref{fig:cdldecomposition} and \ref{fig:coarsegrainedcdl}).
This means the usual TRG leaves short-range correlations in the network under each blocking step and this causes a loss of accuracy.

\subsection{Removal of CDL from network}

The TNR approaches mentioned at the beginning of this section attempt to remove the CDLs from the network.
Some of the authors of this review article has checked the efficiency of the loop-TNR and the gilt-TNR for the Majorana--Wilson fermion system seen in Sec.~\ref{sec:wilsonmajorana} and for the two-flavor Gross--Neveu model~\cite{Asaduzzaman:2022pnw}.
While the loop-TNR comes first in the chronological order, we show here how the gilt-TNR removes the CDL from a network for graphical simplicity.

\begin{figure}[htbp]
  \centering
  \includegraphics[width=0.5\hsize]{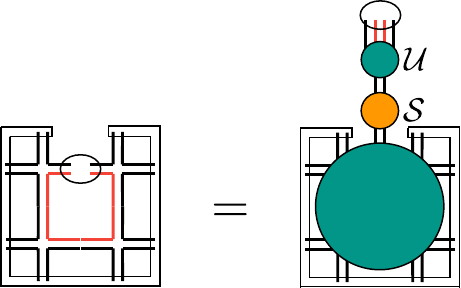}
  \caption{SVD of plaquette. After this decomposition CDL loop is inside $\mathcal{U}$.}
  \label{fig:svdenv}
\end{figure}

In the gilt-TNR, recursive optimization steps are taken place to remove a CDL on a plaquette.
To illustrate this, one can consider an SVD of the plaquette (see Fig.~\ref{fig:svdenv}), which can be seen as an environment of a link.
An important point here is that, by considering the link to be open~\footnote{
  Of course, the open link is to be closed after all.
}, the internal (\textit{i.e.} the CDL) loop is seen to be not enclosed in the plaquette and connected to and only to the open link.
With this logic, the CDL is captured by the unitary matrix ($\mathcal{U}$ in the figure) through the SVD.
After the decomposition, the unitary matrix $\mathcal{U}$ is replaced by another one according to the ``environment spectrum'' $\mathcal{S}$
so that the CDL loop will be truncated down.

While the gilt-TNR is based on local replacements of tensor legs, the loop-TNR consists of an entanglement filtering gauge transformation of tensors and an optimization step that minimizes a bit global cost function compared to the usual TRG.
One can see how is the CDL eliminated by the gauge transformation in the original paper~\cite{yang2017loop},
and \textit{cf.} Ref.~\cite{Gu:2009dr}, where the notion of ``entanglement filtering'' firstly appeared.

In Ref.~\cite{Asaduzzaman:2022pnw}, the performance of the Grassmann version of loop-TNR and gilt-TNR was inspected by computing the free energy and the determination of Fisher’s zeros.
In the following, we review the renormalization group flow of the singular values that are believed to store the information of the system.

\begin{figure}[htbp]
  \centering
  \begin{minipage}{0.32\hsize}
    \includegraphics[width=\hsize]{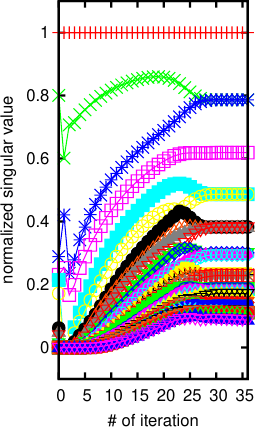}
  \end{minipage}
  \begin{minipage}{0.32\hsize}
    \includegraphics[width=\hsize]{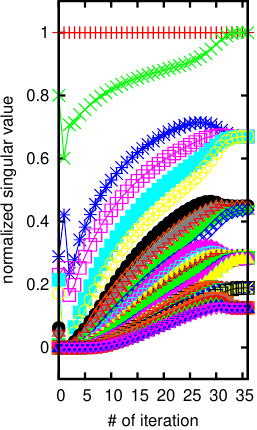}
  \end{minipage}
  \begin{minipage}{0.32\hsize}
    \includegraphics[width=\hsize]{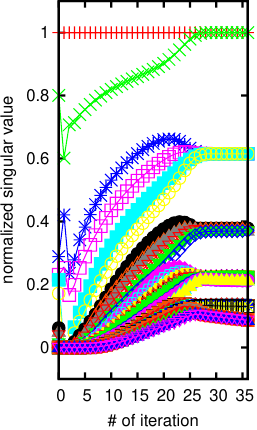}
  \end{minipage}
  \caption{
  Normalized singular values produced by Grassmann TRG at $\kappa=0.9999\kappa_{\mathrm{c}}$ (left), $\kappa=\kappa_{\mathrm{c}}$ (middle), and $\kappa=1.0001\kappa_{\mathrm{c}}$ (right). The bond dimension is set to 64.
  }
  \label{fig:svals_2dwilsonmajorana_bdim64}
\end{figure}

\begin{figure}[htbp]
  \centering
  \begin{minipage}{0.32\hsize}
    \includegraphics[width=\hsize]{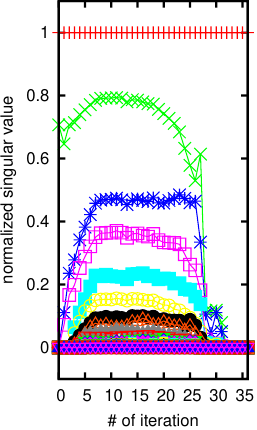}
  \end{minipage}
  \begin{minipage}{0.32\hsize}
    \includegraphics[width=\hsize]{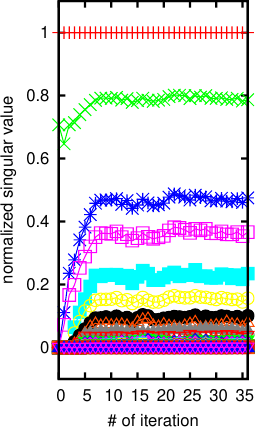}
  \end{minipage}
  \begin{minipage}{0.32\hsize}
    \includegraphics[width=\hsize]{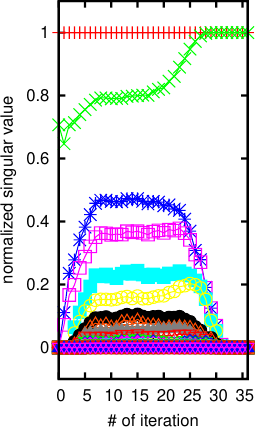}
  \end{minipage}
  \caption{Normalized singular values produced by Grassmann loop-TNR at $\kappa=0.9999\kappa_{\mathrm{c}}$ (left), $\kappa=\kappa_{\mathrm{c}}$ (middle), and $\kappa=1.0001\kappa_{\mathrm{c}}$ (right).
 The bond dimension is set to 16.}
  \label{fig:svals_2dwilsonmajorana_bdim16}
\end{figure}

The renormalization group flow obtained by the plain Grassmann TRG and the loop-TNR are shown in Figs.~\ref{fig:svals_2dwilsonmajorana_bdim64} and \ref{fig:svals_2dwilsonmajorana_bdim16}~\footnote{
In the reference, the result of the gilt-TNR is also shown; however, we omit to show it since the characteristic behavior is quite similar to that of the loop-TNR.
}.
The vertical and the horizontal axes represent the normalized singular value and how many iterations were taken place before that, respectively.
Under these iterative coarse-graining algorithms, the space-time volume of the system grows rapidly along with the number of iterations;
schematically it grows twice at each coarse-graining step.
In Fig.~\ref{fig:svals_2dwilsonmajorana_bdim64} one cannot clearly distinguish the three panels where the reverse temperature is set to $0.9999\kappa_{\mathrm{c}}$, $\kappa_{\mathrm{c}}$, and $1.0001\kappa_{\mathrm{c}}$.
This is on account of a contamination by short-range information that is difficult to properly remove by the normal Grassmann TRG algorithm.
On the other hand, the loop-TNR shows distinguishable fixed point structures at off-critical points such as $0.9999\kappa_{\mathrm{c}}$, $1.0001\kappa_{\mathrm{c}}$.
Also, at the criticality $\kappa_{\mathrm{c}}$, a scale-invariant structure is observed that clearly shows the superiority of the improved renormalization algorithm.
Note that these behaviors of the singular values are qualitatively the same as those observed for the two-dimensional Ising model; the equivalence between the Ising model and the Majorana--Wilson fermion system can be seen in a sense like this. 
For the growth of the entanglement entropy and relationship to the Calabrese--Cardy formula, we refer the reader to the discussion in Ref.~\cite{Asaduzzaman:2022pnw}.

\subsection{Bond-weighting technique}
\label{subsec:btrg}

TRG algorithm can be improved by removing the short-range correlations represented by the loop entanglement. 
Several algorithms are proposed to remove these short-range correlations and achieve much higher accuracy than the original TRG algorithm even at the criticality as demonstrated above.
At the same time, however, these algorithms usually require more computational cost than the original TRG.
Recently, Adachi, Okubo, and Todo have proposed a new idea to improve the accuracy of TRG algorithms without increasing the computational cost~\cite{2022PhRvB.105f0402A}. 
Their idea is to introduce some weight on the edge of the tensor network and construct the coarse-graining transformation including these weights. 
The algorithm is referred to as the bond-weighted TRG (BTRG), which can be regarded as a generalization of the Levin-Nave TRG.
The advantage of the BTRG over the Levin-Nave TRG is demonstrated by benchmarking with the two-dimensional Ising model in Ref.~\cite{2022PhRvB.105f0402A}.
The authors show that the BTRG outperforms the Levin-Nave TRG and the HOTRG at the same bond dimension.
Both for the BTRG and the Levin-Nave TRG, their computational times are proportional to $O(D^5)$ and their memory footprints are $O(D^3)$.
In the two-dimensional HOTRG, the computational time scales with $O(D^7)$, and the scaling of the memory cost is $O(D^4)$.
Therefore, the BTRG has shown the best performance among these three algorithms.
Moreover, the authors numerically demonstrate that non-trivial fixed point tensors can be constructed in the thermodynamic limit by the BTRG.
Recently, the BTRG has been applied to investigate the phase structure of the $CP(1)$ model with a topological $\theta$ term~\cite{Nakayama:2021iyp}.

\begin{figure}[htbp]
  	\centering
	\includegraphics[width=1\hsize]{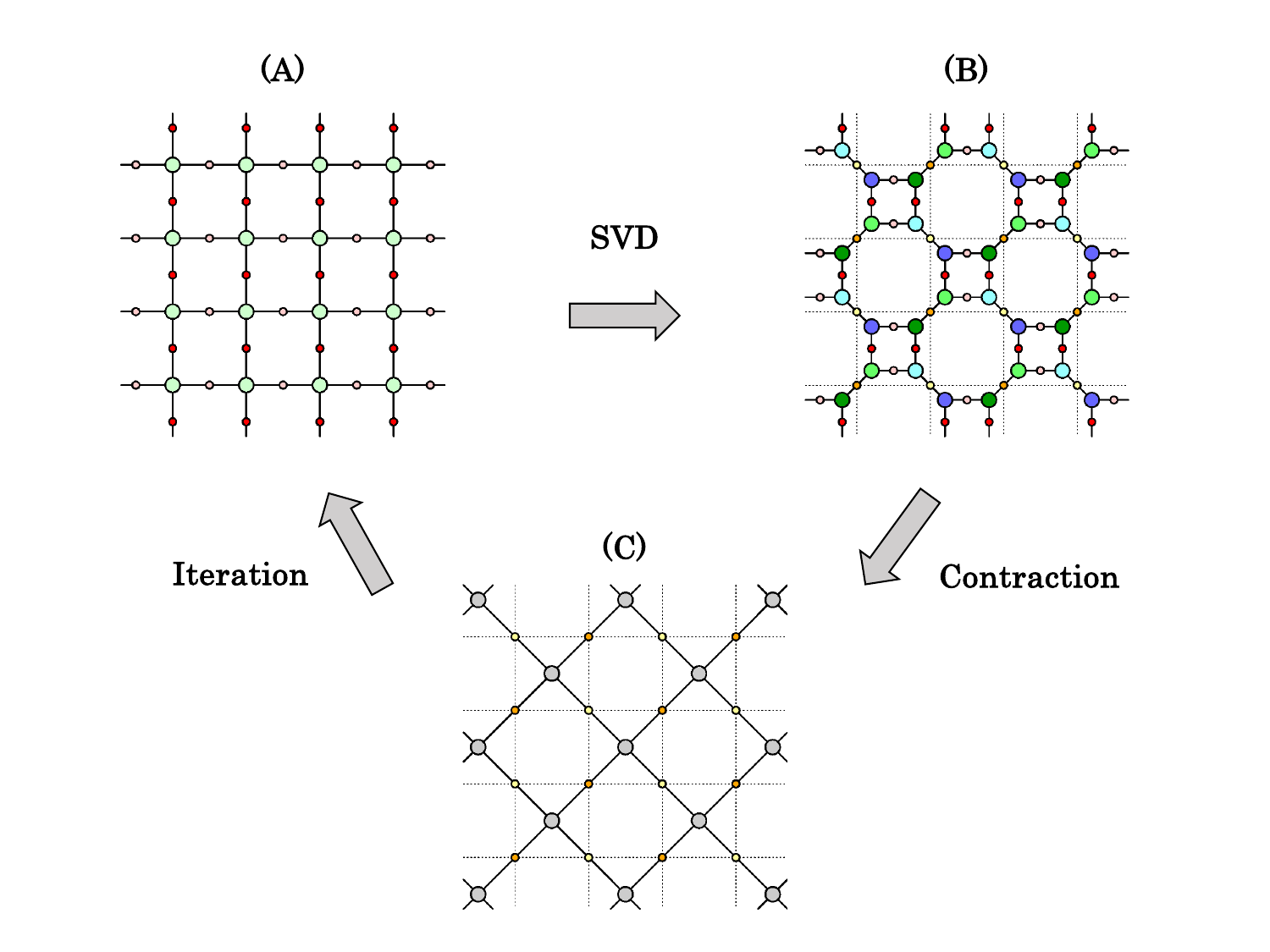}
	\caption{
	Schematic illustration of the BTRG algorithm.
        Background dotted lines denote a real-space square lattice.
        (A) Initial tensor network with bond weights on the lattice.
        (B) SVD introduces three-leg tensors and new bond weights.
        (C) New tensor network with bond weights by contracting four types of three-leg tensors and four bond weights.
	}
  	\label{fig:schematic_btrg}
\end{figure}

We review the BTRG algorithm for the two-dimensional square tensor network which is generated by a four-leg fundamental tensor.
The algorithm is schematically explained in Figure~\ref{fig:schematic_btrg}.
The key step in the BTRG is the low-rank approximation of the four-leg tensor based on the SVD with a hyperparameter such as
\begin{align}
    \label{eq:svd_btrg}
    T_{abcd}
    \simeq
    \sum_{i=1}^{D}
    U_{abi}
    \sigma_{i}^{(1-k)/2}
    \sigma_{i}^{k}
    \sigma_{i}^{(1-k)/2}
    V^{\dag}_{icd}
    ,
\end{align}
where $k\in\mathbb{R}$ denotes the hyperparameter.
If we set $k=0$, Eq.~\eqref{eq:svd_btrg} exactly corresponds to the tensor decomposition employed in the Levin-Nave TRG as shown in Eqs.~\eqref{eq:lntrg_svd_e} and \eqref{eq:lntrg_svd_o}.
With $k\neq0$, we obtain an extra factor $\sigma_{i}^{k}$, which is regarded as a weight on the bond $i$.
The authors in Ref.~\cite{2022PhRvB.105f0402A} have given a stationary condition equation,
\begin{align}
    \label{eq:stationary_eq}
    \left[
        \sigma_{i}^{(1-k)/2}
    \right]^{4}
    \left[
        \sigma_{i}^{k}
    \right]^{4}
    =
    \sigma_{i}
    ,
\end{align}
that determines the optimal choice of the hyperparameter $k$.
Eq.~\eqref{eq:stationary_eq} enforces the singular value spectrum to be invariant under the sequential coarse-graining procedure in the BTRG, assuming that the local four-leg tensors and the bond weights converge after sufficiently many times coarse-graining transformations and the matrices $U$ and $V$ do not affect the spectrum. 
According to Eq.~\eqref{eq:stationary_eq}, the optimal value of the hyperparameter for the two-dimensional square tensor network should be $k=-1/2$, which has been numerically confirmed in Ref.~\cite{2022PhRvB.105f0402A}.
We note that the application of the bond-weighing technique to other TRG algorithms has been discussed in Ref.~\cite{d_adachi_phd}.

\begin{figure}[htbp]
  	\centering
	\includegraphics[width=0.8\hsize]{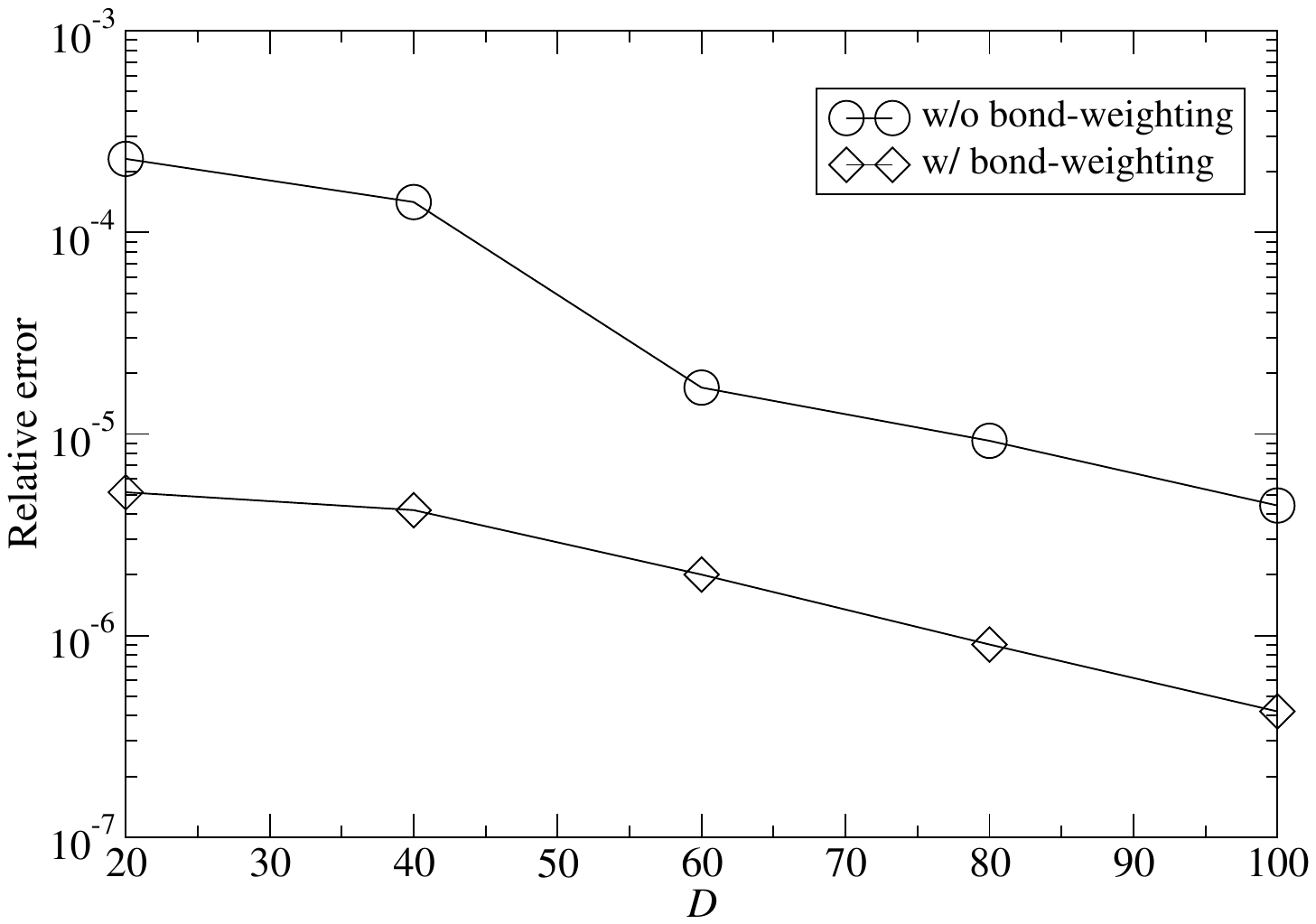}
	\caption{
	Relative error of the free energy for the two-dimensional free massless Wilson fermion. 
        The hyperparameter is set at $k=-1/2$.
	}
  	\label{fig:btrg_vs_trg}
\end{figure}

It should be emphasized that this derivation is completely independent of the details of the lattice theory.
Actually, Ref.~\cite{Akiyama:2022pse} shows that the bond-weighting method improves the accuracy of the Grassmann TRG.
Figure~\ref{fig:btrg_vs_trg} shows the relative error of the free energy for the two-dimensional free massless Wilson fermion. 
With the fixed bond dimension, the bond-weighting method always achieves higher accuracy compared with the normal TRG.
Notice that both computations require the same computational complexity at the same bond dimension.
A sample implementation of the BTRG for the Gross--Neveu model with Wilson fermions at finite density is shown in Ref.~\cite{Akiyama:2023rih}, whose web documentation is also provided.
\footnote{
\url{https://github.com/akiyama-es/Grassmann-BTRG} 
}
Using the code, one can reproduce the result shown in Fig.~\ref{fig:btrg_vs_trg}.

\subsection{Multilayered tensor network formulations for $N_{f}$-flavor fermions}

\begin{figure}[htbp]
    \centering
    \includegraphics[width=1\hsize]{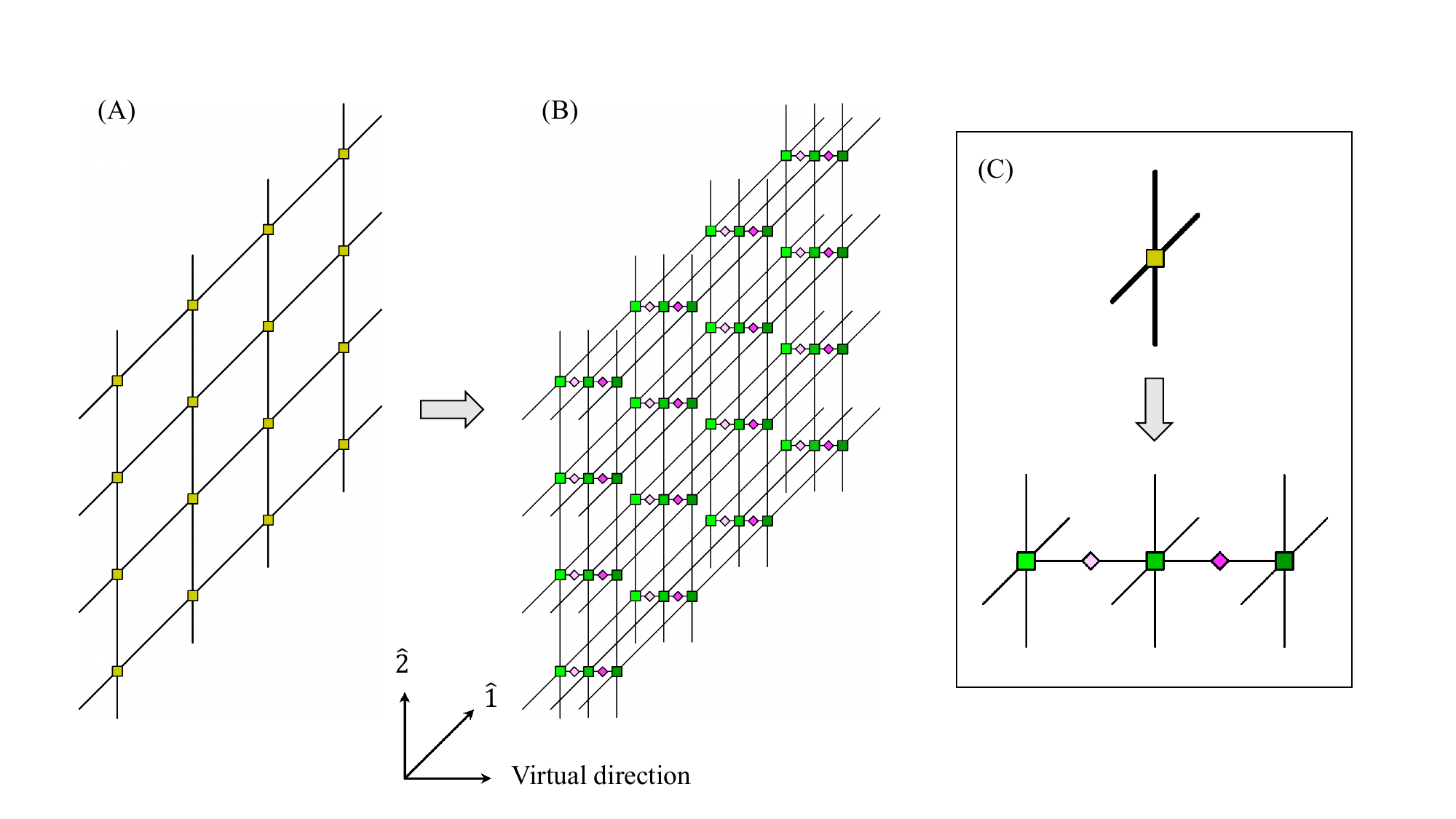}
	\caption{
        Adapted from Ref.~\cite{Akiyama:2023lvr}.
	Multilayered Grassmann tensor network representation for the path integral of the two-dimensional $N_{f}=3$ Gross--Neveu model with Wilson fermions.
    (A) Original Grassmann tensor network representation defined on the square lattice.
    (B) Three-layered Grassmann tensor network.
    (C) MPD rewrites the original fundamental tensor (yellow) with three fundamental tensors (green) and two kinds of singular values (purple).
	}
    \label{fig:mpd_gn_model}
\end{figure}

There is no difficulty in expressing the path integral of the lattice fermion system as the Grassmann tensor network with $N_{f}$ flavors.
In practice, however, the size of the resulting Grassmann tensor scales exponentially for $N_{f}$ and a  $O({\rm e}^N_{f})$ computational memory is required in the numerical computations.
This issue has been started to be addressed recently by Akiyama~\cite{Akiyama:2023lvr} and also by Yosprakob, Nishimura, and Okunishi~\cite{Yosprakob:2023tyr}.

Ref.~\cite{Akiyama:2023lvr} has employed the matrix product decomposition (MPD) to introduce a virtual direction so that each flavor degree of freedom is assigned to the different layers orthogonal to the virtual direction. 
MPD is a common idea in the tensor network methods such as the matrix product state (MPS) and matrix product operator (MPO)~\cite{PhysRevLett.75.3537,Dukelsky_1998}.
Ref.~\cite{Akiyama:2023lvr} has particularly utilized a canonical form of the MPD proposed in Ref.~\cite{PhysRevLett.91.147902}.
Thanks to the MPD, the memory cost for each local Grassmann tensor is reduced from $O({\rm e}^N_{f})$ to $O(N_{f})$, and the technique has been benchmarked with the two-dimensional Gross--Neveu model at finite density with the $N_{f}=2,3$ Wilson fermions.
Although a naive formulation provides the two-dimensional Grassmann tensor network whose bond dimension equals $4^{N_{f}}$, the MPD alternatively gives the ($2+1$)-dimensional network, $N_{f}$ sites along the virtual direction, with the bond dimension four without any approximation.
The schematic picture is shown in Figure~\ref{fig:mpd_gn_model}.
Based on the latter representation, the Silver Blaze phenomenon in the pressure and number density is reproduced with a relatively small bond dimension.

Ref.~\cite{Yosprakob:2023tyr} has proposed a compression scheme for the initial tensor representation of lattice gauge theories with the $N_{f}$-flavor fermions. 
They have introduced replicas of the original gauge field, which allows one to separate the local Grassmann tensor into multiple layers associated with the fermion flavor.
Based on this description, each layer is individually compressed by the isometry insertion; In the case of the $\mathbb{Z}_{K}$ gauge theory with the $N_{f}$-flavor Wilson fermions, the original tensor whose size is $16^4K^{10}$ is converted into the compressed one with $D^4K^2$ elements, where $D$ is the bond dimension introduced by the isometries.
Even with $D=8$, the difference between the resulting $\ln Z$ at finite gauge coupling and chemical potential is suppressed less than $O(10^{-15})$.
Since gauge fields are replicated by the Kronecker deltas, they are maximally entangled along the flavor direction.
Therefore, the tensor contractions along the flavor direction are carried out before the two-dimensional spacetime coarse-graining is performed. 
Using this technique, the chiral susceptibility of the two-dimensional infinite-coupling $\mathbb{Z}_{2}$, $\mathbb{Z}_{4}$, and $U(1)$ gauge theories with the $N_{f}=1,2$ Wilson has been computed. 
The critical hopping parameters have been determined for each case.
The pressure and number density as functions of the chemical potential have also been provided in the case of $\mathbb{Z}_{2}$ gauge theory with $N_{f}=1,2,4$ at finite gauge coupling, where the Silver Blaze phenomenon has been successfully captured.

In addition, Yosprakob has been recently developing a Python package for the Grassmann tensor network computations in Ref.~\cite{Yosprakob:2023flr}, whose web documentation is also provided.
\footnote{
\url{https://ayosprakob.github.io/grassmanntn/} 
}
The Schwinger model is implemented as an example.

\section{Relativistic models with fermion interactions}
\label{sec:relativistic}

\subsection{Gross--Neveu model}

The Gross--Neveu model is a well-known toy model for QCD.
Since it shares several important features of the QCD such as asymptotic freedom and a dynamical mass generation mechanism via symmetry breaking, the model is a good test bed for new computational methods.
Here, we consider the model defined with Wilson fermions. 
The lattice action reads
\begin{align}
\label{eq:action_gnw}
    S&=
    -\frac{1}{2}\sum_{n,\nu}\sum_{f=1}^{N_{f}}
    \left\{
	{\rm e}^{\mu\delta_{\nu,2}}\bar{\psi}^{(f)}(n)(r\mathds{1}-\gamma_{\nu})\psi^{(f)}(n+\hat{\nu})
        +{\rm e}^{-\mu\delta_{\nu,2}}\bar{\psi}^{(f)}(n+\hat{\nu})(r\mathds{1}+\gamma_{\nu})\psi^{(f)}(n)
    \right\}
    \nonumber\\
    &+\sum_{n}\sum_{f=1}(m+2r)\bar{\psi}^{(f)}(n)\psi^{(f)}(n)
    -\frac{g^{2}_{\sigma}}{2N_{f}}\sum_{n}
    \left(\sum_{f}\bar{\psi}^{(f)}(n)\psi^{(f)}(n)\right)^{2}
    \nonumber\\
    &-\frac{g^{2}_{\pi}}{2N_{f}}\sum_{n}
    \left(\sum_{f}\bar{\psi}^{(f)}(n){\rm i}\gamma_{5}\psi^{(f)}(n)\right)^{2},
\end{align}
where $\psi^{(f)}(n)$ and $\bar{\psi}^{(f)}(n)$ are the Wilson fermions with the flavor index $f$.
$g^{2}_{\sigma}$ and $g^{2}_{\pi}$ are the four-fermi coupling constants and $m$, $\mu$, and $r$ represent mass, chemical potential, and the Wilson parameter, respectively.

Takeda and Yoshimura studied the model with $N_{f}=1$ at finite density in Ref.~\cite{Takeda:2014vwa}, which is the first application of the Grassmann TRG to the finite-density system.
They employed the Levin-Nave TRG algorithm with the bond dimension up to $D_{\rm LNTRG}=64$ to compute the fermion number density and its susceptibility as functions of $\mu$.
They successfully observed that the number density saturated to one with sufficiently large chemical potential.
In addition, they considered the model on an anisotropic finite lattice with $(N_{1},N_{2})=(64,32), (96,32)$, where $N_{1}$ and $N_{2}$ denote the spatial and temporal lattice sizes respectively, and found that there were two peaks in the susceptibility under the presence of the finite four-fermi coupling.
Throughout their analysis, they pointed out that the finite bond dimension effect could be enhanced not only at critical points but also in crossover regions.

\begin{figure}[htbp]
    \centering
    \begin{minipage}{0.45\hsize}
        \includegraphics[width=\hsize]{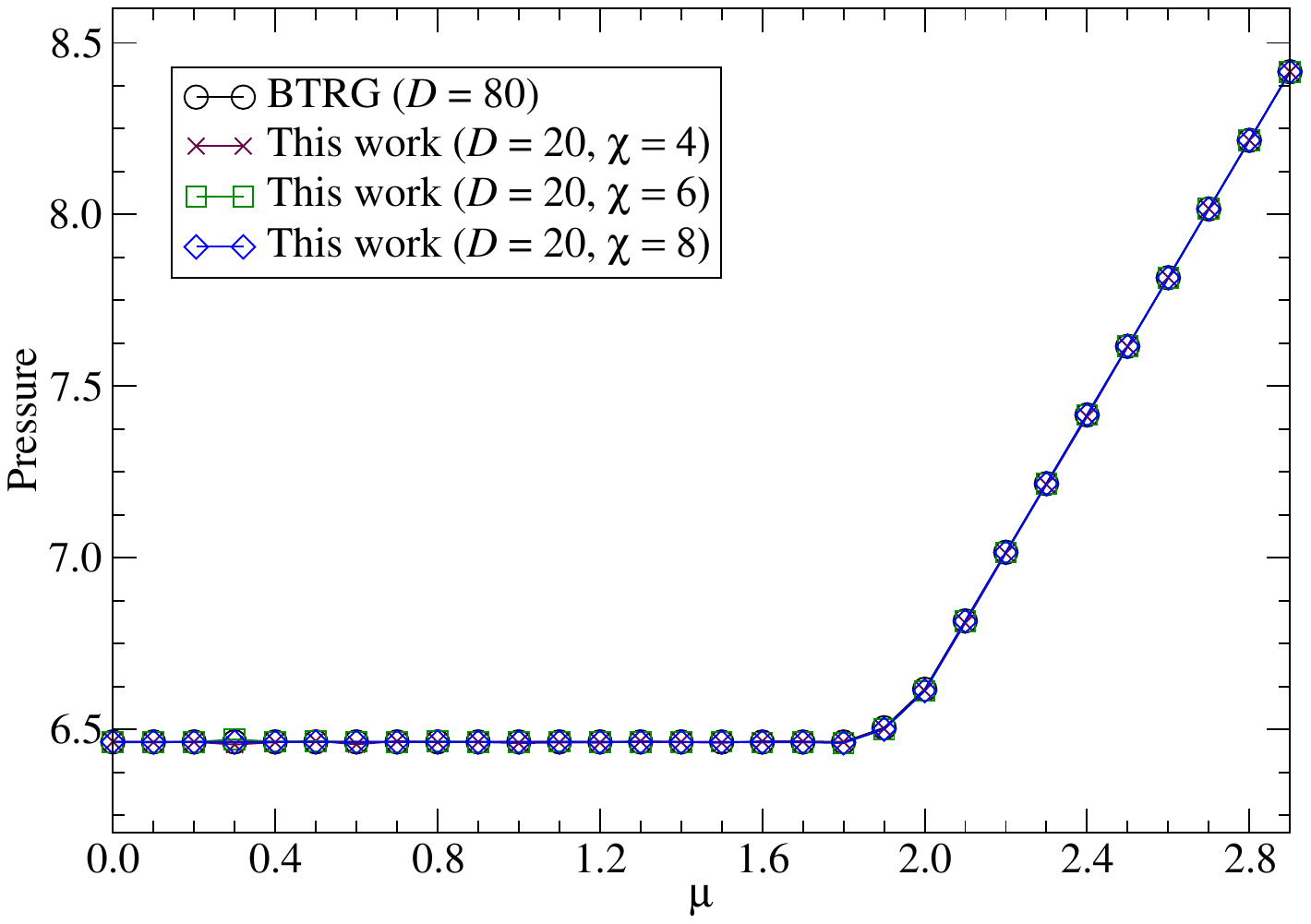}
    \end{minipage}
    \begin{minipage}{0.45\hsize}
        \includegraphics[width=\hsize]{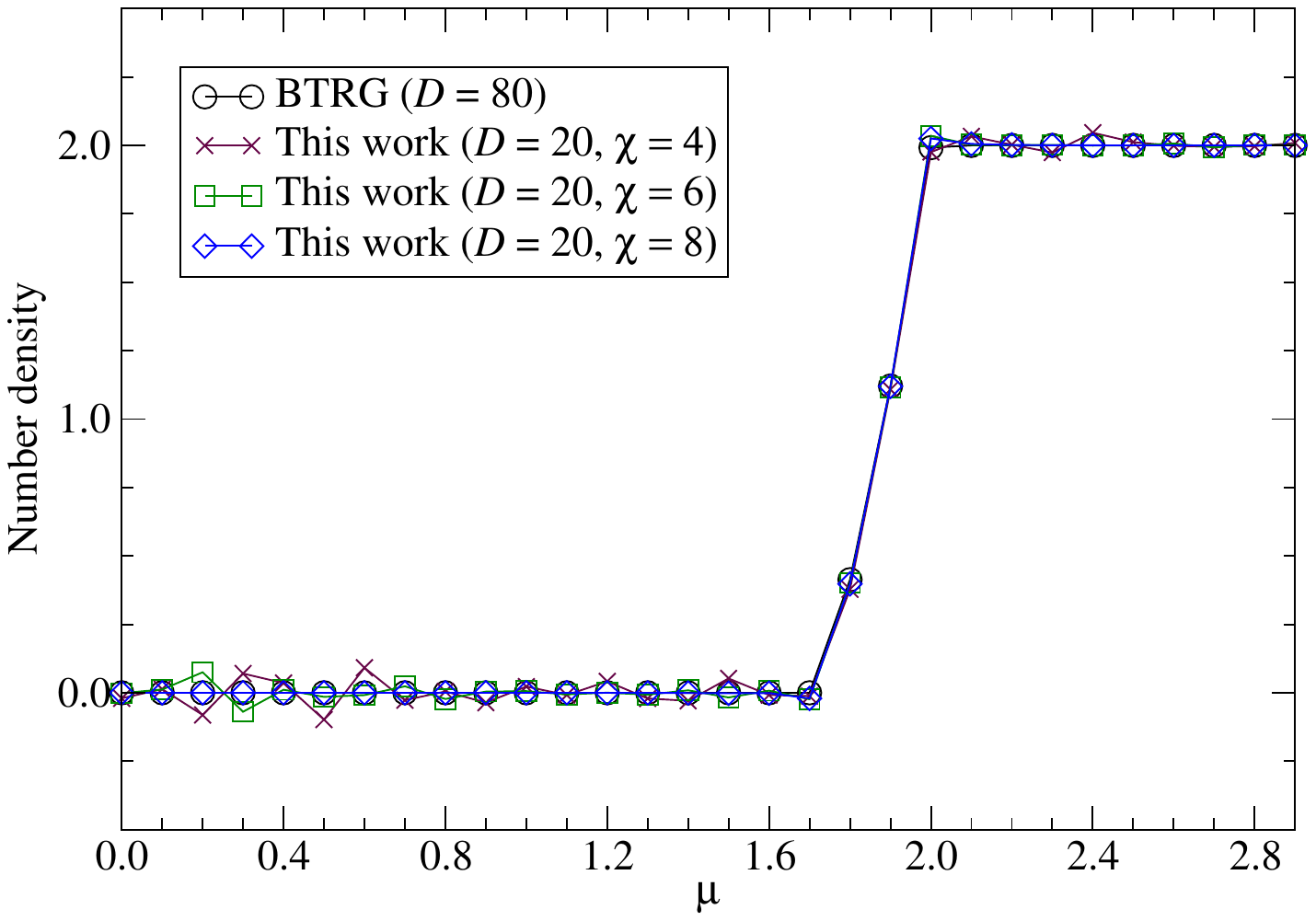}
    \end{minipage}
    \caption{
    Adapted from Ref.~\cite{Akiyama:2023lvr}.
    Pressure (left) and number density (right) of the $N_{f}=2$ Gross--Neveu model with Wilson fermions.
    $D$ denotes the bond dimension for spacetime indices and $\chi$ does the bond dimension along the virtual direction in the multilayered network.
    $m=1$ and $g^2=10$.
    }
  \label{fig:nf2_g10}
\end{figure}

\begin{figure}[htbp]
    \centering
    \begin{minipage}{0.45\hsize}
        \includegraphics[width=\hsize]{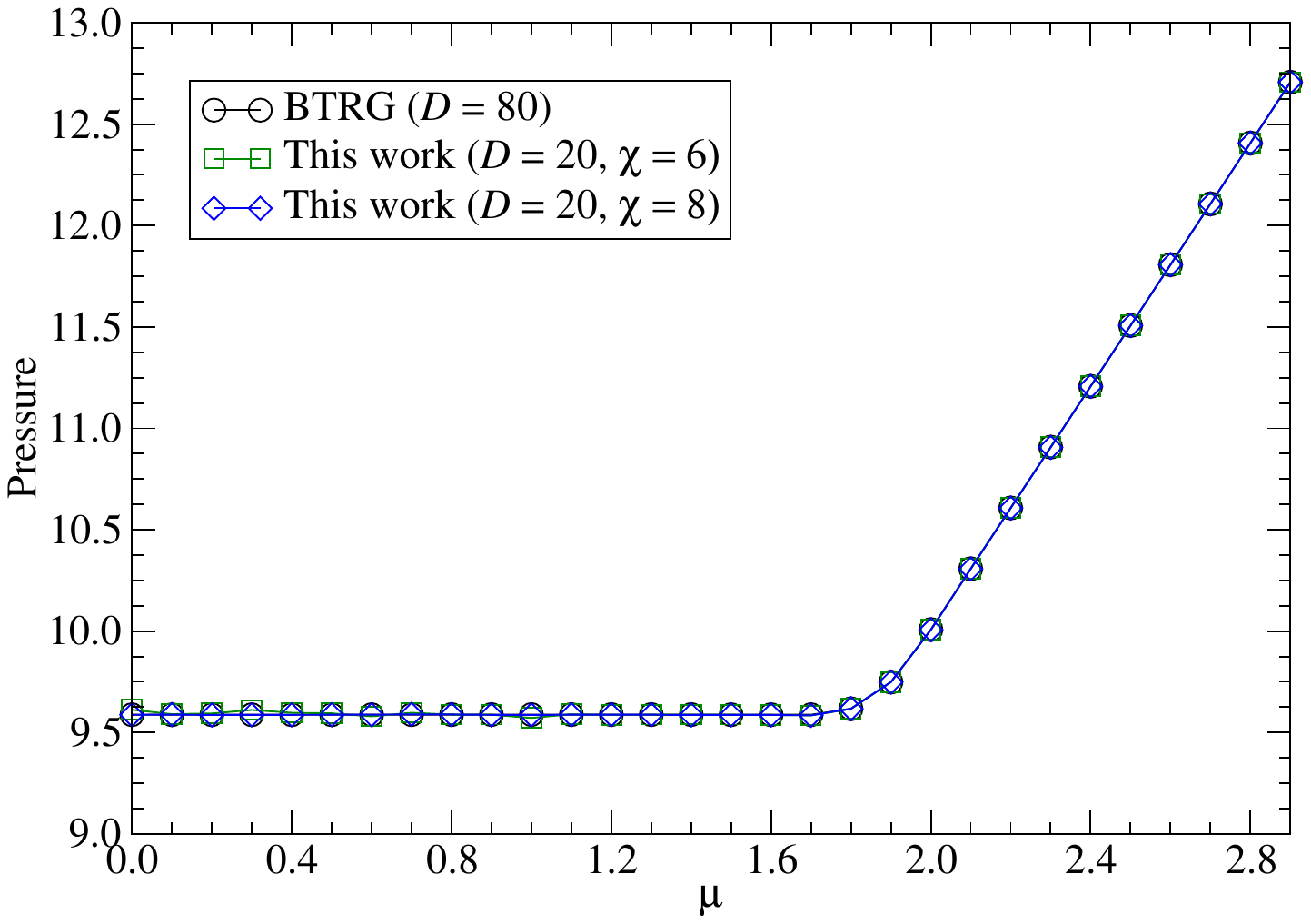}
    \end{minipage}
    \begin{minipage}{0.45\hsize}
        \includegraphics[width=\hsize]{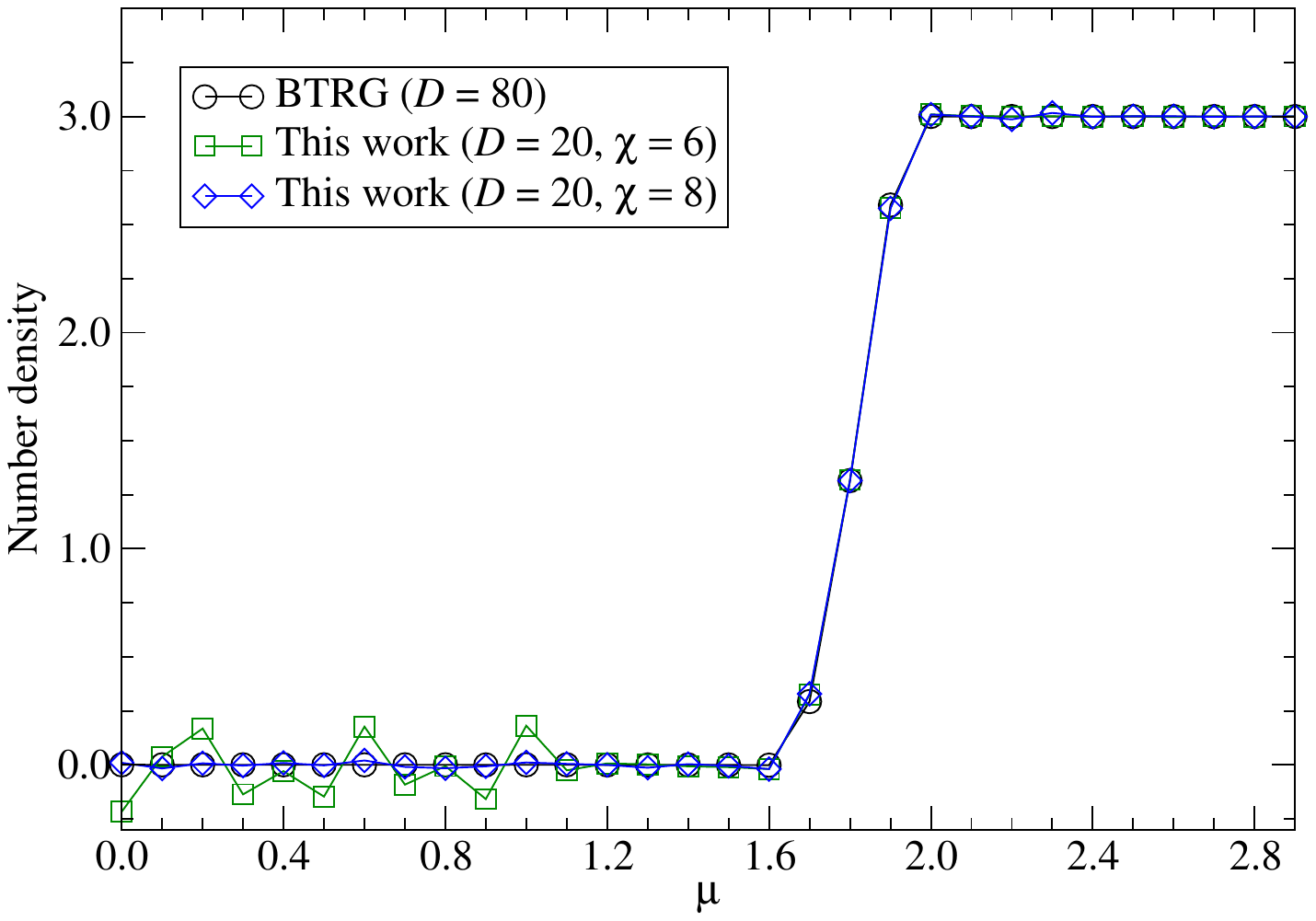}
    \end{minipage}
    \caption{
    Adapted from Ref.~\cite{Akiyama:2023lvr}.
    Pressure (left) and number density (right) of the $N_{f}=3$ Gross--Neveu model with Wilson fermions.
    $D$ denotes the bond dimension for spacetime indices and $\chi$ does the bond dimension along the virtual direction in the multilayered network.
    $m=1$ and $g^2=10$.
    }
  \label{fig:nf3_g10}
\end{figure}

Recently, Akiyama has investigated the model with $N_{f}=2,3$ at finite density~\cite{Akiyama:2023lvr}.
The pressure and number density on a square lattice were computed as functions of chemical potential using two methods, the bond-weighting method and multilayered formulation as shown in Figures~\ref{fig:nf2_g10} and \ref{fig:nf3_g10}.
The results obtained by the two methods are consistent and the number density saturates to two and three for $N_{f}=2,3$, respectively.
They noted that the finite bond dimension effect could be enhanced in the Silver-Blaze regime in the multilayered formulation.

\subsection{QCD in the infinite-coupling limit}

Bloch and Lohmayer made the first study of QCD in the infinite-coupling limit with the Grassmann TRG approach~\cite{Bloch:2022vqz}.
The lattice action with the staggered quarks is given by
\begin{align}   
\label{eq:action_2d_qcd}
    S=
    \sum_{n}
    &\left\{
        \eta_{1}(n)\gamma\bar{\psi}(n)\left[
            {\rm e}^{\mu}U_{1}(n)\psi(n+\hat{1})
            -
            {\rm e}^{\mu}U^{\dag}_{1}(n-\hat{1})\psi(n-\hat{1})
        \right]
    \right.
    \nonumber\\
    &\left.
        +
        \eta_{2}(n)\bar{\psi}(n)\left[
            U_{2}(n)\psi(n+\hat{2})
            -
            U^{\dag}_{2}(n-\hat{2})\psi(n-\hat{2})
        \right]
        +2m\bar{\psi}(n)\psi(n)
    \right\}
    ,
\end{align}
where $m$ and $\mu$ denote the mass and chemical potential.
The lattice site is labelled by $n=(n_{1},n_{2})$ with the temporal coordinate $n_{1}$ and spatial coordinate $n_{2}$.
The action has an anisotropy factor $\gamma$ in the temporal hopping terms.
The link variables $U_{\nu}(n)$ take their value on $SU(3)$ and $\psi(n)$ and $\bar{\psi}(n)$ are the staggered quark fields described by the three-component Grassmann variables.
The staggered sign function is defined via $\eta_{1}(n)=1$ and $\eta_{2}(n)=(-1)^{n_{1}}$.
The (anti-)periodic boundary condition is assumed in the spatial (temporal) direction.
The authors integrate out the $SU(3)$ link variables exactly as shown in Refs.~\cite{Rossi:1984cv,Karsch:1988zx} and the path integral is rewritten by the mesonic and baryonic contributions, following Ref.~\cite{Fromm:2010lga}, characterized by their occupation numbers.
\footnote{
This kind of dual formulation has also been applied in Ref.~\cite{Milde:2021vln} to investigate the infinite-coupling $U(N)$ gauge theories with staggered fermions in three and four dimensions with the variant of HOTRG.
}
Effectively reducing the number of configurations with vanishing contribution in the path integral, they obtain the Grassmann tensor network representation where the bosonic tensor is defined by a four-leg tensor and each subscript is of dimension six.
The resulting Grassmann tensor network is approximately contracted by the HOTRG, where the authors have applied an improved method to derive the projectors in the algorithm~\cite{Bloch:2022qyw}.

The authors calculated the thermodynamic potential $\ln Z/V$ as a function of $\mu$ and $m$ on $2^{2}$ and $4^{2}$ lattices and confirmed that the exact results were reproduced.
Although the convergence for the bond dimension $D_{\rm HOTRG}$ becomes slower with the smaller $m$, it has been shown that the accuracy of the calculations is sufficient to take the chiral limit: a quadratic fit in $1/D_{\rm HOTRG}$ successfully extrapolates $\ln Z/V$ at $m=0$ on a $1024^{2}$ lattice to $D_{\rm HOTRG}\to\infty$ using the results on $40\le D_{\rm HOTRG}\le 128$.

The authors then computed the chiral condensate $\langle\bar{\psi}\psi\rangle$ at vanishing chemical potential. 
They extrapolated $\langle\bar{\psi}\psi\rangle$ at finite mass to $D_{\rm HOTRG}\to\infty$ before taking the infinite-volume limit.
For $m\le0.005$, the condensate was nicely fitted by $\lim_{V\to\infty}\langle\bar{\psi}\psi\rangle=am^{b}$, with $a=2.77$ and $b=0.0414$.
Therefore, it is confirmed that the chiral symmetry is not dynamically broken in infinite-coupling QCD with staggered quarks in two dimensions.
They also studied the number density and chiral condensate at finite chemical potential with $m=0.1$ and $D_{\rm HOTRG}=64$.
In the infinite-volume limit, they have found that a first-order phase transition takes place at $\mu_{\rm c}\simeq0.3508$.
With $\mu>\mu_{\rm c}$, they have observed that the number density saturates to three and the chiral symmetry is restored.
Throughout the study, they employed the stabilized finite-difference method developed in Ref.~\cite{Bloch:2021mjw} to calculate the chiral condensate and number density by the numerical differentiation.

\subsection{Nambu--Jona-Lasinio model}

Akiyama, Kuramashi, Yamashita, and Yoshimura made the first application of the TRG approach to the four-dimensional lattice fermions~\cite{Akiyama:2020soe}.
They investigated the chiral phase transition in the Nambu--Jona-Lasinio (NJL) model at finite density.
Since this model is an effective field theory of the QCD, the efficiency of the TRG approach for the NJL model should be addressed from the viewpoint of the future application of the TRG method toward the QCD at finite density.
The model is defined with the staggered fermions by the following action,
\begin{align}
    \label{eq:action_njl}
    S
    &=
    \frac{1}{2}
    \sum_{n}
    \sum_{\nu=1}^{4}
    \eta_{\nu}(n)
    \left[
        {\rm e}^{\mu\delta_{\nu,4}}\bar{\chi}(n)\chi(n+\hat{\nu})
        -{\rm e}^{-\mu\delta_{\nu,4}}\bar{\chi}(n+\hat{\nu})\chi(n)
    \right]
    \nonumber\\
    &+m\sum_{n}\bar{\chi}(n)\chi(n)
    -g_{0}
    \sum_{n}
    \sum_{\nu}
    \bar{\chi}(n)\chi(n)\bar{\chi}(n+\hat{\nu})\chi(n+\hat{\nu})
    ,
\end{align}
where $\chi(n)$ and $\bar{\chi}(n)$ are the staggered fermions described by the single-component Grassmann variables.
$\eta_{\nu}(n)$ is the staggered sign function $\eta_{\nu}(n)=(-1)^{n_{1}+\cdots+n_{\nu-1}}$ with $\eta_{1}(n)=1$.
$m$, $g_{0}$, and $\mu$ represent the mass, four-fermi coupling constant, and chemical potential.
When $m=0$, Eq.~\eqref{eq:action_njl} is invariant under the continuous transformation,
\begin{align}
    \chi(n)
    \rightarrow
    {\rm e}^{{\rm i}\alpha\eta_{5}(n)}
    \chi(n)
    ,
    \quad\quad
    \bar{\chi}(n)
    \rightarrow
    \bar{\chi}(n)
    {\rm e}^{{\rm i}\alpha\eta_{5}(n)}
    .
\end{align}
This global symmetry is regarded as the chiral symmetry for the staggered NJL model.

Decomposing the hopping terms and four-fermi interaction term, the Grassmann tensor network representation for the path integral is available.
Ref.~\cite{Akiyama:2023lvr} introduces the eight kinds of local tensors to describe the Grassmann tensor network.
This is because the staggered theory defined in Eq.~\eqref{eq:action_njl} a little bit breaks the translational invariance on the lattice.
Note that the site dependence appearing in the local tensor is characterized just by $\eta_{\nu}(n)$ so that the resulting Grassmann tensor network has a periodic structure.
See Refs.~\cite{Akiyama:2023lvr,Akiyama:2022asr} for the detailed derivation of the Grassmann tensor network representation.

The authors in Ref.~\cite{Akiyama:2023lvr} developed the Anisotropic TRG (ATRG) algorithm~\cite{Adachi:2019paf} for fermions.
The ATRG allows us to approximately contract $d$-dimensional tensor networks with the $O\left(D_{\rm ATRG}^{2d+1}\right)$ complexity and $O\left(D_{\rm ATRG}^{d+1}\right)$ memory cost.
These costs should be compared with the $O\left(D_{\rm HOTRG}^{4d-1}\right)$ complexity and $O\left(D_{\rm HOTRG}^{2d}\right)$ memory cost in the HOTRG.
This drastic cost reduction is a result of an additional approximation for fundamental tensors.
The further cost reduction technique for the ATRG has been provided by Oba in Ref.~\cite{Oba:2019csk}.
In addition to the ATRG, several other algorithms have been proposed for the higher-dimensional systems~\cite{Kadoh:2019kqk,Nakayama:2023ytr}.
As a validation of the ATRG for fermionic models, they first considered the NJL model in the heavy-dense limit, where $m\to\infty$ and $\mu\to\infty$ keeping the ratio of ${\rm e}^{\mu}/m$ fixed.
In this limit, the model can be solved analytically~\cite{Bender:1992gn}. 
The number density and fermion condensate as functions of $\mu$ in the thermodynamic and vanishing temperature limits were calculated by the ATRG with the bond dimension $D_{\rm ATRG}\le30$ and the results showed a good agreement with the analytic ones.

The authors then studied the chiral phase transition in the cold and dense regime characterized by $\mu/T=O(10^{3})$.
Due to the sign problem, such a cold and dense regime is inaccessible with the standard Monte Carlo simulation.
They enlarged the bond dimension up to $D_{\rm ATRG}=55$ in the strong coupling regime, where they found a clear signal of the first-order transition in the chiral condensate, and the chiral symmetry was restored with $\mu>\mu_{\rm c}$. 
This is exactly the predicted result by the mean-field theory~\cite{Buballa:2003qv} and the functional renormalization group~\cite{Aoki:2017rjl}.
The authors also computed the pressure and number density which are fundamental ingredients in the equation of state.
All the thermodynamic quantities obtained by the ATRG have shown that the model undergoes the first-order phase transition in the cold and dense region.

\subsection{$\mathcal{N}=1$ Wess--Zumino model}

The interacting two-dimensional $\mathcal{N}=1$ Wess--Zumino model, a simple supersymmetric model, shows a vanishing partition function (Witten index)~\cite{Witten:1982df};
\textit{i.e.} This model suffers from a serious sign problem as in the case of other generic supersymmetric models. (See~\cite{catterall2009} for a review.)

The Euclidean continuum action of the model is defined by
\begin{align}
  \label{eq:wzaction}
  S = \int {\rm d}^{2}x
  \biggl\{
  & \frac{1}{2}\left(\partial_{\mu}\phi\right)^{2}
    + \frac{1}{2} W^\prime \left(\phi\right)^{2}
    + \frac{1}{2}\bar{\psi} \left(\gamma_\mu \partial_\mu
    + W^{\prime\prime}\left(\phi\right)\right)\psi \biggr\},
\end{align}
where $\phi$ and $\psi$ are a one-component real scalar field and a two-component Majorana spinor field, respectively.
The superpotential $W\left(\phi\right)$ is a function of $\phi$
and is the source of the Yukawa- and $\phi^{n}$-interactions.

The Majorana condition for $\psi$ is given by
\begin{align}
  \label{eq:majoranacond}
  \bar{\psi} =-\psi^{\mathrm{T}} C^{-1}
\end{align}
with the charge conjugation matrix $C$,
\begin{align}
  \label{eq:cconj}
  C^{\mathrm{T}}=-C, && C^{\dagger}=C^{-1}, && C^{-1}\gamma_{\mu}C=-\gamma_{\mu}^{\mathrm{T}}.
\end{align}
The continuum action above can be shown to be invariant under the supersymmetry transformation
\begin{align}
  \label{eq:susytrans}
  &\delta \phi
    = \bar{\epsilon} \psi, \\
  &\delta \psi
    = \left(\gamma_\mu \partial_\mu \phi - W^{\prime}\left(\phi\right)\right) \epsilon,
\end{align}
where $\epsilon$ is a two-component Grassmann number that satisfies the Majorana condition~\eqref{eq:majoranacond}.

\begin{figure}[htbp]
  \centering
  \includegraphics[width=0.8\hsize]{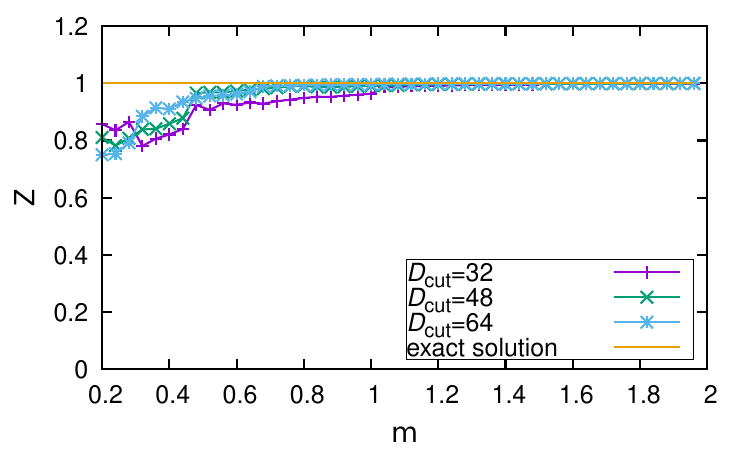}
  \caption{
  Adapted form Ref.~\cite{Meurice:2020pxc}.
    The partition function of the $\mathcal{N}=1$ free Wess--Zumino model as a function of $m$ on $V=2 \times 2$ lattice.
  }
  \label{fig:wittenindex}
\end{figure}

Figure~\ref{fig:wittenindex} shows the partition function of the free $\mathcal{N}=1$ Wess--Zumino model, whose superpotential is defined by $W(\phi) = (1/2)m\phi^{2}$ with the mass parameter $m$, on $V = 2 \times 2$ lattice.
(Note that the sign problem occurs even in this free case.)
The periodic boundary conditions are imposed in all directions for both fermions and bosons.
For this specific case, the analytical solution can be shown to be $1$ for the $m>0$ region shown in the figure.
The Grassmann TRG results show better agreement with the exact solution at larger masses although smaller masses seem to be difficult.
As for the difficulty in the small mass region, the authors of Ref.~\cite{Kadoh:2018hqq} concluded that it is due to the lack of damping factor in the local Boltzmann weight;
in other words, the reason is that the Gauss--Hermite quadrature applied to the scalar boson part does not converge.

Even though they showed results only for the non-interacting case, their construction of the tensor network does not depend on the shape of the superpotential,
so that further studies of the interacting Wess--Zumino model are awaited.

\subsection{Non-abelian lattice gauge theories coupled to fermions}

Non-abelian gauge theories have huge internal degrees of freedom, 
and this fact prevents one from building a tensor network representation in a non-expensive way.
Indeed, tensor network studies for gauge theories are limited to abelian cases when considering coupling to fermions.
Recently a subgroup of the authors reported a way to construct a tensor network representation of non-abelian gauge theories with a reasonable numerical complexity~\cite{Asaduzzaman:2023pyz}.
In their work, parameterized group elements are discretized via a Gaussian quadrature method, and a further reduction of the degrees of freedom is achieved by applying the higher-order orthogonal iteration (HOOI) algorithm~\cite{de2000multilinear,de2000best} to plaquette tensors.
Moreover, for the fermion part, they adopt the reduced staggered formulation~\cite{vandenDoel:1983mf} to completely eliminate the redundancies for the fermion part.

They numerically showed the accuracy of their approximation by checking the partition function and the average plaquette for the $SU(2)$ case.
Surprisingly, the convergence of the HOOI algorithm is so fast, and the loss of accuracy by the truncation is shown to be quite mild.
In other words, the main source of the error is the quadrature method with a few number of Gaussian nodes; they adopted up to $5$ for the quadrature at each angle.
It would be interesting to see how such a drastic approximation is tolerable for more complicated cases.

\section{The Hubbard model}
\label{sec:hubbard}

The Hubbard model is one of the most fundamental lattice fermion models describing the itinerant spin-$1/2$ electrons via the repulsive Coulomb interaction.
The model is defined by the following Hamiltonian,
\begin{align}
\label{eq:h_hubbard}
    H
    =
    -t
    \sum_{\langle ij\rangle}
    \sum_{s=\uparrow,\downarrow}
    \left(
        c^{\dag}_{i,s}c_{j,s}
        +
        c^{\dag}_{j,s}c_{i,s}
    \right)
    +U
    \sum_{i\in\Lambda}
    n_{i,\uparrow}n_{i,\downarrow}
    .
\end{align}
The first term, the tight-binding Hamiltonian, represents the kinetic energy of electrons and the second one shows the repulsive Coulomb potential with $U\ge0$. 
$\langle ij\rangle$ labels the nearest-neighbor sites in the $d$-dimensional hypercubic lattice $\Lambda$.
The interesting aspect of the Hubbard model originates from the fact that each term is diagonalizable in a different space; the kinetic term is diagonalized in the momentum space but the Coulomb potential term is diagonalized in the real space.
Exact solutions of the Hubbard model are available only in the cases of $d=1$~\cite{Lieb:1968zza} and $d=\infty$~\cite{PhysRevLett.62.324,MllerHartmann1989CorrelatedFO}, despite its simplicity.
In the condensed matter community, the model in general dimensions and on various lattice geometries have been extensively studied with mean-field approaches, field-theoretical ways, and numerical methods.
We recommend for interested readers to see several recent reviews~\cite{1994RvMP...66..763D,RevModPhys.68.13,Hal_Tasaki_1998,PhysRevX.5.041041,doi:10.1146/annurev-conmatphys-031620-102024,doi:10.1146/annurev-conmatphys-090921-033948,Ostmeyer:2022gwi} and references therein.

Let us review the Hubbard model very briefly.
The creation and annihilation operators $c^{\dag}_{i,s}$ and $c_{i,s}$ satisfy the anti-commutation relation,
\begin{align}
\label{eq:anti_comm}
    \left\{
        c_{i,s}, c^{\dag}_{i',s'}
    \right\}
    =
    \delta_{ii'}
    \delta_{ss'}
    ,
    \quad\quad
    \left\{
        c_{i,s}, c_{i',s'}
    \right\}
    =
    0
    ,
    \quad\quad
    \left\{
        c^{\dag}_{i,s}, c^{\dag}_{i',s'}
    \right\}
    =
    0
    .
\end{align}
The number operator is defined via 
\begin{align}
\label{eq:n_hubbard}
    N
    =
    \sum_{i,s}
    n_{i,s}
\end{align}
with $n_{i,s}=c^{\dag}_{i,s}c_{i,s}$. 
The spin operator at the site $i$ is given by $\bm{S}_{i}=\left(S^{(x)}_{i},S^{(y)}_{i},S^{(z)}_{i}\right)$, where
\begin{align}
\label{eq:spin_z_hubbard}
    S^{(\nu)}_{i}
    =
    \frac{1}{2}
    \sum_{s,s'}
    c^{\dag}_{i,s}
    \sigma^{(\nu)}_{ss'}
    c_{i,s'}
    .
\end{align}
$\sigma^{(\nu)}~(\nu=x,y,z)$ is the Pauli matrices.
The total spin operator is 
\begin{align}
\label{eq:spin_hubbard}
    \bm{S}
    =
    \sum_{i}
    \bm{S}_{i}
    .
\end{align}
The Hamiltonian in Eq.~\eqref{eq:h_hubbard} has the global $U(2)=U(1)\times SU(2)$ symmetry. 
With $R\in U(2)$, Eq.~\eqref{eq:h_hubbard} is invariant under the transformation,
\begin{align}
    c_{i,s}
    \rightarrow
    \sum_{s'}
    R_{s,s'}
    c_{i,s'}
    .
\end{align}
This symmetry results in the global charge conservation and the spin isotropy.
The global charge conservation originates from the global $U(1)$ symmetry and the total particle number $N$ is a good quantum number.
The spin isotropy is a result of the global $SU(2)$ symmetry and $S^{(z)}$ and $\bm{S}^{2}$ are good quantum numbers.

The grand canonical Hamiltonian is given by $H-\mu N$, where $\mu$ is the chemical potential.
Since the model describes the spin-$1/2$ electrons, $\langle N\rangle=2|\Lambda|$ with the total lattice sites $|\Lambda|$.
When the filling is a single electron per site, that is $\langle N\rangle=|\Lambda|$, the situation is referred to as half-filling.
At half-filling, the system acquires the so-called particle-hole symmetry.
To see this, let us consider the transformation such that
\begin{align}
\label{eq:ph_symm}
    c_{i,s}
    \rightarrow
    \eta_{i}
    c^{\dag}_{i,s}
    ,
\end{align}
where $\eta_{i}=\pm1$ on the even and odd lattice sites, respectively.
The Hamiltonian in Eq.~\eqref{eq:h_hubbard} is then transformed as
\begin{align}
    H
    \rightarrow
    H+U(|\Lambda|-N)
    ,
\end{align}
which means that Eq.~\eqref{eq:ph_symm} is the symmetry if the system is at half-filing $\langle N\rangle=|\Lambda|$.
From the viewpoint of the grand canonical Hamiltonian,
\begin{align}
    H-\mu N
    \rightarrow
    H+U(|\Lambda|-N)
    +\mu N
    -2\mu|\Lambda|
    ,
\end{align}
and Eq.~\eqref{eq:ph_symm} does describe the symmetry setting $\mu=U/2$.
For a detailed explanation of particle-hole symmetry, see Ref.~\cite{doi:10.1146/annurev-conmatphys-031620-102024}, for instance.

Let us now formulate the Hubbard model within the path-integral formalism.
The path-integral representation of the grand partition function is
\begin{align}
\label{eq:z_hubbard}
    Z=
    &\int
    \left[
        {\rm d}\bar{\psi}
        {\rm d}\psi
    \right]
    \exp
    \left[
        -\int^{\beta}_{0}{\rm d}\tau
        \left\{
            \sum_{n\in\Lambda}
            \bar{\psi}(n,\tau)
            \left(
                \frac{\partial}{\partial\tau}
                -
                \mu
            \right)
            \psi(n,\tau)
        \right.
    \right.
    \nonumber\\
    &\left.
        \left.
            -t
            \sum_{n}
            \sum_{\sigma=1}^{d}
            \left(
                \bar{\psi}(n+\hat{\sigma},\tau)\psi(n,\tau)
                +
                \bar{\psi}(n,\tau)\psi(n+\hat{\sigma},\tau)
            \right)
            +\frac{U}{2}
            \left(\bar{\psi}(n,\tau)\psi(n,\tau)\right)^{2}
        \right\}
    \right]
    ,
\end{align}
where the two-component Grassmann variables $\psi$ and $\bar{\psi}$ have been defined via
\begin{align}
\label{eq:def_psi}
    \psi(n,\tau)
    =
    \begin{bmatrix}
        \psi_{\uparrow}(n,\tau) \\
        \psi_{\downarrow}(n,\tau)
    \end{bmatrix}
    ,
    \quad\quad
    \bar{\psi}(n,\tau)
    =
    \begin{bmatrix}
        \bar{\psi}_{\uparrow}(n,\tau),~\bar{\psi}_{\downarrow}(n,\tau)
    \end{bmatrix}
    .
\end{align}
The imaginary time is parameterized by $\tau$ and $\beta$ is the inverse temperature.
The Grassmann fields obey the anti-periodic boundary condition along the imaginary time direction.
When the imaginary time direction is discretized by $\beta=N_{\tau}\epsilon$, we can identify
\begin{align}
\label{eq:s_hubbard}
    S=&
    \sum_{n\in\Lambda'}
    \epsilon
    \left[
        \bar{\psi}(n)
        \frac{\psi(n+\hat{\tau})-\psi(n)}{\epsilon}
    \right.
    \nonumber\\
    &
    \left.
        -t
        \sum_{\sigma=1}^{d}
        \left(
            \bar{\psi}(n+\hat{\sigma})\psi(n)
            +
            \bar{\psi}(n)\psi(n+\hat{\sigma})
        \right)
        +\frac{U}{2}
        \left(\bar{\psi}(n)\psi(n)\right)^{2}
        -\mu
        \bar{\psi}(n)\psi(n)
    \right]
\end{align}
as the action of the Hubbard model on the $(d+1)$-dimensional anisotropic lattice $\Lambda'$.
The unit vector along the imaginary time direction is denoted as $\hat{\tau}$ in Eq.~\eqref
{eq:s_hubbard}.
The Grassmann fields $\psi(n)$ and $\bar{\psi}(n)$ on $\Lambda'$ are defined in the same manner with Eq.~\eqref{eq:def_psi}.
With $\mu=U/2$, the system is at half-filling as before.

\subsection{(1+1)-dimensional model}

\begin{figure}[htbp]
  	\centering
	\includegraphics[width=1\hsize]{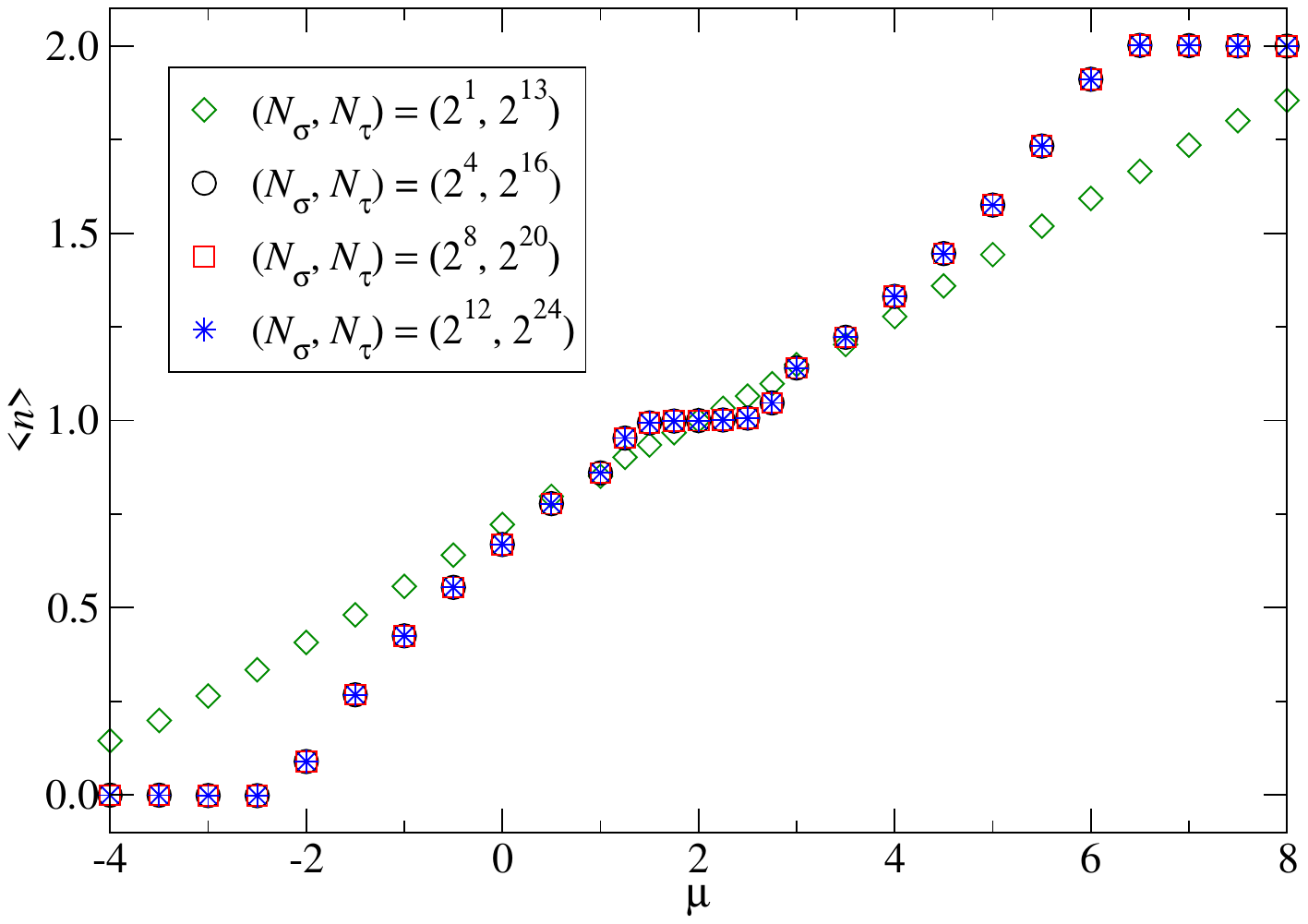}
	\caption{
        Adapted from Ref.~\cite{Akiyama:2021xxr}.
	Electron number density $\langle n\rangle$ at $(U,t)=(4,1)$ with $D_{\rm HOTRG}=80$ and $\epsilon=10^{-4}$.
    $N_{\sigma}$ and $N_{\tau}$ denote the spatial and temporal lattice sizes, respectively.
	}
  	\label{fig:hubbard_number_beta}
\end{figure}

\begin{figure}[htbp]
    \centering
    \begin{minipage}{0.8\hsize}
        \includegraphics[width=\hsize]{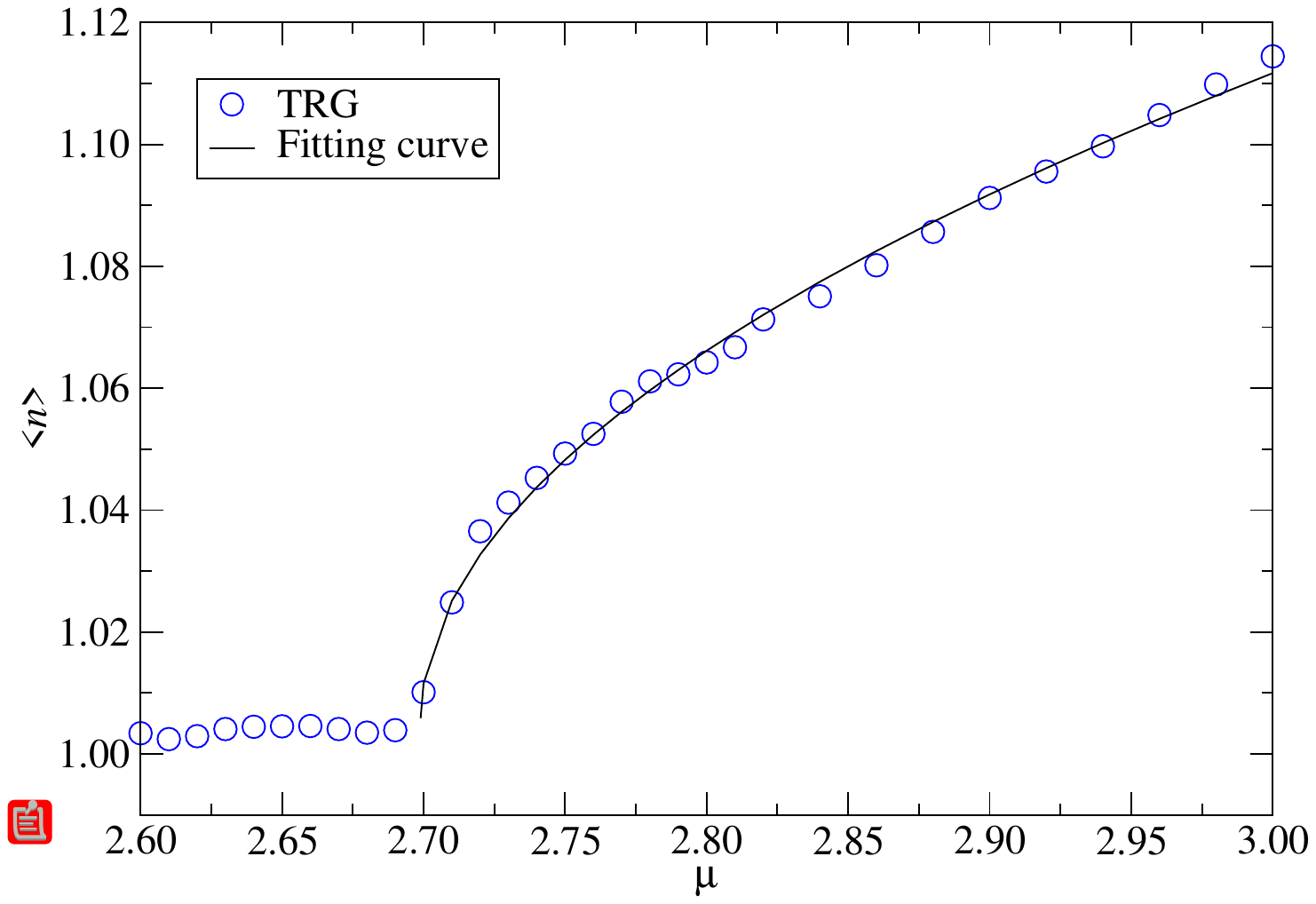}
    \end{minipage}
    \begin{minipage}{0.8\hsize}
        \includegraphics[width=\hsize]{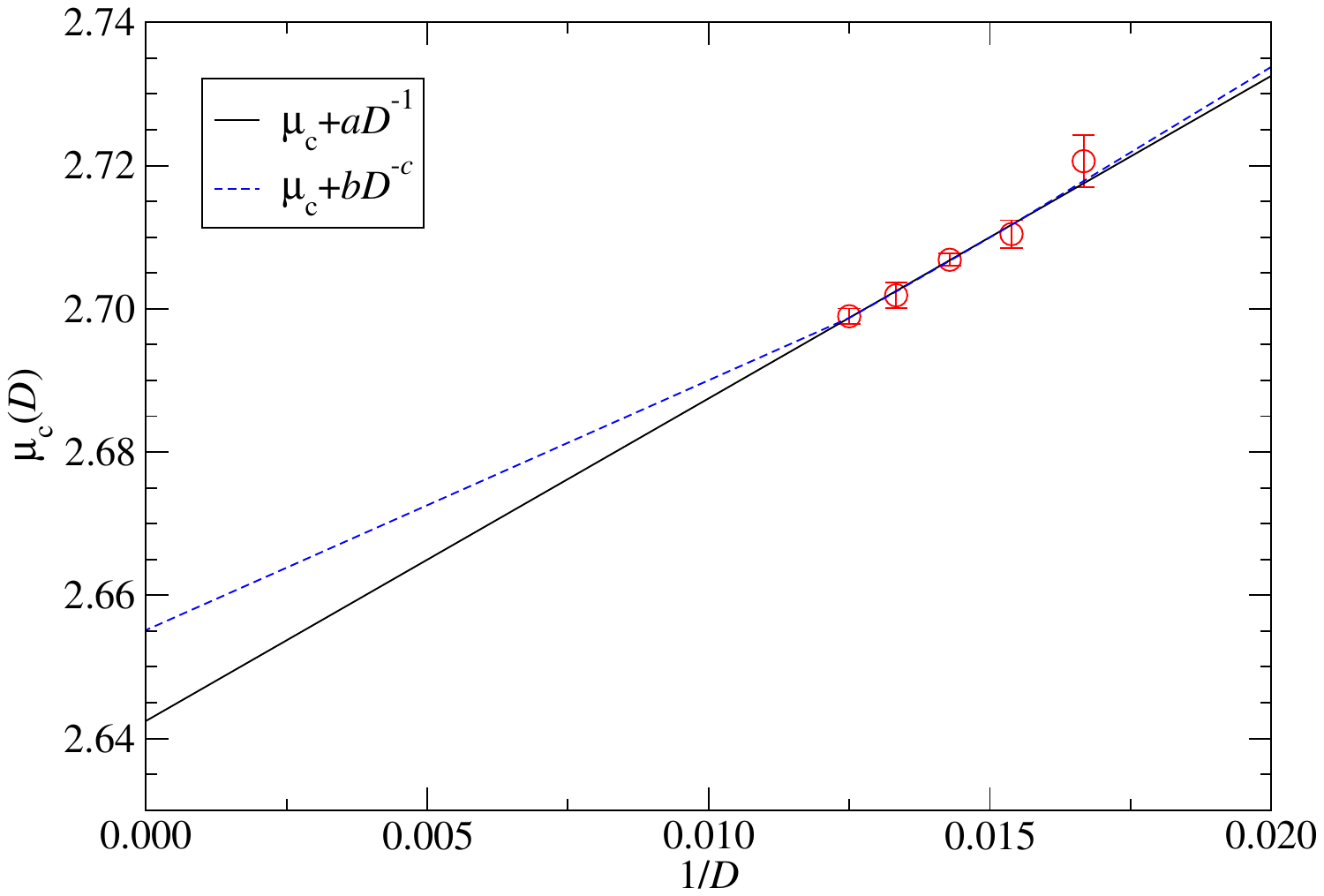}
    \end{minipage}
    \caption{
    Adapted from Ref.~\cite{Akiyama:2021xxr}.
    (top) Electron number density $\langle n\rangle$ at $\beta=N_{\tau}\epsilon=1677.7216$ with $\epsilon=10^{-4}$ and $D_{\rm HOTRG}=80$. 
    The fit ansatz in Eq.~\eqref{eq:fit_n} results in $\mu_{\rm c}(D_{\rm HOTRG})=2.698(1)$ and $\nu=0.51(2)$.
    (bottom) $\mu_{\rm c}(D_{\rm HOTRG})$ as a function of $1/D_{\rm HOTRG}$.
    The solid line shows the fit result of Eq.~\eqref{eq:fit_1d} and the dotted curve gives the fit result of Eq.~\eqref{eq:fit_1d_error}.
    }
  \label{fig:hubbard_mu_c}
\end{figure}

Akiyama and Kuramashi provided a benchmark study of the TRG method for the ($1+1$)-dimensional Hubbard model in Ref.~\cite{Akiyama:2021xxr}.
They also derived the Grassmann tensor network representation of the path integral defined by Eq.~\eqref{eq:s_hubbard} in general spatial dimensions.
In Eq.~\eqref{eq:s_hubbard}, the time slice $\epsilon$ should be $\epsilon\ll1$ to reproduce the grand partition function in Eq.~\eqref{eq:z_hubbard}.
In other words, the contribution from the temporal hopping terms ($O(1)$) becomes much larger than that from the spatial ones ($O(\epsilon)$) in the original action.
Consequently, the initial Grassmann tensor network becomes extremely anisotropic in the temporal direction.
The authors in Ref.~\cite{Akiyama:2021xxr} applied the HOTRG algorithm along the temporal direction in advance before the spacetime coarse-graining took place to investigate the ground state.

Throughout their study, the number density $\langle n\rangle$ is computed as a function of the chemical potential $\mu$ on the three points $(U,t)=(4,0),(0,1),(4,1)$.
At finite $U$, the Mott plateau $\langle n\rangle=1$ is reproduced as shown in Figure~\ref{fig:hubbard_number_beta}.
At $(U,t)=(4,1)$, they provided the numerical fit of the number density via
\begin{align}
\label{eq:fit_n}
    \langle n\rangle
    =
    A+B|\mu-\mu_{\rm c}(D_{\rm HOTRG})|^{\nu}
    .
\end{align}
This fit gives the pseudo critical point $\mu_{\rm c}(D_{\rm HOTRG})$ and the exponent $\nu$.
With the bond dimension $D_{\rm HOTRG}\in[60,80]$, the resulting $\nu$ is consistent with the exact value $\nu=0.5$.
Extrapolating $\mu_{\rm c}(D_{\rm HOTRG})$ to the limit $D_{\rm HOTRG}\to\infty$, they used the two fitting ansatz,
\begin{align}
\label{eq:fit_1d}
    \mu_{\rm c}(D_{\rm HOTRG})
    =
    \mu_{\rm c}
    +
    aD_{\rm HOTRG}^{-1}
    ,
\end{align}
\begin{align}
\label{eq:fit_1d_error}
    \mu_{\rm c}(D_{\rm HOTRG})
    =
    \mu_{\rm c}
    +
    bD_{\rm HOTRG}^{-c}
    ,
\end{align}
where the latter fit was employed to estimate uncertainty in the fitting assumption.
These numerical fits are shown in Figure~\ref{fig:hubbard_mu_c}.
The extrapolation has given $\mu_{\rm c}=2.642(05)(13)$ which is in agreement with the exact value $\mu_{\rm c}=2.643\cdots$ based on the Bethe ansatz.

 \subsection{(2+1)-dimensional model}

Akiyama, Kuramashi, and Yamashita investigated the metal-insulator transition in the (2+1)-dimensional Hubbard model~\cite{Akiyama:2021glo}.
They used the ATRG algorithm to investigate the ground state and applied the same strategy with Ref.~\cite{Akiyama:2021xxr}; the spacetime coarse-graining after the imaginary-time evolution.
As a validation of the numerical strategy, the number density at $(U,t)=(8,0)$ is computed.
The TRG result is consistent with the exact one and the Mott plateau is reproduced around $\mu=4$ as expected.
The number density is also computed at $(U,t)=(80,1),(8,1),(2,1)$.
With $(U,t)=(80,1)$, the TRG gives the smooth number density as a function of the chemical potential and $\mu_{\rm c}/U\neq1$ in contrast to $(U,t)=(8,0)$. 
These results imply that the TRG calculation captures the spatial hopping effects even with the large repulsion parameter $U$.
At $(U,t)=(8,1),(2,1)$, the transition point $\mu_{\rm c}$ is determined by the global fit using the quadratic function,
\begin{align}
    \langle n\rangle
    =
    1
    +
    a(\mu-\mu_{\rm c}(D_{\rm ATRG}))
    +
    b(\mu-\mu_{\rm c}(D_{\rm ATRG}))^{2}
    ,
\end{align}
with
\begin{align}
    \mu_{\rm c}(D_{\rm ATRG})
    =
    \mu_{\rm c}+cD_{\rm ATRG}^{-1}
    .
\end{align}
$D_{\rm ATRG}$ is varied as $56,64,72,80$.
Their estimates are $\mu_{\rm c}=6.43(4)$ at $(U,t)=(8,1)$ and $\mu_{\rm c}=1.30(6)$ at $(U,t)=(2,1)$.
Based on these estimates, the authors expect that $|\mu_{\rm c}-U/2|$ vanishes only at $U=0$, that is the metal-insulator transition could take place with any finite repulsion.

\section{Conclusions}

We have reviewed the two formulations to derive the Grassmann tensor network representations for fermionic path integrals.
We have shown that both formulations can result in the same ordinary tensor (bosonic tensor or coefficient tensor) by properly ordering the auxiliary Grassmann variables.
Exact contractions are defined as the integration of these auxiliary Grassmann variables.
These formulations immediately allow us to extend any TRG algorithms for fermions.
By introducing the Grassmann parity functions, we can immediately extend the algorithms, such as the Levin-Nave TRG and HOTRG, for fermionic path integrals under explicit correspondence with the original ones. In particular, the geometric representation and the connectivities remain identical. 

These Grassmann TRG algorithms have been applied to various lattice theories including the relativistic models not only in two but also in four dimensions and the Hubbard model at finite density.
Some of these models include examples where the Monte Carlo method is extremely difficult to apply due to the sign problem.

We have also reviewed several TNR algorithms, the bond-weighting method, and multilayered Grassmann tensor networks for $N_{f}$-flavor fermions.
Although some of these improved TRG methods were originally proposed for spin systems, recent numerical calculations have shown that these methods are also efficient for fermionic systems.
Research on such improved algorithms has continued to progress in recent years; a new algorithm of loop-TNR~\cite{homma2023nuclear}, combinations of the Monte Carlo method and TRG~\cite{ferris2015unbiased,Huggins:2017yzz,Arai:2022uee}, and application of the machine-learning techniques to the TRG~\cite{PhysRevX.9.031041,PhysRevB.101.220409,Jha:2023bpn}.
The extension of these novel improved methods to fermionic systems is considered important in examining whether these methods are also valid for more general physical systems including fermions.

\section{Acknowledgments}
We thank the members of the QuLAT Collaboration for valuable discussions.
Y. M. was supported in part by the U.S. Department of Energy (DOE) under Award Number DE-SC0019139 and DE-SC0010113.
This research used resources of the Syracuse University HTC Campus Grid and NSF award ACI-1341006 and the National Energy Research Scientific Computing Center (NERSC), a U.S. Department of Energy Office of Science User Facility located at Lawrence Berkeley National Laboratory, operated under Contract No. DE-AC02-05CH11231 using NERSC awards HEP-ERCAP0020659 and HEP-ERCAP0023235.
S. A. acknowledges the support from the Endowed Project for Quantum Software Research and Education, the University of Tokyo (\url{https://qsw.phys.s.u-tokyo.ac.jp/}) 
and JSPS KAKENHI Grant Number JP23K13096.
\newline

\noindent References

\begin{thebibliography}{100}
\expandafter\ifx\csname url\endcsname\relax
  \def\url#1{{\tt #1}}\fi
\expandafter\ifx\csname urlprefix\endcsname\relax\def\urlprefix{URL }\fi
\providecommand{\eprint}[2][]{\url{#2}}

\bibitem{leo66}
Kadanoff L~P 1966 {\em Physics Physique Fizika\/} {\bf 2} 263--272

\bibitem{wilson73pr}
Wilson K~G and Kogut J~B 1974 {\em Phys. Rept.\/} {\bf 12} 75--199

\bibitem{wilson74rmp}
Wilson K~G 1975 {\em Rev. Mod. Phys.\/} {\bf 47} 773

\bibitem{cardy1996scaling}
Cardy J 1996 {\em Scaling and Renormalization in Statistical Physics\/}
  Cambridge Lecture Notes in Physics (Cambridge University Press) ISBN
  9780521499590

\bibitem{Migdal:1975zf}
Migdal A~A 1975 {\em Sov. Phys. JETP\/} {\bf 42} 743

\bibitem{kadanoff76}
Kadanoff L~P 1976 {\em Annals Phys.\/} {\bf 100} 359--394

\bibitem{dyson68}
Dyson F~J 1969 {\em Commun. Math. Phys.\/} {\bf 12} 91--107

\bibitem{baker72}
Baker G~A 1972 {\em Phys. Rev. B\/} {\bf 5} 2622--2633

\bibitem{hmreview}
Meurice Y 2007 {\em J. Phys.\/} {\bf A40} R39 (\textit{Preprint}
  \eprint{hep-th/0701191})

\bibitem{berges2000}
Berges J, Tetradis N and Wetterich C 2002 {\em Phys. Rept.\/} {\bf 363}
  223--386 (\textit{Preprint} \eprint{hep-ph/0005122})

\bibitem{bervillier2007}
Bervillier C, Juttner A and Litim D~F 2007 {\em Nucl. Phys. B\/} {\bf 783}
  213--226 (\textit{Preprint} \eprint{hep-th/0701172})

\bibitem{bervillier2013}
Bervillier C 2013 {\em Nucl. Phys. B\/} {\bf 876} 587--604 (\textit{Preprint}
  \eprint{1307.3679})

\bibitem{white92}
White S~R 1992 {\em Phys. Rev. Lett.\/} {\bf 69} 2863--2866

\bibitem{schollwock2005}
Schollw{\"o}ck U 2005 {\em Rev. Mod. Phys.\/} {\bf 77} 259--315
  (\textit{Preprint} \eprint{cond-mat/0409292})

\bibitem{uli2011}
{Schollw{\"o}ck} U 2011 {\em Annals of Physics\/} {\bf 326} 96--192
  (\textit{Preprint} \eprint{1008.3477})

\bibitem{vidal2007}
Vidal G 2007 {\em Phys. Rev. Lett.\/} {\bf 99} 220405 (\textit{Preprint}
  \eprint{cond-mat/0512165})

\bibitem{cirac2009}
Cirac J~I and Verstraete F 2009 {\em J. Phys. A\/} {\bf 42} 504004
  (\textit{Preprint} \eprint{0910.1130})

\bibitem{schollwock2011}
Schollw{\"o}ck U 2011 {\em Phil. Trans. Roy. Soc. Lond.\/} {\bf A369}
  2643--2661

\bibitem{orus2013}
Or{\'u}s R 2014 {\em Annals Phys.\/} {\bf 349} 117--158 (\textit{Preprint}
  \eprint{1306.2164})

\bibitem{2015PhRvL.115r0405E}
{Evenbly} G and {Vidal} G 2015 {\em Phys. Rev. Lett.\/} {\bf 115} 180405
  (\textit{Preprint} \eprint{1412.0732})

\bibitem{silvi2017}
Silvi P, Tschirsich F, Gerster M, J\"unemann J, Jaschke D, Rizzi M and
  Montangero S 2019 {\em SciPost Phys. Lect. Notes\/} {\bf 8} 1
  (\textit{Preprint} \eprint{1710.03733})

\bibitem{hv2017}
Haegeman J and Verstraete F 2017 {\em Ann. Rev. Condensed Matter Phys.\/} {\bf
  8} 355--406 (\textit{Preprint} \eprint{1611.08519})

\bibitem{montangero2018}
Montangero S 2018 {\em Introduction to Tensor Network Methods: Numerical
  simulations of low-dimensional many-body quantum systems\/} (Springer
  International Publishing) ISBN 9783030014094

\bibitem{ran2020}
Ran S~J, Tirrito E, Peng C, Chen X, Tagliacozzo L, Su G and Lewenstein M 2020
  {\em Tensor Network Contractions: Methods and Applications to Quantum
  Many-Body Systems\/} (Springer International Publishing) ISBN 9783030344894

\bibitem{RevModPhys.93.045003}
Cirac J~I, Perez-Garcia D, Schuch N and Verstraete F 2021 {\em Rev. Mod.
  Phys.\/} {\bf 93} 045003 (\textit{Preprint} \eprint{2011.12127})

\bibitem{nishinoctm}
{Nishino} T and {Okunishi} K 1996 {\em Journal of the Physical Society of
  Japan\/} {\bf 65} 891 (\textit{Preprint} \eprint{cond-mat/9507087})

\bibitem{Levin:2006jai}
Levin M and Nave C~P 2007 {\em Phys. Rev. Lett.\/} {\bf 99} 120601
  (\textit{Preprint} \eprint{cond-mat/0611687})

\bibitem{Gu:2009dr}
Gu Z~C and Wen X~G 2009 {\em Phys. Rev. B\/} {\bf 80} 155131 (\textit{Preprint}
  \eprint{0903.1069})

\bibitem{PhysRevLett.103.160601}
Xie Z~Y, Jiang H~C, Chen Q~N, Weng Z~Y and Xiang T 2009 {\em Phys. Rev.
  Lett.\/} {\bf 103} 160601 (\textit{Preprint} \eprint{0809.0182})

\bibitem{Gu:2010yh}
Gu Z~C, Verstraete F and Wen X~G  (\textit{Preprint} \eprint{1004.2563})

\bibitem{Gu:2013gba}
Gu Z~C 2013 {\em Phys. Rev. B\/} {\bf 88} 115139 (\textit{Preprint}
  \eprint{1109.4470})

\bibitem{2012PhRvB..86d5139X}
{Xie} Z~Y, {Chen} J, {Qin} M~P, {Zhu} J~W, {Yang} L~P and {Xiang} T 2012 {\em
  Phys. Rev. B\/} {\bf 86} 045139 (\textit{Preprint} \eprint{1201.1144})

\bibitem{efratirmp}
{Efrati} E, {Wang} Z, {Kolan} A and {Kadanoff} L~P 2014 {\em Reviews of Modern
  Physics\/} {\bf 86} 647--667 (\textit{Preprint} \eprint{1301.6323})

\bibitem{prb87}
Meurice Y 2013 {\em Phys. Rev. B\/} {\bf 87} 064422 (\textit{Preprint}
  \eprint{1211.3675})

\bibitem{Lyu:2021qlw}
Lyu X, Xu R~G and Kawashima N 2021 {\em Phys. Rev. Res.\/} {\bf 3} 023048
  (\textit{Preprint} \eprint{2102.08136})

\bibitem{PhysRevB.104.165132}
{Ueda} A and {Oshikawa} M 2021 {\em Phys. Rev. B\/} {\bf 104} 165132
  (\textit{Preprint} \eprint{2105.11460})

\bibitem{Ueda:2022nur}
Ueda A and Oshikawa M 2022 {\em Phys. Rev. E\/} {\bf 106} 014104
  (\textit{Preprint} \eprint{2202.07042})

\bibitem{PhysRevB.108.024413}
{Ueda} A and {Oshikawa} M 2023 {\em Phys. Rev. B\/} {\bf 108} 024413
  (\textit{Preprint} \eprint{2302.06632})

\bibitem{Huang:2023vwp}
Huang C~Y, Chan S~H, Kao Y~J and Chen P 2023 {\em Phys. Rev. B\/} {\bf 107}
  205123 (\textit{Preprint} \eprint{2302.02585})

\bibitem{guo2023tensor}
Guo W and Wei T~C 2023  (\textit{Preprint} \eprint{2305.09899})

\bibitem{Ueda:2023ihr}
Ueda A and Yamazaki M 2023  (\textit{Preprint} \eprint{2307.02523})

\bibitem{Meurice:2011wy}
Meurice Y, Perry R and Tsai S~W 2011 {\em Phil. Trans. Roy. Soc. Lond.\/} {\bf
  A369} 2602--2611 (\textit{Preprint} \eprint{1102.5717})

\bibitem{Byrnes:2002nv}
Byrnes T, Sriganesh P, Bursill R~J and Hamer C~J 2002 {\em Phys. Rev. D\/} {\bf
  66} 013002 (\textit{Preprint} \eprint{hep-lat/0202014})

\bibitem{Banuls:2013jaa}
Ba\~nuls M, Cichy K, Jansen K and Cirac J 2013 {\em JHEP\/} {\bf 11} 158
  (\textit{Preprint} \eprint{1305.3765})

\bibitem{buyens2014}
Buyens B, Haegeman J, Van~Acoleyen K, Verschelde H and Verstraete F 2014 {\em
  Phys. Rev. Lett.\/} {\bf 113} 091601 (\textit{Preprint} \eprint{1312.6654})

\bibitem{buyens2016}
Buyens B, Haegeman J, Verschelde H, Verstraete F and Van~Acoleyen K 2016 {\em
  Phys. Rev. X\/} {\bf 6} 041040 (\textit{Preprint} \eprint{1509.00246})

\bibitem{Funcke:2019zna}
Funcke L, Jansen K and K\"uhn S 2020 {\em Phys. Rev. D\/} {\bf 101} 054507
  (\textit{Preprint} \eprint{1908.00551})

\bibitem{Dempsey:2022nys}
Dempsey R, Klebanov I~R, Pufu S~S and Zan B 2022 {\em Phys. Rev. Res.\/} {\bf
  4} 043133 (\textit{Preprint} \eprint{2206.05308})

\bibitem{Okuda:2022hsq}
Okuda T 2023 {\em Phys. Rev. D\/} {\bf 107} 054506 (\textit{Preprint}
  \eprint{2210.00297})

\bibitem{Honda:2022edn}
Honda M, Itou E and Tanizaki Y 2022 {\em JHEP\/} {\bf 11} 141
  (\textit{Preprint} \eprint{2210.04237})

\bibitem{Itou:2023img}
Itou E, Matsumoto A and Tanizaki Y 2023 {\em JHEP\/} {\bf 11} 231
  (\textit{Preprint} \eprint{2307.16655})

\bibitem{kuhn2015}
K\"uhn S, Cirac J~I and Ba\~nuls M~C 2015 {\em JHEP\/} {\bf 07} 130
  (\textit{Preprint} \eprint{1505.04441})

\bibitem{banuls2017}
Ba\~nuls M~C, Cichy K, Cirac J~I, Jansen K and K\"uhn S 2017 {\em Phys. Rev.
  X\/} {\bf 7} 041046 (\textit{Preprint} \eprint{1707.06434})

\bibitem{Hayata:2023pkw}
Hayata T, Hidaka Y and Nishimura K 2023  (\textit{Preprint}
  \eprint{2311.11643})

\bibitem{Liu:2023lsr}
Liu H, Bhattacharya T, Chandrasekharan S and Gupta R 2023  (\textit{Preprint}
  \eprint{2312.17734})

\bibitem{bruckmann2018}
Bruckmann F, Jansen K and K\"uhn S 2019 {\em Phys. Rev. D\/} {\bf 99} 074501
  (\textit{Preprint} \eprint{1812.00944})

\bibitem{celi2014}
Tagliacozzo L, Celi A and Lewenstein M 2014 {\em Phys. Rev. X\/} {\bf 4} 041024
  (\textit{Preprint} \eprint{1405.4811})

\bibitem{silvi2016}
Silvi P, Rico E, Dalmonte M, Tschirsich F and Montangero S 2017 {\em Quantum\/}
  {\bf 1} 9 (\textit{Preprint} \eprint{1606.05510})

\bibitem{silvi2019}
Silvi P, Sauer Y, Tschirsich F and Montangero S 2019 {\em Phys. Rev. D\/} {\bf
  100} 074512 (\textit{Preprint} \eprint{1901.04403})

\bibitem{rico2014}
{Rico} E, {Pichler} T, {Dalmonte} M, {Zoller} P and {Montangero} S 2014 {\em
  Phys. Rev. Lett.\/} {\bf 112} 201601 (\textit{Preprint} \eprint{1312.3127})

\bibitem{pichler2016}
{Pichler} T, {Dalmonte} M, {Rico} E, {Zoller} P and {Montangero} S 2016 {\em
  Phys. Rev. X\/} {\bf 6} 011023 (\textit{Preprint} \eprint{1505.04440})

\bibitem{zohar2015b}
Zohar E and Burrello M 2015 {\em Phys. Rev. D\/} {\bf 91} 054506
  (\textit{Preprint} \eprint{1409.3085})

\bibitem{banuls2018}
Ba\~nuls M~C, Cichy K, Cirac J~I, Jansen K and K\"uhn S 2018 {\em PoS\/} {\bf
  LATTICE2018} 022 (\textit{Preprint} \eprint{1810.12838})

\bibitem{cinchy2019}
Ba\~nuls M~C and Cichy K 2020 {\em Rept. Prog. Phys.\/} {\bf 83} 024401
  (\textit{Preprint} \eprint{1910.00257})

\bibitem{balian75}
Balian R, Drouffe J~M and Itzykson C 1975 {\em Phys. Rev. D\/} {\bf 11}(8)
  2104--2119

\bibitem{drouffe92}
Itzykson C and Drouffe J 1992 {\em Statistical Field Theory\/} Cambridge
  monographs on mathematical physics (Cambridge University Press)

\bibitem{Liu:2013nsa}
Liu Y, Meurice Y, Qin M~P, Unmuth-Yockey J, Xiang T, Xie Z~Y, Yu J~F and Zou H
  2013 {\em Phys. Rev. D\/} {\bf 88} 056005 (\textit{Preprint}
  \eprint{1307.6543})

\bibitem{pre89}
Yu J~F, Xie Z~Y, Meurice Y, Liu Y, Denbleyker A, Zou H, Qin M~P and Chen J 2014
  {\em Phys. Rev. E\/} {\bf 89} 013308 (\textit{Preprint} \eprint{1309.4963})

\bibitem{prd89}
Denbleyker A, Liu Y, Meurice Y, Qin M~P, Xiang T, Xie Z~Y, Yu J~F and Zou H
  2014 {\em Phys. Rev. D\/} {\bf 89} 016008 (\textit{Preprint}
  \eprint{1309.6623})

\bibitem{Zou:2014rha}
Zou H, Liu Y, Lai C~Y, Unmuth-Yockey J, Bazavov A, Xie Z~Y, Xiang T,
  Chandrasekharan S, Tsai S~W and Meurice Y 2014 {\em Phys. Rev. A\/} {\bf 90}
  063603 (\textit{Preprint} \eprint{1403.5238})

\bibitem{Akiyama:2023fyk}
Akiyama S, Jha R~G and Unmuth-Yockey J 2023  (\textit{Preprint}
  \eprint{2312.11649})

\bibitem{meurice2019}
Meurice Y 2019 {\em Phys. Rev. D\/} {\bf 100} 014506 (\textit{Preprint}
  \eprint{1903.01918})

\bibitem{meurice2020}
Meurice Y 2020 {\em Phys. Rev. D\/} {\bf 102} 014506 (\textit{Preprint}
  \eprint{2003.10986})

\bibitem{Meurice:2020pxc}
Meurice Y, Sakai R and Unmuth-Yockey J 2022 {\em Rev. Mod. Phys.\/} {\bf 94}
  025005 (\textit{Preprint} \eprint{2010.06539})

\bibitem{PhysRevA.80.042333}
{Barthel} T, {Pineda} C and {Eisert} J 2009 {\em Phys. Rev. A\/} {\bf 80}
  042333 (\textit{Preprint} \eprint{0907.3689})

\bibitem{corboz2010}
{Corboz} P, {Evenbly} G, {Verstraete} F and {Vidal} G 2010 {\em Phys. Rev. A\/}
  {\bf 81} 010303 (\textit{Preprint} \eprint{0904.4151})

\bibitem{2001PThPh.105..409N}
{Nishino} T, {Hieida} Y, {Okunishi} K and {Maeshima} N 2001 {\em Progress of
  Theoretical Physics\/} {\bf 105} 409--417 (\textit{Preprint}
  \eprint{cond-mat/0011103})

\bibitem{2003PThPh.110..691G}
{Gendiar} A, {Maeshima} N and {Nishino} T 2003 {\em Progress of Theoretical
  Physics\/} {\bf 110} 691--699 (\textit{Preprint} \eprint{cond-mat/0303376})

\bibitem{verstraete2004}
Verstraete F and Cirac J~I 2004  (\textit{Preprint} \eprint{cond-mat/0407066})

\bibitem{Kraus:2010zak}
{Kraus} C~V, {Schuch} N, {Verstraete} F and {Cirac} J~I 2010 {\em Phys. Rev.
  A\/} {\bf 81} 052338 (\textit{Preprint} \eprint{0904.4667})

\bibitem{Zohar:2015jnb}
Zohar E and Burrello M 2016 {\em New J. Phys.\/} {\bf 18} 043008
  (\textit{Preprint} \eprint{1511.08426})

\bibitem{Zohar:2017yxl}
Zohar E and Cirac J~I 2018 {\em Phys. Rev. D\/} {\bf 97} 034510
  (\textit{Preprint} \eprint{1710.11013})

\bibitem{Zapp:2017fcr}
Zapp K and Or{\'u}s R 2017 {\em Phys. Rev. D\/} {\bf 95} 114508
  (\textit{Preprint} \eprint{1704.03015})

\bibitem{Felser:2019xyv}
Felser T, Silvi P, Collura M and Montangero S 2020 {\em Phys. Rev. X\/} {\bf
  10} 041040 (\textit{Preprint} \eprint{1911.09693})

\bibitem{Magnifico:2020bqt}
Magnifico G, Felser T, Silvi P and Montangero S 2021 {\em Nature Commun.\/}
  {\bf 12} 3600 (\textit{Preprint} \eprint{2011.10658})

\bibitem{Felser:2020ntd}
Felser T, Notarnicola S and Montangero S 2021 {\em Phys. Rev. Lett.\/} {\bf
  126} 170603 (\textit{Preprint} \eprint{2011.08200})

\bibitem{Emonts:2020drm}
Emonts P, Ba\~nuls M~C, Cirac I and Zohar E 2020 {\em Phys. Rev. D\/} {\bf 102}
  074501 (\textit{Preprint} \eprint{2008.00882})

\bibitem{Robaina:2020aqh}
Robaina D, Ba\~nuls M~C and Cirac J~I 2021 {\em Phys. Rev. Lett.\/} {\bf 126}
  050401 (\textit{Preprint} \eprint{2007.11630})

\bibitem{Emonts:2022yom}
Emonts P, Kelman A, Borla U, Moroz S, Gazit S and Zohar E 2023 {\em Phys. Rev.
  D\/} {\bf 107} 014505 (\textit{Preprint} \eprint{2211.00023})

\bibitem{Bender:2023gwr}
Bender J, Emonts P and Cirac J~I 2023 {\em Phys. Rev. Res.\/} {\bf 5} 043128
  (\textit{Preprint} \eprint{2304.05916})

\bibitem{Cataldi:2023xki}
Cataldi G, Magnifico G, Silvi P and Montangero S 2023  (\textit{Preprint}
  \eprint{2307.09396})

\bibitem{Emonts:2023ttz}
Emonts P and Zohar E 2023 {\em Phys. Rev. D\/} {\bf 108} 014514
  (\textit{Preprint} \eprint{2304.06744})

\bibitem{emontspepslectures}
Emonts P and Zohar E 2020 {\em SciPost Phys. Lect. Notes\/} {\bf 12} 1
  (\textit{Preprint} \eprint{1807.01294})

\bibitem{Vanhecke:2019pez}
Vanhecke B, Haegeman J, Van~Acoleyen K, Vanderstraeten L and Verstraete F 2019
  {\em Phys. Rev. Lett.\/} {\bf 123} 250604 (\textit{Preprint}
  \eprint{1907.08603})

\bibitem{yang2017loop}
Yang S, Gu Z~C and Wen X~G 2017 {\em Phys. Rev. Lett.\/} {\bf 118} 110504
  (\textit{Preprint} \eprint{1512.04938})

\bibitem{Hauru:2017tne}
Hauru M, Delcamp C and Mizera S 2018 {\em Phys. Rev. B\/} {\bf 97} 045111
  (\textit{Preprint} \eprint{1709.07460})

\bibitem{Takeda:2014vwa}
Takeda S and Yoshimura Y 2015 {\em PTEP\/} {\bf 2015} 043B01 (\textit{Preprint}
  \eprint{1412.7855})

\bibitem{Akiyama:2023lvr}
Akiyama S 2023 {\em Phys. Rev. D\/} {\bf 108} 034514 (\textit{Preprint}
  \eprint{2304.01473})

\bibitem{Bloch:2022vqz}
Bloch J and Lohmayer R 2023 {\em Nucl. Phys. B\/} {\bf 986} 116032
  (\textit{Preprint} \eprint{2206.00545})

\bibitem{Akiyama:2020soe}
Akiyama S, Kuramashi Y, Yamashita T and Yoshimura Y 2021 {\em JHEP\/} {\bf 01}
  121 (\textit{Preprint} \eprint{2009.11583})

\bibitem{Kadoh:2018hqq}
Kadoh D, Kuramashi Y, Nakamura Y, Sakai R, Takeda S and Yoshimura Y 2018 {\em
  JHEP\/} {\bf 03} 141 (\textit{Preprint} \eprint{1801.04183})

\bibitem{Asaduzzaman:2023pyz}
Asaduzzaman M, Catterall S, Meurice Y, Sakai R and Toga G~C 2023
  (\textit{Preprint} \eprint{2312.16167})

\bibitem{1994RvMP...66..763D}
{Dagotto} E 1994 {\em Rev. Mod. Phys.\/} {\bf 66} 763--840 (\textit{Preprint}
  \eprint{cond-mat/9311013})

\bibitem{RevModPhys.68.13}
Georges A, Kotliar G, Krauth W and Rozenberg M~J 1996 {\em Rev. Mod. Phys.\/}
  {\bf 68} 13--125

\bibitem{Hal_Tasaki_1998}
Tasaki H 1998 {\em Journal of Physics: Condensed Matter\/} {\bf 10} 4353

\bibitem{PhysRevX.5.041041}
{LeBlanc} J~P~F, {Antipov} A~E, {Becca} F, {Bulik} I~W, {Chan} G~K~L, {Chung}
  C~M, {Deng} Y, {Ferrero} M, {Henderson} T~M, {Jim{\'e}nez-Hoyos} C~A, {Kozik}
  E, {Liu} X~W, {Millis} A~J, {Prokof'ev} N~V, {Qin} M, {Scuseria} G~E, {Shi}
  H, {Svistunov} B~V, {Tocchio} L~F, {Tupitsyn} I~S, {White} S~R, {Zhang} S,
  {Zheng} B~X, {Zhu} Z, {Gull} E and {Simons Collaboration on the Many-Electron
  Problem} 2015 {\em Phys. Rev. X\/} {\bf 5} 041041 (\textit{Preprint}
  \eprint{1505.02290})

\bibitem{doi:10.1146/annurev-conmatphys-031620-102024}
{Arovas} D~P, {Berg} E, {Kivelson} S~A and {Raghu} S 2022 {\em Annual Review of
  Condensed Matter Physics\/} {\bf 13} 239--274 (\textit{Preprint}
  \eprint{2103.12097})

\bibitem{doi:10.1146/annurev-conmatphys-090921-033948}
{Qin} M, {Sch{\"a}fer} T, {Andergassen} S, {Corboz} P and {Gull} E 2022 {\em
  Annual Review of Condensed Matter Physics\/} {\bf 13} 275--302
  (\textit{Preprint} \eprint{2104.00064})

\bibitem{Ostmeyer:2022gwi}
Ostmeyer J 2023 {\em PoS\/} {\bf LATTICE2022} 230 (\textit{Preprint}
  \eprint{2210.06874})

\bibitem{Akiyama:2021xxr}
Akiyama S and Kuramashi Y 2021 {\em Phys. Rev. D\/} {\bf 104} 014504
  (\textit{Preprint} \eprint{2105.00372})

\bibitem{Akiyama:2021glo}
Akiyama S, Kuramashi Y and Yamashita T 2022 {\em PTEP\/} {\bf 2022} 023I01
  (\textit{Preprint} \eprint{2109.14149})

\bibitem{Shimizu:2014uva}
Shimizu Y and Kuramashi Y 2014 {\em Phys. Rev. D\/} {\bf 90} 014508
  (\textit{Preprint} \eprint{1403.0642})

\bibitem{Shimizu:2014fsa}
Shimizu Y and Kuramashi Y 2014 {\em Phys. Rev. D\/} {\bf 90} 074503
  (\textit{Preprint} \eprint{1408.0897})

\bibitem{Shimizu:2017onf}
Shimizu Y and Kuramashi Y 2018 {\em Phys. Rev. D\/} {\bf 97} 034502
  (\textit{Preprint} \eprint{1712.07808})

\bibitem{Asaduzzaman:2022pnw}
Asaduzzaman M, Catterall S, Meurice Y, Sakai R and Toga G~C 2023 {\em JHEP\/}
  {\bf 01} 024 (\textit{Preprint} \eprint{2210.03834})

\bibitem{Sakai:2017jwp}
Sakai R, Takeda S and Yoshimura Y 2017 {\em PTEP\/} {\bf 2017} 063B07
  (\textit{Preprint} \eprint{1705.07764})

\bibitem{Yoshimura:2017jpk}
Yoshimura Y, Kuramashi Y, Nakamura Y, Takeda S and Sakai R 2018 {\em Phys. Rev.
  D\/} {\bf 97} 054511 (\textit{Preprint} \eprint{1711.08121})

\bibitem{Meurice:2018fky}
Meurice Y 2018 {\em PoS\/} {\bf LATTICE2018} 231

\bibitem{Bao:2019hfc}
Bao C 2019 {\em {Loop Optimization of Tensor Network Renormalization:
  Algorithms and Applications}\/} Ph.D. thesis U. Waterloo (main)

\bibitem{Akiyama:2020sfo}
Akiyama S and Kadoh D 2021 {\em JHEP\/} {\bf 10} 188 (\textit{Preprint}
  \eprint{2005.07570})

\bibitem{Akiyama:2022pse}
Akiyama S 2022 {\em JHEP\/} {\bf 11} 030 (\textit{Preprint}
  \eprint{2208.03227})

\bibitem{Yosprakob:2023tyr}
Yosprakob A, Nishimura J and Okunishi K 2023 {\em JHEP\/} {\bf 11} 187
  (\textit{Preprint} \eprint{2309.01422})

\bibitem{Akiyama:2023rih}
Akiyama S 2023  (\textit{Preprint} \eprint{2311.17691})

\bibitem{Yosprakob:2023flr}
Yosprakob A 2023  (\textit{Preprint} \eprint{2309.07557})

\bibitem{Okunishi:2021but}
Okunishi K, Nishino T and Ueda H 2022 {\em J. Phys. Soc. Jap.\/} {\bf 91}
  062001 (\textit{Preprint} \eprint{2111.12223})

\bibitem{Akiyama:2021nhe}
Akiyama S, Kuramashi Y and Yoshimura Y 2022 {\em PoS\/} {\bf LATTICE2021} 530
  (\textit{Preprint} \eprint{2111.04240})

\bibitem{Kadoh:2022loj}
Kadoh D 2022 {\em PoS\/} {\bf LATTICE2021} 633

\bibitem{Kuramashi:2019cgs}
Kuramashi Y and Yoshimura Y 2020 {\em JHEP\/} {\bf 04} 089 (\textit{Preprint}
  \eprint{1911.06480})

\bibitem{Fukuma:2021cni}
Fukuma M, Kadoh D and Matsumoto N 2021 {\em PTEP\/} {\bf 2021} 123B03
  (\textit{Preprint} \eprint{2107.14149})

\bibitem{Hirasawa:2021qvh}
Hirasawa M, Matsumoto A, Nishimura J and Yosprakob A 2021 {\em JHEP\/} {\bf 12}
  011 (\textit{Preprint} \eprint{2110.05800})

\bibitem{Dittrich:2014mxa}
Dittrich B, Mizera S and Steinhaus S 2016 {\em New J. Phys.\/} {\bf 18} 053009
  (\textit{Preprint} \eprint{1409.2407})

\bibitem{Kuwahara:2022ubg}
Kuwahara T and Tsuchiya A 2022 {\em PTEP\/} {\bf 2022} 093B02
  (\textit{Preprint} \eprint{2205.08883})

\bibitem{Luo:2022eje}
Luo X and Kuramashi Y 2023 {\em Phys. Rev. D\/} {\bf 107} 094509
  (\textit{Preprint} \eprint{2208.13991})

\bibitem{Akiyama:2022eip}
Akiyama S and Kuramashi Y 2022 {\em JHEP\/} {\bf 05} 102 (\textit{Preprint}
  \eprint{2202.10051})

\bibitem{2018PhRvE..97c3310M}
{Morita} S, {Igarashi} R, {Zhao} H~H and {Kawashima} N 2018 {\em Phys. Rev.
  E\/} {\bf 97} 033310 (\textit{Preprint} \eprint{1712.01458})

\bibitem{wang2011cluster}
Wang L and Verstraete F 2011  (\textit{Preprint} \eprint{1110.4362})

\bibitem{PhysRevLett.113.046402}
{Corboz} P, {Rice} T~M and {Troyer} M 2014 {\em Phys. Rev. Lett.\/} {\bf 113}
  046402 (\textit{Preprint} \eprint{1402.2859})

\bibitem{PhysRevB.100.035449}
Iino S, Morita S and Kawashima N 2019 {\em Phys. Rev. B\/} {\bf 100} 035449
  (\textit{Preprint} \eprint{1905.02351})

\bibitem{PhysRevB.81.174411}
{Zhao} H~H, {Xie} Z~Y, {Chen} Q~N, {Wei} Z~C, {Cai} J~W and {Xiang} T 2010 {\em
  Phys. Rev. B\/} {\bf 81} 174411 (\textit{Preprint} \eprint{1002.1405})

\bibitem{2015arXiv150907484E}
{Evenbly} G 2017 {\em Phys. Rev. B\/} {\bf 95} 045117 (\textit{Preprint}
  \eprint{1509.07484})

\bibitem{doi:10.1143/JPSJ.65.891}
{Nishino} T and {Okunishi} K 1996 {\em Journal of the Physical Society of
  Japan\/} {\bf 65} 891 (\textit{Preprint} \eprint{cond-mat/9507087})

\bibitem{PhysRevB.78.155117}
{Or{\'u}s} R and {Vidal} G 2008 {\em Phys. Rev. B\/} {\bf 78} 155117
  (\textit{Preprint} \eprint{0711.3960})

\bibitem{Wolff:2020oky}
Wolff U 2020 {\em Nucl. Phys. B\/} {\bf 955} 115061 (\textit{Preprint}
  \eprint{2003.01579})

\bibitem{Gausterer:1995np}
Gausterer H and Lang C~B 1995 {\em Nucl. Phys. B\/} {\bf 455} 785--795
  (\textit{Preprint} \eprint{hep-lat/9506028})

\bibitem{Butt:2019uul}
Butt N, Catterall S, Meurice Y, Sakai R and Unmuth-Yockey J 2020 {\em Phys.
  Rev. D\/} {\bf 101} 094509 (\textit{Preprint} \eprint{1911.01285})

\bibitem{Gattringer:2015nea}
Gattringer C, Kloiber T and Sazonov V 2015 {\em Nucl. Phys. B\/} {\bf 897}
  732--748 (\textit{Preprint} \eprint{1502.05479})

\bibitem{Goschl:2017kml}
G\"oschl D, Gattringer C, Lehmann A and Weis C 2017 {\em Nucl. Phys. B\/} {\bf
  924} 63--85 (\textit{Preprint} \eprint{1708.00649})

\bibitem{eckart1936approximation}
Eckart C and Young G 1936 {\em Psychometrika\/} {\bf 1} 211--218

\bibitem{2022PhRvB.105f0402A}
{Adachi} D, {Okubo} T and {Todo} S 2022 {\em Phys. Rev. B\/} {\bf 105} L060402
  (\textit{Preprint} \eprint{2011.01679})

\bibitem{Nakayama:2021iyp}
Nakayama K, Funcke L, Jansen K, Kao Y~J and K\"uhn S 2022 {\em Phys. Rev. D\/}
  {\bf 105} 054507 (\textit{Preprint} \eprint{2107.14220})

\bibitem{d_adachi_phd}
Adachi D 2020 {\em {High-accuracy tensor renormalization group algorithms and
  their applications}\/} Ph.D. thesis The University of Tokyo

\bibitem{PhysRevLett.75.3537}
{{\"O}stlund} S and {Rommer} S 1995 {\em Phys. Rev. Lett.\/} {\bf 75}
  3537--3540 (\textit{Preprint} \eprint{cond-mat/9503107})

\bibitem{Dukelsky_1998}
{Dukelsky} J, {Mart{\'\i}n-Delgado} M~A, {Nishino} T and {Sierra} G 1998 {\em
  Europhysics Letters\/} {\bf 43} 457--462 (\textit{Preprint}
  \eprint{cond-mat/9710310})

\bibitem{PhysRevLett.91.147902}
{Vidal} G 2003 {\em Phys. Rev. Lett.\/} {\bf 91} 147902 (\textit{Preprint}
  \eprint{quant-ph/0301063})

\bibitem{Rossi:1984cv}
Rossi P and Wolff U 1984 {\em Nucl. Phys. B\/} {\bf 248} 105--122

\bibitem{Karsch:1988zx}
Karsch F and Mutter K~H 1989 {\em Nucl. Phys. B\/} {\bf 313} 541--559

\bibitem{Fromm:2010lga}
Fromm M 2010 {\em {Lattice QCD at strong coupling}\/} Ph.D. thesis Zurich, ETH

\bibitem{Milde:2021vln}
Milde P, Bloch J and Lohmayer R 2022 {\em PoS\/} {\bf LATTICE2021} 462
  (\textit{Preprint} \eprint{2112.01906})

\bibitem{Bloch:2022qyw}
Bloch J, Lohmayer R, Meister M and Nunhofer M 2023 {\em Nucl. Phys. B\/} {\bf
  987} 116107 (\textit{Preprint} \eprint{2210.02266})

\bibitem{Bloch:2021mjw}
Bloch J, Jha R~G, Lohmayer R and Meister M 2021 {\em Phys. Rev. D\/} {\bf 104}
  094517 (\textit{Preprint} \eprint{2105.08066})

\bibitem{Akiyama:2022asr}
Akiyama S 2022 {\em {Tensor renormalization group approach to
  higher-dimensional lattice field theories}\/} Ph.D. thesis University of
  Tsukuba

\bibitem{Adachi:2019paf}
Adachi D, Okubo T and Todo S 2020 {\em Phys. Rev. B\/} {\bf 102} 054432
  (\textit{Preprint} \eprint{1906.02007})

\bibitem{Oba:2019csk}
Oba H 2020 {\em PTEP\/} {\bf 2020} 013B02 (\textit{Preprint}
  \eprint{1908.07295})

\bibitem{Kadoh:2019kqk}
Kadoh D and Nakayama K 2019  (\textit{Preprint} \eprint{1912.02414})

\bibitem{Nakayama:2023ytr}
Nakayama K 2023  (\textit{Preprint} \eprint{2307.14191})

\bibitem{Bender:1992gn}
Bender I, Hashimoto T, Karsch F, Linke V, Nakamura A, Plewnia M, Stamatescu I~O
  and Wetzel W 1992 {\em Nucl. Phys. B Proc. Suppl.\/} {\bf 26} 323--325

\bibitem{Buballa:2003qv}
Buballa M 2005 {\em Phys. Rept.\/} {\bf 407} 205--376 (\textit{Preprint}
  \eprint{hep-ph/0402234})

\bibitem{Aoki:2017rjl}
Aoki K~I, Kumamoto S~I and Yamada M 2018 {\em Nucl. Phys. B\/} {\bf 931}
  105--131 (\textit{Preprint} \eprint{1705.03273})

\bibitem{Witten:1982df}
Witten E 1982 {\em Nucl. Phys. B\/} {\bf 202} 253

\bibitem{catterall2009}
Catterall S, Kaplan D~B and Unsal M 2009 {\em Phys. Rept.\/} {\bf 484} 71--130
  (\textit{Preprint} \eprint{0903.4881})

\bibitem{de2000multilinear}
De~Lathauwer L, De~Moor B and Vandewalle J 2000 {\em SIAM J. Matrix Anal.
  Appl.\/} {\bf 21} 1253--1278

\bibitem{de2000best}
De~Lathauwer L, De~Moor B and Vandewalle J 2000 {\em SIAM J. Matrix Anal.
  Appl.\/} {\bf 21} 1324--1342

\bibitem{vandenDoel:1983mf}
van~den Doel C and Smit J 1983 {\em Nucl. Phys. B\/} {\bf 228} 122--144

\bibitem{Lieb:1968zza}
Lieb E~H and Wu F~Y 1968 {\em Phys. Rev. Lett.\/} {\bf 20} 1445--1448

\bibitem{PhysRevLett.62.324}
Metzner W and Vollhardt D 1989 {\em Phys. Rev. Lett.\/} {\bf 62} 324--327

\bibitem{MllerHartmann1989CorrelatedFO}
M{\"u}ller-Hartmann E 1989 {\em Zeitschrift f{\"u}r Physik B Condensed
  Matter\/} {\bf 74} 507--512

\bibitem{homma2023nuclear}
Homma K and Kawashima N 2023  (\textit{Preprint} \eprint{2306.17479})

\bibitem{ferris2015unbiased}
Ferris A~J 2015  (\textit{Preprint} \eprint{1507.00767})

\bibitem{Huggins:2017yzz}
Huggins W, Freeman C~D, Stoudenmire M, Tubman N~M and Whaley K~B 2017
  (\textit{Preprint} \eprint{1710.03757})

\bibitem{Arai:2022uee}
Arai E, Ohki H, Takeda S and Tomii M 2023 {\em Phys. Rev. D\/} {\bf 107} 114515
  (\textit{Preprint} \eprint{2211.13107})

\bibitem{PhysRevX.9.031041}
{Liao} H~J, {Liu} J~G, {Wang} L and {Xiang} T 2019 {\em Phys. Rev. X\/} {\bf 9}
  031041 (\textit{Preprint} \eprint{1903.09650})

\bibitem{PhysRevB.101.220409}
{Chen} B~B, {Gao} Y, {Guo} Y~B, {Liu} Y, {Zhao} H~H, {Liao} H~J, {Wang} L,
  {Xiang} T, {Li} W and {Xie} Z~Y 2020 {\em Phys. Rev. B\/} {\bf 101} 220409
  (\textit{Preprint} \eprint{1912.02780})

\bibitem{Jha:2023bpn}
Jha R~G and Samlodia A 2024 {\em Comput. Phys. Commun.\/} {\bf 294} 108941
  (\textit{Preprint} \eprint{2306.00358})

\end{thebibliography}

\providecommand{\newblock}{}

\end{document}